\documentclass[12pt]{article}
\usepackage[a4paper, bindingoffset=0cm, hmargin={2.5cm, 2cm},vmargin={2.5cm, 2.5cm}]{geometry}

\usepackage{graphicx}
\usepackage[english]{babel}
\usepackage[utf8]{inputenc}

\usepackage[T1]{fontenc}
\usepackage{lmodern}
\usepackage[bookmarks]{hyperref}
\usepackage{multirow}
\usepackage{float}
\usepackage{appendix}

\usepackage{amsmath}
\usepackage{amssymb}
\usepackage{amsthm}
\usepackage{amsfonts}
\usepackage{mathtools}
\usepackage{dsfont}
\usepackage{braket}

\usepackage{tikz}
\usetikzlibrary{positioning}
\usetikzlibrary{patterns}
\usetikzlibrary{decorations.markings}
\usetikzlibrary{decorations.pathreplacing}
\usetikzlibrary{decorations.text}
\usetikzlibrary{perspective}
\usetikzlibrary{topaths}
\usetikzlibrary{shapes.misc}

\tikzset{->-/.style={decoration={
			markings,
			mark=at position .5 with {\arrow{>}}},postaction={decorate}}}

\tikzset{-<-/.style={decoration={
			markings,
			mark=at position .5 with {\arrow{<}}},postaction={decorate}}}

\tikzset{
	on each segment/.style={
		decorate,
		decoration={
			show path construction,
			moveto code={},
			lineto code={
				\path [#1]
				(\tikzinputsegmentfirst) -- (\tikzinputsegmentlast);
			},
			curveto code={
				\path [#1] (\tikzinputsegmentfirst)
				.. controls
				(\tikzinputsegmentsupporta) and (\tikzinputsegmentsupportb)
				..
				(\tikzinputsegmentlast);
			},
			closepath code={
				\path [#1]
				(\tikzinputsegmentfirst) -- (\tikzinputsegmentlast);
			},
		},
	},
	mid arrow/.style={postaction={decorate,decoration={
				markings,
				mark=at position .5 with {\arrow[#1]{latex}}
	}}}
}

\usepackage{color}
\definecolor{MPGgreen}{RGB}{0,108,102}
\definecolor{MPGgreenlight}{RGB}{198,211,37}
\definecolor{MPGgreendark}{RGB}{0,85,85}
\definecolor{MPGorange}{RGB}{239,124,0}
\definecolor{MPGbluelight}{RGB}{0,177,234}
\definecolor{MPGbluedark}{RGB}{41,72,93}

\let\originalleft\left
\let\originalright\right
\renewcommand{\left}{\mathopen{}\mathclose\bgroup\originalleft}
\renewcommand{\right}{\aftergroup\egroup\originalright}

\DeclareMathOperator{\KN}{KN}
\newcommand{\h}{\mathrel{\phantom{=}}}
\newcommand{\n}{\nonumber\\}

\makeatletter
\newsavebox{\@brx}
\newcommand{\llangle}[1][]{\savebox{\@brx}{\(\m@th{#1\langle}\)}%
	\mathopen{\copy\@brx\mkern2mu\kern-0.9\wd\@brx\usebox{\@brx}}}
\newcommand{\rrangle}[1][]{\savebox{\@brx}{\(\m@th{#1\rangle}\)}%
	\mathclose{\copy\@brx\mkern2mu\kern-0.9\wd\@brx\usebox{\@brx}}}
\makeatother

\allowdisplaybreaks

\newcommand{\req}[1]{(\ref{#1})}

\def\fc#1#2{\frac{#1}{#2}}

\newcommand{\nwc}{\newcommand}
\nwc{\ba}  {\begin{array}}
\nwc{\ea}  {\end{array}}
\nwc{\bdm} {\begin{displaymath}}
\nwc{\edm} {\end{displaymath}}

\nwc{\bea} {\begin{equation}\ba{lcl}}
\nwc{\eea} {\ea\end{equation}}

\nwc{\be} {\begin{equation}}
\nwc{\ee} {\end{equation}}

\nwc{\bda} {\bdm\ba{lcl}}
\nwc{\eda} {\ea\edm}

\nwc{\bc}  {\begin{center}}
\nwc{\ec}  {\end{center}}

\nwc{\ds}  {\displaystyle}

\nwc{\nn} {\nonumber}
\nwc{\nnn} {\nonumber \vspace{.2cm} \\ }
\nwc{\ra}{\rightarrow}
\nwc{\lra}{\longrightarrow}
\def\lf{\left}\def\ri{\right}

\nwc{\p} {\partial}

\def\ap{\alpha'}

\def\Ic{{\cal I}}

\def\eps{\epsilon}
\def\z{\zeta}

\def\Oc{{\cal O}}

\def\eps{\epsilon}

\def\si{\sigma}
\def\Om{\Omega}

\numberwithin{equation}{section}

\begin{document}
	
	\pagestyle{empty}
	
	\begin{center}
		{\bf\LARGE
			Superstring scattering \\[3mm] on the real projective plane}
		
		\vspace{1cm}
		
		{\large\sc
			{ Andreas Bischof\ and\ Stephan Stieberger}
			
			\vspace{1cm}

			{\it\small
				Max-Planck Institut f\"ur Physik,\\Werner--Heisenberg--Institut, 
85748 Garching, Germany
			}
		}
		
	\end{center}
	\vspace{1.5cm}
	
	\begin{center}
		{\bf Abstract}\\
		\end{center}
Superstring scattering from orientifold planes requires considering   string amplitudes  on world--sheets with crosscaps with the lowest order case (in string coupling constant)  having the topology of the real projective plane. While amplitudes on the latter have been formulated  for the trivial one-- and two--point cases in this work we go beyond these cases thereby solving various technicalities. The latter include reducing the complex world--sheet integration of closed string insertions over the real projective plane to pure real open string integrals. As a result we find that scattering of $n$ closed strings on the  real projective plane can be expressed in terms of disk amplitudes involving $2n$ open strings. In this work 
we explicitly work out in pure spinor formalism the case $n=3$ which can be written as a linear combination of two   (gauge--invariant) six open string amplitudes. We also present the low--energy expansion of this result necessary to construct closed string couplings on orientifold planes.

	\vspace{1cm}
	\begin{flushright}
{\small MPP--2025--1}
\end{flushright}
	
	\clearpage
	\pagestyle{plain}
	\tableofcontents
	\break
	
	\section{Introduction}

In addition to fundamental strings (consistent) superstring theories also possess extended objects like D$p$--branes and orientifold O$p$--planes, cf.\ e.g.\ \cite{Polchinski:1996na,Angelantonj:2002ct}. In fact, four--dimensional string compactifications with D-branes, orientifolds and fluxes play a major role in finding realistic string vacua in four space--time dimensions \cite{Blumenhagen:2006ci}. Interactions of open strings with 
D$p$--branes are described by world--sheets with boundaries. While the dynamics of D$p$--branes is described by open strings ending on them, orientifold O$p$--planes are localized at space--time fixed points and are non--dynamical objects. The latter are required for consistency\footnote{While the amplitude for a single closed string on the sphere is zero, there is a non--vanishing amplitude  for a single closed string (with zero momentum) to appear from the
vacuum, either through the disk or the projective plane producing the potentials of different signs.}, with their negative tension to cancel the positive tension of D$p$--branes. Furthermore they describe the world--sheet orientation reversal of closed oriented\footnote{In the unoriented
theory, strings are invariant under the action of world--sheet parity.} strings. More precisely, the action of the world--sheet parity reversal is the combined action of a world--sheet and space-time parity operation with the latter producing space--time $\mathbb Z_2$ orbifold fixed points (planes)   giving rise to the orientifold planes. Interactions of closed strings with 
O$p$--planes are described by world--sheets with a crosscap instead of the boundary for the open string case. The crosscap is attached to the surface at the position of the O$p$--plane.

Perturbative closed string scattering is formulated by compact two--dimensional Riemann surfaces without boundaries, while
open string scattering  is described by compact Riemann surfaces with boundaries. The latter may be obtained from the former by involution. Furthermore, in the case of unoriented (oriented) strings we deal with unoriented (oriented) surfaces. Since an arbitrary two--dimensional surface $M$ is topologically equivalent to a
sphere with $b$ holes, $c$ cross--caps, and $g$ handles these parameters specify the  underlying perturbative order in the string coupling given by 
$g_{\rm string}^{-\chi}$,
with the Euler number $\chi$
\be
\chi= 2-b-c-2g\ ,
\ee
 of the Riemann manifold $M$ and $g_{\rm string}=e^{\Phi}$ the string coupling constant determined by the vacuum expectation value of an universal 
massless scalar mode (dilaton) $\Phi$ of closed strings. There are three Riemann surfaces of positive Euler number $\chi$: the sphere,
the disk, and the projective plane. The latter is a disk with opposite points identified.
At the order $g_{\rm string}^{-1}$, i.e.\ $\chi=1$ we have two surfaces, namely the real projective plane (with $b=0,c=1$) and the disk (with $b=1,c=0$). While the  real projective plane is non--orientable, the disk is an oriented world--sheet.

For amplitude computations on the disk involving both open and closed strings there are by now many results in the literature albeit only recently pure closed string scattering on the disk beyond two--point level has been elaborated in full generality \cite{Bischof:2023uor}.
The same is not true for amplitude computations on the real projective plane involving  closed strings. In fact, only the trivial cases of one and two closed strings on the real projective plane have been  discussed so far by conformal field theory methods \cite{Garousi:2006zh,Aldi:2020dvw}. The purpose of this work is to go beyond the two--point case and overcome 
the various technicalities. The latter include discussing vertex operators, which are invariant under the orientifold action, tackling the complex world--sheet integration of closed string insertions over the real projective plane and reducing them to pure real open string integrals. We use the pure spinor formalism for our closed string three--point disk amplitude computation. Thus, a large part of our results holds for massless NSNS, RR, RNS and NSR states (using the language of the RNS formalism), even though at some point we will specify our findings to the scattering of only NSNS states. Apart from the broad validity of its results, the pure spinor formalism offers further advantages.  The pure spinor formalism has proven to be very powerful for  open string disk amplitudes, as its BRST cohomology structure allows to obtain very compact expressions \cite{6pt,npt_1,npt_2}, for a review cf.\ \cite{Mafra:2022wml}.
 As a consequence there is an extended literature on open string disk amplitudes in the pure spinor formalism that we can build on. 
Via contour deformations of the corresponding world--sheet integrals disk amplitudes involving $n$ closed and $m$ open strings can be reduced to pure open string amplitudes involving $2n+m$ open strings only \cite{Stieberger:2009hq,Stieberger:2015vya,Stieberger:2016lng,Stieberger:2015qja,Stieberger:2014cea}. These manipulations are generalizations of the famous KLT relations on the sphere \cite{KLT}  by also taking into account mixings of left-- and right--movers.
In particular,  we shall make use of the relation between closed string $n$--point disk amplitudes  and open string $2n$--point disk amplitudes.

Our work is structured  as follows: In Section \ref{sec::psf} we give a short introduction into the world--sheet degrees of freedom of the pure spinor formalism and review the massless vertex operators of closed string states on the disk. In Section \ref{sec::open_string_correlator} we review the computation of open string amplitudes on the disk. We analyze how these amplitudes decompose into a minimal basis of kinematic building blocks, expressed through Berends--Giele supercurrents and hypergeometric integrals. 
In Section \ref{sec::rp2_doubling_trick} we propose a scattering amplitude prescription for the real projective plane. We introduce unintegrated and integrated vertex operators, which are invariant under the combined action of world--sheet parity and a space--time orbifold (orientifold). We  are able to reduce the scattering amplitude of $n$ closed strings 
on the real projective plane to a scattering  amplitude on the complex sphere.
In Section \ref{chap::plane} we systematically  analyse our  amplitude prescription for the real projective plane by analytic continuing the complex coordinates  to real coordinates describing open string 
insertions. In the end we are able to write the scattering amplitude of three closed strings 
on the real projective plane as linear combination of disk amplitudes involving six open strings.
In Section \ref{alphaprime_expand} we compute the low energy expansion in the inverse string tension of the  scattering amplitude of three closed strings on the real projective plane.
Furthermore, in Appendix \ref{sec::KLT} we derive monodromy relations which are used to relate open string amplitudes with different color orderings to reduce the number of independent components.

	\section{The Pure Spinor Formalism} \label{sec::psf} This section introduces the pure spinor formalism, outlining its world-sheet degrees of freedom and the calculation of closed string tree-level amplitudes on the disk $D_2$. We closely follow the introduction provided in \cite{2pt}.


	\subsection{Matter and Ghost CFT of the Pure Spinor Formalism} \label{sec::pdf_matter} In the pure spinor formalism, the type IIB action of the world-sheet degrees of freedom is given by: \begin{equation} 
		S=\frac{1}{2\pi}\int\mathrm d^2z\,\left(\frac12\partial X^m\overline\partial X_m+p_\alpha\overline\partial\theta^\alpha+\overline p_\alpha\partial\overline\theta^\alpha - w_\alpha\overline\partial\lambda^\alpha-\overline w_\alpha\partial\overline\lambda^\alpha\right)\ ,\label{eq::psfaction2} 
	\end{equation}
where $X^m(z,\overline z),\theta^\alpha(z),p_\alpha(z)$ and $\overline\theta^\alpha(\overline z), \overline p_\alpha(\overline z)$ are the matter variables \cite{GS_cov,Siegel:1985xj} and $\lambda^\alpha(z), w_\alpha(z)$ and $\overline\lambda^\alpha(\overline z), \overline w_\alpha(\overline z)$ are the pure spinor ghosts introduced by Berkovits \cite{Berkovits}. To ensure the theory has vanishing central charge in $D=10$ space-time dimensions, the ghost field $\lambda^\alpha$, which is a bosonic spinor, has to obey the pure spinor constraint: 
	\begin{equation} 
		\lambda^\alpha\gamma_{\alpha\beta}^m\lambda^\beta=0\ ,\qquad m=0,\ldots,9\ ,\quad \alpha,\beta=1,\ldots,16\ ,\label{eq::ps_const} 
	\end{equation}
where $\gamma^m_{\alpha\beta}$ are the symmetric $16\times16$ Pauli matrices in $D=10$. For type IIA the right-moving spinorial fields have opposite chirality.\par It is convenient to introduce the supersymmetric momentum and the Green-Schwarz constraint: 
	\begin{align} 
		\Pi^m&=\partial X^m+\frac 1 2 (\theta\gamma^m\partial\theta)\ , \nonumber \\ 
		d_\alpha&=p_\alpha-\frac 1 2 \left(\partial X^m+\frac 1 4 (\theta\gamma^m\partial\theta)\right)(\gamma_m\theta)_\alpha \ , \label{eq::composite_fields} 
	\end{align}
which are conformal primaries of weight $h=1$ that appear in the vertex operators of massless fields and thus play an important role in the calculation of scattering amplitudes in the pure spinor formalism. Since $\lambda^{\alpha}$ is a commuting $SO(1,9)$ Weyl spinor, there is an additional $h=1$ field $N^{mn}(z)=\frac12 (\lambda\gamma^{mn}w)$, which arises form the ghost contribution to the Lorentz current.\par Furthermore, in the pure spinor formalism the holomorphic part of the energy momentum tensor $T$ can be written as 
	\begin{equation} 
		T(z)=-\frac12\partial X^m\partial X_m-p_\alpha\partial\theta^\alpha+w_\alpha\partial\lambda^\alpha \ , \label{eq::psfemt} 
	\end{equation}
with a similar expression for the anti-holomorphic energy momentum tensor.\footnote{All formulas in this subsection have an obvious antiholomorphic counterpart.}\par For the computation of scattering amplitudes, we will need the following OPEs between the fields in the action \eqref{eq::psfaction2} and the composite fields \eqref{eq::composite_fields}\cite{Berkovits, Berkovits:2002zk}: 
	\begin{gather} 
		\begin{align*} 
			X^m(z,\overline z)X^n(w,\overline w)&=-\eta^{mn}\ln{|z-w|^2}\ , & p_\alpha(z)\theta^\beta(w) & = \frac{\delta_\alpha^{\hphantom\alpha\beta}}{z-w}\ , \\ \Pi^m(z)\Pi^n(w) & = \frac{-\eta^{mn}}{(z-w)^2}\ , & d_\alpha(z) d_\beta(w) & = -\frac{\gamma^m_{\alpha\beta}\Pi_m(w)}{z-w}\ , \\ d_\alpha(z)\Pi^m(w) & = \frac{(\gamma^m\partial\theta)_{\alpha}(w)}{z-w}\ , & d_\alpha(z)\theta^\beta(w) & = \frac{\delta^{\hphantom\alpha\beta}_\alpha}{z-w}\ , 
		\end{align*}\\ 
		N^{mn}(z)N^{pq}(w)= 2 \frac{\eta^{p[n}N^{m]q}(w)-\eta^{q[n}N^{m]p}(w)}{z-w}-6\frac{\eta^{m[q}\eta^{p]n}}{(z-w)^2}\ . \label{psfOPE} 
	\end{gather} 
We can obtain the physical spectrum of the pure spinor superstring from the cohomology of the BRST operator, which takes the simple form \cite{Berkovits}: 
	\begin{equation} 
		Q=\oint\frac{\mathrm{d}z}{2\pi i}\,\lambda^\alpha(z) d_\alpha(z)\ . \label{BRST} 
	\end{equation}
The BRST charge is nilpotent $Q^2=0$ as can be verified using the OPE of $d_\alpha$ with $d_\beta$ in \eqref{psfOPE} and the fact that the ghost field $\lambda$ satisfies the pure spinor constraint \eqref{eq::ps_const}. We can write the action of the conformal weight one fields $\Pi^m$ and $d_\alpha$ on a generic superfield $\mathcal V$ as: 
	\begin{align} 
		d_\alpha(z) \mathcal V (X(w),\theta(w))&=\frac{D_\alpha \mathcal V(X(w),\theta(w))}{z-w}\ ,\nonumber\\
		\Pi^m(z)\mathcal V(X(w),\theta(w))&=-\frac{ik^m\mathcal V(X(w),\theta(w))}{z-w}\ , 
	\end{align}
where $D_\alpha=\partial_\alpha+\frac12(\gamma^m\theta)_\alpha\partial_m$. Therefore, the BRST charge acts on a superfield $\mathcal V(X,\theta)$ as $Q\mathcal V=\lambda^\alpha D_\alpha\mathcal V$.


\subsection{Massless Vertex Operators in the Pure Spinor Formalism}
\label{sec_Vops}

In string theory, a scattering process can be described by a punctured Riemann surface, where each puncture represents a vertex operator position corresponding to the creation or annihilation of a string state. In this section, we focus on the scattering amplitudes of closed strings on the disk $D_2$, which can be mapped to the upper half plane $\mathbb H_+$ by a conformal transformation. \par
We are interested in the scattering amplitudes involving only massless closed string excitations, which are subject to:
\begin{equation}
	k^2=0\ ,
\end{equation}
where $k$ is the momentum of the string state. These massless modes are described by the space-time superfields  $A_\alpha, A_m, W^\alpha$ and $F_{mn}$ (the superfields of super-Maxwell theory) living on superspace spanned by $X$ and $\theta$. The super Yang-Mills (SYM) fields $A_m, W^\alpha$ and $F_{mn}$ are not independent: Rather, they are the field strengths given by \cite{Berkovits:2002zk}
\begin{align}
	A_m & =  \frac18\gamma^{\alpha\beta}_mD_\alpha A_\beta \ ,\nonumber\\
	W^\alpha & = \frac1{10}\gamma_m^{\alpha\beta}(D_\beta A^m-\partial^mA_\beta)\ ,\nonumber\\
	F_{mn} & =  \partial_mA_n-\partial_nA_m\ .
\end{align}
The equations of motion of the superfields are given by the following set of equations \cite{Berkovits:2002zk,Witten:1985nt}
\begin{align}
	2 D_{(\alpha} A_{\beta)} & =  \gamma^m_{\alpha \beta} A_m\ , &  D_\alpha A_m & =  (\gamma_m W)_\alpha + \partial_m A_\alpha\ , \nonumber\\
	D_\alpha {\cal F}_{mn} & =  2 \partial_{[m} ( \gamma_{n]} W)_\alpha\ , &  D_\alpha W^\beta & =  \frac14 (\gamma^{mn})_\alpha \, \! ^\beta  {\cal F}_{mn} \ .	\label{eq::eoms_superfields}
\end{align}
Moreover, each of the superfields is a functional of the superspace coordinates $X$ and $\theta$. The $X^m$-dependence of the superfields can be organized into plane waves with momentum $k^m$ and $\theta^\alpha$-dependence is given by a power series in $\theta$. 
Using the gauge choice $\theta^\alpha A_\alpha(X,\theta)=0$, the expansions can be found for instance in \cite{Schlotterer:2011psa}. For example, for the bosonic space-time degrees of freedom (of a vector field with polarization vector $e_m$) it takes the following form:\footnote{The expansion in \cite{Schlotterer:2011psa} is more general including also fermionic space-time degrees of freedom. Moreover, the momenta in \eqref{expansion} are real, i.e$.$ they differ from the corresponding momenta of \cite{Schlotterer:2011psa} by a factor of $i$.}
\begin{align}
	A_{\alpha}(X,\theta)&=e^{ik\cdot X}\left\{\frac{e_m}{2}(\gamma^m\theta)_\alpha-\frac{1}{16}(\gamma_p\theta)_\alpha(\theta\gamma^{mnp}\theta)i k_{[m}e_{n]} +\mathcal O(\theta^5)\right\}\ ,\nonumber\\
	A_m(X,\theta)&=e^{ik\cdot X}\left\{e_m-\frac14 ik_p(\theta\gamma_m^{\hphantom{m}pq}\theta)e_q +\mathcal O(\theta^4)\right\} \ ,\nonumber\\
	W^\alpha(X,\theta)&=e^{ik\cdot X}\left\{-\frac12ik_{[m}e_{n]}(\gamma^{mn}\theta)^\alpha +\mathcal O(\theta^3)\right\}\ ,\nonumber\\
	F_{mn}(X,\theta)&=e^{ik\cdot X}\left\{2ik_{[m}e_{n]}-\frac12ik_{[p}e_{q]}ik_{[m}(\theta\gamma_{n]}^{\hphantom{n]}pq}\theta) +\mathcal O(\theta^4)\right\}\ .
	\label{expansion}
\end{align}
It is sufficient to display the expansions only up to a certain order in $\theta$ that is relevant for the computation of a scattering amplitude. All higher orders will not contribute, because they drop out due to the zero mode prescription of the pure spinor formalism.\par
The corresponding closed string Hilbert space is a tensor product of two open string Hilbert spaces
\begin{equation}
	H_{\text{closed}}=H_{\text{open}}\otimes H_{\text{open}}\ .\label{eq::hilbert_space_factorization}
\end{equation}
Immediately one can imagine that this property carries over to the vertex operators describing the states in the Hilbert space $H_{\text{closed}}$. Indeed, using the operator state correspondence and the factorisation of the Hilbert space \eqref{eq::hilbert_space_factorization} the closed string vertex operators $V(z,\overline z)=V(z)\otimes \overline V_{i}(\overline z)$ and $U(z,\overline z)=U(z)\otimes \overline U_{i}(\overline z)$ are double copies of open string vertex operators. Hence, the vertex operator for such a closed string state at position $z$ and $\overline z$ splits into a direct product of left- and right-moving open string vertex operators:
\begin{align}
	V(z,\overline z)&=|\lambda^\alpha A_{\alpha}|^2(z,\overline z)\ ,\n
	U_{i}(z,\overline z)&=\left|\partial\theta^\alpha A_\alpha+\Pi^mA_m+d_\alpha W^\alpha+{\textstyle \frac12}N^{mn}F_{mn}(z,\overline z)\right|^2(z,\overline z)\ ,\label{eq::closed_string_vertex_operators}
\end{align}
where we used the notation $|\lambda^\alpha A_{\alpha}|^2(z,\overline z)=\big(\lambda^\alpha A_{\alpha}(\theta)\big)(z)\big(\overline\lambda^{\hat\alpha} \overline A_{\hat\alpha}(\overline\theta)\big)(\overline z)e^{ik\cdot X(z,\overline z)}$.\footnote{Here, we have used \eqref{eq::SYM_field_splitting} to separate the plane wave factor from the $\theta$-dependent part of the superfields. Alternatively, we could separate $X^m(z,\overline z)=X^m(z)+\overline X^m(\overline z)$ into left- and right movers. This would imply that the plane wave factor for the left-moving part of a vertex operator only depends holomorphically on $z$ via $X^m=X^m(z)$ in \eqref{expansion}. Nevertheless, the full closed string vertex operator \eqref{eq::closed_string_vertex_operators} contains the plane wave factor $e^{ik\cdot X(z,\overline z)}$, because the plane wave factors of the holomorphic and antiholomorphic sector in \eqref{eq::closed_string_vertex_operators} can be combined into $e^{ik\cdot X(z,\overline z)}$ \cite{Bischof:2023uor}.} Overlined SYM or world-sheet fields are the antiholomorphic counterparts of the corresponding holomorphic fields, whose spinorial indices with hats $\hat\alpha,\hat\beta,\ldots$ have the opposite (same) chirality as $\alpha,\beta,\ldots$ for type IIA (type IIB) superstring theory. Moreover, we have used that the superfields and therefore also vertex operators can be split as follows
	\begin{equation}
		\mathcal{V}(X,\theta)=\mathcal{V}(\theta)e^{ik{\cdot}X}\label{eq::SYM_field_splitting}
	\end{equation}
to emphasise that the closed string vertex operator depends on a plane wave factor  $e^{ik\cdot X(z,\overline z)}$. The $\overline \theta$-expansions for the antiholomorphic fields are as in $\eqref{expansion}$ but with independent polarizations $\overline e^m,\overline\chi^\alpha$ instead of $e^m,\chi^\alpha$, whose tensor product forms the closed string polarization. For example, the purely bosonic (NSNS) closed string polarization tensor is given by $\epsilon^{mn}=e^m\otimes \overline e^n$.\par
The vertex operator $V$ is BRST closed, which means that
\begin{equation}
	Q V = 0\ .
	\label{QV0}
\end{equation}
This is equivalent to putting the superfield $A_\alpha$ on-shell. Moreover, the vertex operator $U$ is BRST exact and therefore fulfils 
\begin{equation}
	QU=\partial V\ .
	\label{QV1}
\end{equation}
Hence, it is in the BRST cohomology once we integrate it over the world-sheet. Thus, we call $V$ and $U$ the unintegrated and integrated vertex operator, respectively.\footnote{In the literature the unintegrated and integrated vertex operator are often denoted by $V$ and $U$, respectively, but in \cite{2pt} a different notation was introduced. The closed string vertex operator can simultaneously contain an integrated and unintegrated factor after gauge fixing, e.g.\ $V\otimes \overline U$ or $U\otimes \overline V$. \label{VUV0V1}} Analogous statements hold for the right-moving part of \eqref{eq::closed_string_vertex_operators}.\par
Due to the boundary of the disk the left- and right-moving part of a closed string vertex operator are not independent any more, i.e$.$  the boundary of the disk imposes an interaction between the holomorphic and antiholomorphic fields. For the computation of the three-point function of closed strings on the disk the doubling trick was used to rewrite the right-moving part of the vertex operator \eqref{eq::closed_string_vertex_operators} in order to allow for a unified treatment of the left- and right-movers \cite{Bischof:2023uor}.
Concretely, in the following we consider type IIB theory in a flat ten dimensional space-time, which contains a D$p$-brane that is spanned in the $X_1\times X_2\times\ldots\times X_p$ plane. As usual, we use the fact that the D-brane is infinitely heavy in the small coupling regime, i.e$.$ it can absorb an arbitrary amount of momentum in the $X_{p+1},\ldots,X_9$ directions transversal to the D-brane. Thus, momentum is effectively only conserved along the D-brane in the perturbative regime that we are working in. 

Left- and right-movers separately have the standard correlators on the upper half plane 
\begin{align}
		\langle X^m(z)X^n(w)\rangle&=-\eta^{mn}\ln(z-w)\ ,\nonumber\\
		\langle p_{\alpha}(z)\theta^\beta(w)\rangle&=\frac{\delta_\alpha^{\hphantom{\alpha}\beta}}{z-w}\ ,\nonumber\\
		\langle w_{\alpha}(z)\lambda^\beta(w)\rangle&=\frac{\delta_\alpha^{\hphantom{\alpha}\beta}}{z-w}\ ,
	\label{eq::correlators}
\end{align}
where the antiholomorphic part is analogous. At the boundary of $\mathbb H_+$, i.e$.$ at the real axis, the first $p+1$ components of the world-sheet fields satisfy Neumann  boundary conditions and the remaining $9-p$ components Dirichlet boundary conditions. Both of these boundary conditions impose non-vanishing correlators between the holomorphic and antiholomorphic parts of the fields. We can simplify the calculations by employing the doubling trick, i.e$.$ we replace the right moving space-time vectors and space-time spinors by
\begin{align}
		\text{vectors: }&\overline X^m(\overline z)=D^m_{\hphantom mn}X^n(\overline z)\ ,\nonumber\\ 
		\text{spinors: }&\overline\Psi^\alpha(\overline z) = M^\alpha_{\hphantom{\alpha}\beta}\Psi^\beta(\overline z) \quad \text{ or } \quad \overline{\Psi}_\alpha(\overline z) = ((M^T)^{-1}) _{\alpha}^{\hphantom{\alpha}\beta}\Psi_\beta(\overline z)\ ,\label{eq::replace}
\end{align}
with $\Psi^\alpha \in \{ \theta^\alpha, \lambda^\alpha\}$ and $\Psi_\alpha \in \{ p_\alpha, w_\alpha\}$ and constant matrices $D$ and $M$.\footnote{A priori, one would introduce two matrices $M$ and $N$ that account for the boundary conditions of the fermions, one for each chirality. Here we already used the result of \cite{2pt} that $N=(M^T)^{-1}$.} Effectively, this corresponds to extending the world-sheet fields to the entire complex plane. This allows us to use only the correlators in \eqref{eq::correlators}, leading to
\begin{gather}
	\langle X^m(z)\overline X^n(\overline w)\rangle=-D^{mn}\ln(z-\overline w)\ ,\nonumber\\
	\begin{align}
		\langle p_{\alpha}(z)\overline\theta^\beta(\overline w)\rangle&=\frac{M^{\beta}_{\hphantom{\beta} \alpha}}{z-\overline w}\ ,&\langle \overline 	p_{\alpha}(\overline z)\theta^\beta(w)\rangle&=\frac{((M^T)^{-1})_\alpha^{\hphantom{\alpha}\beta}}{\overline z-w}\ ,\nonumber\\
		\langle w_\alpha(z)\overline \lambda^\beta(\overline w)\rangle&=\frac{M^{\beta}_{\hphantom{\beta} \alpha}}{z-\overline w}\ ,&\langle \overline 	w_\alpha(\overline z)\lambda^\beta(w)\rangle&=\frac{((M^T)^{-1})_\alpha^{\hphantom\alpha\beta}}{\overline z-w}\ .\label{eq::correlator2ptdisk}
	\end{align}
\end{gather}
The matrix
\begin{align}
	D^{mn}=\left\{\begin{matrix} \eta^{mn} \quad&m,n\in\{0,1,\ldots,p\} \\
		-\eta^{mn}\quad&m,n\in\{p+1,\ldots,9\} \\
		0 \quad& {\rm otherwise} \end{matrix}\right. 
	\label{eq::boundary_matrix}
\end{align} 
is the same as in the RNS formalism \cite{Garousi:1996ad,Hashimoto:1996bf}, but the matrix $M$ here  differs from the RNS version due to the different spinor representations of the RNS and the PS formalisms. A detailed discussion of the matrix $M$ and its properties can be found in \cite{2pt} and a similar analysis in the context of the RNS formalism can be found for instance in \cite{Garousi:1996ad}. 
As described above, only the momentum parallel to the brane is conserved, because D-branes are infinitely heavy objects, that can absorb momentum in the transverse direction. We can introduce a parallel and transverse momentum
\begin{equation}
	{k_i}_\parallel=\frac12(k_i+D{\cdot}k_i)\ ,\qquad	{k_i}_\perp=\frac12(k_i-D{\cdot}k_i)\ ,
\end{equation}
so that for momentum conservation we have
\begin{equation}
	\sum_{i=1}^n {k_i}_\parallel=0\ .\label{eq::mom_con}
\end{equation}
Making the replacements \eqref{eq::replace} in the right-moving parts of the vertex operators \eqref{eq::closed_string_vertex_operators} one can show that \cite{2pt,Bischof:2024vpd}
\begin{align}
	\overline V(\overline z)&=\biggl[ \lambda^\alpha A_\alpha[D{\cdot}\overline e,M^{-1}\overline\chi,D{\cdot}k](X,\theta)\biggl](\overline z)\ ,
	\label{eq::redefinedvertex1}\\
	\overline U(\overline z)&=\biggl[ \overline\partial\theta^\alpha A_\alpha[D{\cdot}\overline e,M^{-1}\overline\chi,D{\cdot}k](X,\theta)+\Pi^mA_m[D{\cdot}\overline e,M^{-1}\overline\chi,D{\cdot}k](X,\theta)\nonumber\\
	&\h + d_\alpha W^\alpha[D{\cdot}\overline e,M^{-1}\overline\chi,D{\cdot}k](X,\theta)+\frac12N^{mn}{\mathcal F}_{mn}[D{\cdot}\overline e,M^{-1}\overline\chi,D{\cdot}k](X,\theta)\biggl](\overline z)\ ,
	\label{eq::redefinedvertex2}
\end{align}
i.e.\ the doubling trick amounts to substituting every antiholomorphic superfield by its holomorphic counterpart and simultaneously multiplying the polarisation vector and momentum with $D$ and the polarization spinors with $M^{-1}$ to account for the boundary conditions.
To simplify the notation we will drop the explicit dependence of the superfields on the polarisation and momentum, but instead introduce the following notation
\begin{equation}
	\mathcal{V}_{\overline \imath}(\overline z)\equiv\overline{\mathcal{V}}_i[\overline e_i,\overline\chi_i,k_i](\overline X(\overline z),\overline \theta(\overline z))=\mathcal{V}_i[D{\cdot}\overline e_i,M^{-1}\overline\chi_i,D{\cdot}k_i](X(\overline z),\theta(\overline z)) \ ,\label{eq::doubling_trick_superfield}
\end{equation}
where ${\mathcal{V}} \in \{A_\alpha, A_m, W^\alpha, {F}_{mn}\}$ and $i$ denotes the label of an external string state. Moreover, we will also use this notation for the vertex operators. To summarize this: An overlined label indicates that the field or vertex operator originates from the right-moving part of a string state after employing the doubling trick \eqref{eq::replace}.

	\section{Open string amplitudes on the disk}\label{sec::open_string_correlator}

The prescription to calculate open string amplitudes on the disk is well known and tested in the pure spinor formalism \cite{5pt,Mafra:2010gj,Mafra:2011nv} and therefore also closed string amplitudes on the sphere are straightforwardly calculated using the KLT relations \cite{KLT}. Both world-sheets have no moduli and therefore we only have to take care of the residual symmetry of the conformal Killing group (CKG) of the world-sheet topology. Because we have three (six) conformal Killing vectors (CKV) on the disk (sphere), we have to fix the position of three (six) real world-sheet positions \cite{BLT}. Fixing the reparametrization invariance of the Möbius group of the world-sheet leads to the insertion of unintegrated vertex operators at these positions, while the other vertex operator are integrated over. It is a convenient choice to fix the vertex operators $i=1,n-1$ and $n$, where $n$ is the number of external states, to some arbitrary positions $z_1, z_{n-1}$ and $z_n$. In principle, the amplitude is independent of the assignment of the integrated and unintegrated vertex operators. For $n$ massless open strings we find that the scattering amplitude prescription is given by the following correlation function of vertex operators\footnote{As discussed in \cite{Mafra:2022wml} the computation of gauge theory scattering amplitudes simplifies tremendously when considering ordered gauge invariants that depend only on kinematics \cite{Berends:1987cv,Mangano:1987xk}, i.e.\ color stripped or color ordered amplitudes. The color dressed S-matrix elements can be recovered by summing over color ordered open string amplitudes with appropriate color weights. For partial open string amplitudes the number of local diagrams grows factorial instead of exponentially \cite{Mafra:2022wml}. Moreover, we are interested in closed string amplitudes, where the Chan-Paton factors of open string amplitudes are irrelevant. Therefore, we are only considering color stripped or ordered open string amplitudes.} \cite{Berkovits:2000fe}
\begin{align}
	\mathcal{A}(\sigma)&=\int_{D_2(\sigma)}\mathrm{d}z_2\,\mathrm{d}z_3\,\cdots\mathrm{d}z_{n-2}\,\llangle V_1(z_1)U_2(z_2)U_3(z_3)\cdots U_{n-2}(z_{n-2})V_{n-1}(z_{n-1})V_{n}(z_n)\rrangle\ ,\label{eq::open_string_scattering_prescription}
\end{align}
where $\llangle\ldots\rrangle$ denotes the path integral over the variables in the pure spinor action \eqref{eq::psfaction2}. The integration domain, which is the boundary of the disk, can be parametrized by (parts of) the compactified real line 
\begin{equation}
	D_2(\sigma)=\left\{(z_1,z_2,\ldots,z_n)\in\mathbb R^n|-\infty<z_{\sigma(1)}<z_{\sigma(2)}<\ldots<z_{\sigma(n)}<\infty\right\}\ ,
\end{equation}
where $\sigma\equiv\sigma(1,2,\ldots, n)$ is the permutation of the labels that corresponds to the color ordering of the $n$ external string states. Note that three vertex operator positions $(z_1,z_{n-2},z_n)$ in $D_2(\sigma)$ are position fixed due to the $PSL(2,\mathbb R)$ invariance of the world-sheet. A convenient choice is $(z_1,z_{n-1},z_n)=(0,1,\infty)$.\par
The correlation function $\llangle\ldots\rrangle$ in \eqref{eq::open_string_scattering_prescription} is evaluated by integrating out the non-zero modes of the $h=1$ conformal primaries $\partial\theta^\alpha(z_i),\Pi^m(z_i),d_\alpha(z_i)$ and $N^{mn}(z_i)$, which is done by applying Wick's theorem and using the OPEs in  \eqref{psfOPE}. Similarly, one has to compute the contractions of non-zero modes of plane wave factors $e^{ik_i\cdot X(z_i,\overline z_i)}$, which gives the Koba-Nielsen factor of the corresponding amplitude. Thereby, one replaces the conformal primaries by their singularities with the other fields in the correlator and obtains \cite{Berkovits:2006ik}
\begin{align}
	\mathcal{A}(\sigma)&=\int_{D_2(\sigma)}\mathrm{d}z_2\,\mathrm{d}z_3\,\cdots\mathrm{d}z_{n-2}\, \llangle V_1(z_1)U_2(z_2)U_3(z_3)\cdots U_{n-2}(z_{n-2})V_{n-1}(z_{n-1})V_{n}(z_n)\rrangle\n
	&=\int_{D_2(\sigma)}\mathrm{d}z_2\,\mathrm{d}z_3\,\cdots\mathrm{d}z_{n-2}\,\KN(\{z_i\})\left\langle\mathcal{K}_n(\{z_i\})\right\rangle\ ,\label{eq::open_string_amplitude_K}
\end{align} 
where we have used \eqref{eq::SYM_field_splitting} to strip off the open string Koba-Nielsen factor from the vertex operators and introduced the zero mode correlator $\langle\ldots\rangle$. The contraction of the plane wave factors is given by
\begin{equation}
	\KN(\{z_i\})=\prod_{i<j}^n|z_{ij}|^{s_{ij}}\ ,
\end{equation}
where $s_{ij}=\frac{1}{2}(k_i+k_j)^2=k_i{\cdot} k_j$ for massless states, i.e.\ $k_i^2=0$. Moreover, the exact dependence of $\mathcal{K}_n(\{z_i\})$ on $z_i$ and in terms of the superfields $A^i_\alpha,A^i_m,W^\alpha_i$ and $F^i_{mn}$ is determined by the OPE contractions of the vertex operators, while the dependence on $\alpha'$ follows from the momentum expansion. The exact expression can be obtained by following the steps in the next subsection  and can be expressed in terms of SYM amplitudes. Nevertheless, this determines the correlator as a unique function of the world-sheet coordinates $z_i$ on the disk \cite{Polchinski:1998rq}. Note that $\mathcal{K}_n(\{z_i\})$ contains all the information of the external states like momenta and polarization vectors/spinors. These enter $\mathcal{K}_n(\{z_i\})$ via the $\theta$-expansions in \eqref{expansion}.\par
Moreover, $\mathcal{K}_n(\{z_i\})$ contains the zero modes of $\lambda$ and $\theta$. At tree level, only the fields $\lambda^\alpha, \theta^\alpha$ and $X^m$ contain zero modes due to their conformal dimension being zero. All other fields, which have a conformal weight of $h=1$, do not have zero modes on the disk \cite{DHoker:1988pdl}. To evaluate the correlator in \eqref{eq::open_string_amplitude_K}, we first integrate out the non-zero modes. The evaluation of the $X^m$ zero modes results in a momentum-preserving $\delta$-function. Consequently, we are left with an expression in the pure spinor superspace that contains only the zero modes of $\lambda^\alpha$ and $\theta^\alpha$:
\begin{equation}
	\left\langle\mathcal{K}_n(\{z_i\})\right\rangle=\left\langle\lambda^\alpha\lambda^\beta\lambda^\gamma f_{\alpha\beta\gamma}(\theta;\{z_i\})\right\rangle\ .\label{eq::zero_modes_f}
\end{equation}
The functional $f_{\alpha\beta\gamma}(\theta;\{z_i\})$ is a composite superfield of the external states, containing their kinematic content and thus being $\alpha'$ dependent.   
The argument of $\langle\ldots\rangle$ in \eqref{eq::zero_modes_f} has a finite power series expansion of the enclosed superfields in $\theta^\alpha$. It was argued in \cite{Berkovits} that only terms involving five powers of $\theta$ and three powers of $\lambda$ contribute, i.e.\
\begin{equation}
	\left\langle\lambda^\alpha\lambda^\beta\lambda^\gamma f_{\alpha\beta\gamma}(\theta;\{z_i\})\right\rangle=\left\langle\left.\lambda^\alpha\lambda^\beta\lambda^\gamma f_{\alpha\beta\gamma}(\theta;\{z_i\})\right|_{\theta^5}\right\rangle\ .\label{eq::zero_modef_5}
\end{equation}
Given that the tensor product of three $\lambda$ and five $\theta$ contains a unique scalar, which is the unique element of the cohomology of the BRST operator in the pure spinor formalism at $\mathcal{O}(\lambda^3\theta^5)$, all terms of this type are proportional to each other \cite{Mafra:2008gkx} and are determined by:
\begin{equation}
	\langle(\lambda\gamma^m\theta)(\lambda\gamma^n\theta)(\lambda\gamma^p\theta)(\theta\gamma_{mnp}\theta)\rangle_0 = 2880\ .\label{eq::zmp}
\end{equation}
Even though only five $\theta^\alpha$ out of 16 are present in \eqref{eq::zmp}, the zero mode prescription can be shown to be supersymmetric \cite{Berkovits}. Because there is only one unique element \eqref{eq::zmp} of the cohomology of $Q$ at order $\mathcal{O}(\lambda^3\theta^5)$, we can evaluate any zero-mode correlator using symmetry arguments together with the normalization condition in equation \eqref{eq::zmp} \cite{Mafra:2022wml}.

\subsection*{Evaluating the  open string correlator on the disk}\label{sec::open_string}

The correlator of $n$ open strings on the disk can be systematically evaluated by following the steps presented in \cite{Mafra:2010gj,Mafra:2011nv,Mafra:2009bz}: Integrating out the non-zero modes of the $h=1$ fields corresponds to summing over their OPE singularities. For $n$ external states this results in a sum over $(n-2)!$ single pole terms and a number of double pole integrands that will be used as corrections to the single pole integrands to form the BRST building blocks $T_{ijk\ldots p}$ from the associated OPE residue $L_{jiki\ldots pi}$. The composite superfields $L_{2131\ldots p1}$ of the single pole residues are derived from the contraction of vertex operators, when the integrated vertex operators $U_{2}U_{3}\cdots U_{p}$ approach an unintegrated vertex operator $V_1$, where we have chosen the external states $1,n-1$ and $n$ to be position fixed. The final result is independent on the order of integrating out the $h=1$ primaries and we can choose for example the order $z_1\to z_3\to \ldots\to z_{p}\to z_1$, which is reflected in the $z_{ij}$ in the denominator:
\begin{equation}
	V_1(z_1)U_2(z_2)U_3(z_3)\ldots U_p(z_p)\sim\frac{2^{p-2}L_{[p1,[(p-1)1,[\ldots,[41,[31,21]]\ldots]]]}}{z_{23}z_{34}\cdots z_{(p-1)p}z_{p1}}\ .
\end{equation}
Performing these steps to express all single pole composite superfields in terms of $L_{jiki\ldots pi}$ and after the double pole correction are absorbed to transform the composite superfields in their according building blocks $T_{ijk\ldots p}$ the scattering amplitude of $n$ open strings on the disk takes the form
\begin{align}
	&\mathcal{A}(\rho(1,2,\ldots,n))=\n
	&=\prod_{q=2}^{n-2}\int_{D_2(\rho)} \mathrm{d}z_q\, \langle V_1(0)U_2(z_2)U_3(z_3)\cdots U_{n-2}(z_{n-2})V_{n-1}(1)V_n(\infty)\rangle\n
	&=\prod_{q=2}^{n-2}\int_{D_2(\rho)} \mathrm{d}z_q\,\prod_{i<j}|z_{ij}|^{s_{ij}}\sum_{p=1}^{n-2}\Bigg\langle\frac{ T_{12\ldots p}T_{(n-1)(n-2)\ldots p+1}V_n}{(z_{12}z_{23}\cdots z_{(p-1)p})(z_{(n-1)(n-2)}z_{(n-2)(n-3)}\cdots z_{(p+2)(p+1)})}\n
	&\h+\mathcal{P}(2,3,\ldots, n-2)\Bigg\rangle\ , \label{eq::open_string_n_point}
\end{align}
which is manifestly symmetric in the labels $2,3,\ldots,n-2$ of the integrated vertex operators $U_i$. This is denote above by the sum over all $(n-3)!$ permutations of these labels. Remarkably, the poles in $z_{ij}$ associated to the BRST building block in the numerator follow a pattern. From the above correlator we find that 
\begin{equation}
	T_{12\ldots p} \longleftrightarrow \frac{1}{z_{12}z_{23}\cdots z_{(p-1)p}}\ .\label{eq::pattern_BRST_building_block}
\end{equation}
Because of the $(n-3)!$ permutations of the labels of the integrated vertex operators and the sum over $p$ that collects $n-2$ distinct permutation orbits, this special structure \eqref{eq::pattern_BRST_building_block} of the open string amplitude allows to express \eqref{eq::open_string_n_point} in terms of $(n-2)!$ kinematic numerators and hypergeometric integrals.\par
To simplify the scattering of $n$ open strings further we exchange the BRST building blocks $T_{12\ldots p}$ for the Berends-Giele supercurrents $M_{12\ldots p}$, which is possible due to the pattern \eqref{eq::pattern_BRST_building_block}, i.e.\ the $z_{ij}$-dependence in the denominator of the associated $T_{12\ldots p}$. This connection relies on the synergy of different terms in the sum over permutations and the BRST symmetries 
\begin{align}
	p=2n+1 &: & 0&=T_{12\ldots n+1[n+2[\ldots[2n-1[(2n)(2n+1)]]\ldots ]]}-2T_{2n+1\ldots n+2[n+1[\ldots[3[21]]\ldots]]}\ ,\nonumber\\
	p=2n &: & 0&=T_{12\ldots n[n+1[\ldots[2n-2[(2n-1)(2n)]]\ldots]]}+T_{2n\ldots n+1[n[\ldots[3[21]]\ldots]]}\ ,\label{eq::BRST_symmetry_building_block}
\end{align}
of the building blocks. At level $p$ this interplay reduced the number of independent building blocks $T_{ijk\ldots p}$ down to $(p-1)!$. Therefore, we find that the building blocks and currents are related inside the amplitude as
\begin{align}
	\frac{ T_{12\ldots p}}{z_{12}z_{23}\cdots z_{(p-1)p}}+\mathcal{P}(2,3,\ldots, p)&=(-1)^{p-1}\prod_{k=2}^p\sum_{i=1}^{k-1}\frac{s_{ik}}{z_{ik}}M_{12\ldots p}+\mathcal{P}(2,3,\ldots, p)\ ,\n
	\frac{T_{(n-1)(n-2)\ldots p+1}}{z_{(n-1)(n-2)}\cdots z_{(p+2)(p+1)}}+\mathcal{P}(2,3,\ldots, p)&=(-1)^{n-p-1}\prod_{k=p+1}^{n-2}\sum_{j=k+1}^{n-1}\frac{s_{ik}}{z_{ik}}M_{(n-1)(n-2)\ldots p+1}\n&\h+\mathcal{P}(2,3,\ldots, p)\ ,\n
	&=(-1)^{n-p-1}\prod_{k=p+1}^{n-2}\sum_{j=k+1}^{n-1}\frac{s_{kj}}{z_{kj}}M_{(p+1)(p+2)\ldots n-1}\n&\h+\mathcal{P}(2,3,\ldots, p)\ ,\label{eq::T_to_M}
\end{align}
where we have used the reflection symmetry 
\begin{equation}
	M_{12\ldots p}=(-1)^pM_{p(p-q)\ldots 21}
\end{equation}
of the rank $n-1-p$ Berends-Giele currents in the last line to rewrite $M_{(n-1)(n-2)\ldots p+1}=(-1)^{n-p-2}M_{(p+1)(p+2)\ldots n-1}$.\par
The structure of the substitution \eqref{eq::T_to_M} is precisely of the form that we can apply integration by parts relations to the chain of $\frac{s_{ik}}{z_{ik}}$ sums, which arise after exchanging $T_{12\ldots p}$ for $M_{12\ldots p}$ using \eqref{eq::T_to_M}. The basic idea is that the integration boundaries correspond to zeros in the Koba-Nielson factor, i.e.\ vertex operator positions. Hence, the boundary terms in the world-sheet integrals vanish:
\begin{equation}
	\int_{D_2(\rho)}\mathrm{d}z_2\,\cdots\mathrm{d}z_{n-2}\,\frac{\partial}{\partial z_k}\frac{\prod_{i<j}|z_{ij}|^{s_{ij}}}{z_{i_1 j_1}z_{i_2 j_2}\cdots z_{i_{n-4} j_{n-4}}}=0\ ,
\end{equation}
which relates different integrals in the $n$ open string amplitude with $n-3$ powers of $z_{i_m j_m}$ in the denominator. In the case where the differentiation variable $z_k$ does not appear in any $z_{i_m j_m}$, i.e.\ $k\notin\{i_m,j_m\}$ for $m=2,3,\ldots n-4$, the derivative $\frac{\partial}{\partial z_k}$ acts only on the Koba-Nielsen factor such that
\begin{equation}
	\int_{D_2(\rho)}\mathrm{d}z_2\,\cdots\mathrm{d}z_{n-2}\,\frac{\prod_{i<j}|z_{ij}|^{s_{ij}}}{z_{i_1 j_1}z_{i_2 j_2}\cdots z_{i_{n-4} j_{n-4}}}\sum_{\substack{i=1\\i\neq k}}^{n-1}\frac{s_{ik}}{z_{ik}}=0\ .
\end{equation}
Leaving the first $\frac{n}{2}-1$ factors $\sum_{i=1}^{k-1}\frac{s_{ik}}{z_{ik}}$ untouched and using these relations to integrate the other $\frac{n-3}{2}$ factors by parts yields
\begin{align}
	&\prod_{q=2}^{n-2}\int\mathrm{d}z_q\,\prod_{i<j}|z_{ij}|^{s_{ij}}\frac{s_{12}}{z_{12}}\left(\frac{s_{13}}{z_{13}}+\frac{s_{23}}{z_{23}}\right)\cdots\left(\frac{s_{1(n-2)}}{z_{1(n-2)}}+\ldots+\frac{s_{(n-1)(n-2)}}{z_{(n-1)(n-2)}}\right)\n
	&=\prod_{q=2}^{n-2}\int\mathrm{d}z_q\,\prod_{i<j}|z_{ij}|^{s_{ij}}\biggl(\prod_{k=2}^{\frac{n}{2}}\sum_{i=1}^{k-1}\frac{s_{ik}}{z_{ik}}\biggr)\biggl(\prod_{k=\frac{n}{2}+1}^{n-2}\sum_{j=k+1}^{n-1}\frac{s_{kj}}{z_{kj}}\biggr)\ .
\end{align}
After introducing Berends-Giele currents in the $n$-point function the above relation can be used to write \eqref{eq::open_string_n_point} as
\begin{align}
	\mathcal{A}(1,2,\ldots, n)&=\prod_{q=2}^{n-2} \int_{D_2(\rho)} \mathrm{d}z_q\, \prod_{i<j} |z_{ij}|^{s_{ij}} \biggl\langle
	\sum_{p=1}^{n-2} \biggl(\prod_{k=2}^{p}  \sum_{i=1}^{k-1} {s_{ik} \over z_{ik}} M_{12\ldots p} \biggr) \n
	&\h \times \biggl( \prod_{k=p+1}^{n-2} \sum_{j=k+1}^{n-1} {s_{kj} \over z_{kj}} M_{ p+1,\ldots , n-2,n-1} \biggr) V_n 
	+ \mathcal P(2,3,\ldots,n-2) \biggr\rangle\n
	&= \prod_{q=2}^{n-2} \int_{D_2(\rho)} \mathrm{d}z_q\, \prod_{i<j} |z_{ij}|^{s_{ij}} \biggl\{ \biggl( \prod_{k=2}^{\frac{n}{2}}  
	\sum_{m=1}^{k-1} {s_{mk} \over z_{mk}}  \biggr) \biggl( \prod_{k=\frac{n}{2}+1}^{n-2} \sum_{n=k+1}^{n-1} {s_{kn} \over z_{kn}}  \biggr) \n
	&\h \times \sum_{p=1}^{n-2} \langle M_{12\ldots p} M_{p+1\ldots n-2,n-1} V_n \rangle+\mathcal P(2,3,\ldots,n-2) \biggr\} \n
	&= \prod_{q=2}^{n-2} \int_{D_2(\rho)} \mathrm{d}z_q\, \prod_{i<j} |z_{ij}|^{s_{ij}} \biggl\{ \biggl( \prod_{k=2}^{\frac{n}{2}}  
	\sum_{m=1}^{k-1} {s_{mk} \over z_{mk}}  \biggr) \biggl( \prod_{k=\frac{n}{2}+1}^{n-2} \sum_{n=k+1}^{n-1} {s_{kn} \over z_{kn}}  \biggr) \n
	&\h \times A_{\text{SYM}}(1,2,3,\ldots,n-1,n)+\mathcal P(2,3,\ldots,n-2) \biggr\}\ .\label{eq::open_string_n_point_v2}
\end{align}
In the final result of these manipulations the kinematic building blocks can be expressed as a linear combination of $(n-3)!$ field theory amplitudes, which are each multiplied by a hypergeometric integral $F^{\sigma(2,3,\ldots,n-2)}_{\mathcal{I}_\rho}$ with $\mathcal I_\rho=D_2(\rho)$, which is given by the world-sheet dependent part of \eqref{eq::open_string_n_point_v2} and $\sigma\in S_{n-3}$ refers to the sum over permutations. The string integrands are integrated over parts of the real line given by $D_2(\rho)$ depending on the color ordering $\rho$.
	\section{Amplitude prescription for the real projective plane}\label{sec::rp2_doubling_trick}

Tree--level scattering of closed strings off an O$p$-plane is conformally mapped to a world--sheet described by  the real projective plane. The real projective plane can be defined as the quotient $S^2/\mathbb Z_2$ where the $\mathbb Z_2$--action  identifies antipodal points. The action $\mathbb Z_2:z\mapsto-\frac{1}{\overline z}$ is the identification  to construct the real projective plane as the quotient $S^2/\mathbb Z_2$ from the sphere $S^2$. In addition, this identification preserves the real projective plane and acts without fixed points such that the real projective plane has no boundary. Topologically, $\mathbb{RP}^2$ is equivalent to a sphere with a crosscap. The conformal Killing group of this configuration is $SU(2)$ with three real parameters and corresponds to the subgroup of $PSL(2,\mathbb C)$, which commutes with the $\mathbb Z_2$ action \cite{BLT}. Therefore, the real projective plane has three conformal Killing vectors, which allow to fix one and a half (i.e.\ three real) vertex operator positions as on the disk leaving all others integrated. Hence for the computation of  scattering amplitudes of closed strings on the real projective plane we allow  for a vertex operator $V_i\otimes\overline U_i$ or $U_i\otimes \overline V_i$, because the conformal Killing group $SU(2)$ of the real projective plane does not allow to  completely fix the positions of two closed string vertex operators. This setup is also discussed in \cite{Bischof:2023uor} in the context of closed strings scattering off a D$p$-brane.

With similar arguments as in \cite{Bischof:2023uor} after gauge fixing the $n$-point closed string amplitude reads
\begin{align}
	\mathcal{A}^{\mathbb{RP}^2}_n&=2i g_c^n T'_p \int_0^1\mathrm{d}y\,\llangle[\Big] V^{\mathbb{RP}^2}_1(i,-i)(V\otimes \overline U)^{\mathbb{RP}^2}_2(iy,-iy)\prod_{j=3}^{n}\int_{\mathbb{H}_+}\mathrm{d}^2z_j\,U^{\mathbb{RP}^2}_j(z_j,\overline{z}_j)\rrangle[\Big]\ ,\label{eq::treepres_rp2}
\end{align}
where $T'_p$ is the tension of the O$p$-plane representing the crosscap state and $(V\otimes \overline U)^{\mathbb{RP}^2}$ is the  vertex operator with half of its (complex) position fixed. 
Since the fundamental domain of the real projective plane is the disk, which can be mapped to the upper half plane $\mathbb{H}_+$, we integrate the remaining integrated vertex operators over $\mathbb{H}_+$.  This implies that for the amplitude of $n$ closed strings on the real projective plane we can parameterise  the first two vertex operator insertions on the disk as  $z_1=x_1+iy_1$ and $z_2=x_2+iy_2$ with the choice $x_1=0,y_1=1$ and $x_2=0$ while keeping the integration over $y_2=y$. For more details about the gauge fixing of closed strings on the disk we refer to \cite{Bischof:2023uor}.\par

In the original type I theory open unoriented strings are considered with Neumann boundary conditions along all $D$ space--time directions.
In addition, closed unoriented strings can appear through interactions of open unoriented strings.
Unoriented states are invariant under the world-sheet parity $\Omega$
\begin{equation}\label{Parity}
	\Omega:\quad X^m(z)\longleftrightarrow X^m(\overline z)\ ,
\end{equation}
exchanging the left--moving $X^m(z)$ and right--moving fields $X^m(\overline z)$ of  the space--time coordinate field $X^m(z,\overline z)=X^m(z)+X^m(\overline z)$.
Only $\Om=+1$ states are kept to form unoriented strings. 
This can be interpreted as gauging $\Omega$ including also an orientation reversal of the transition functions which build the world--sheet \cite{Polchinski:1996na,Polchinski:1998rq}. When gauging a world--sheet symmetry, a string transported around a closed curve on the
world--sheet needs only come back to itself up to a gauge transformation. Hence, gauging 
the world--sheet parity implies the inclusion of unoriented world--sheets like the real projective plane at tree--level or Klein bottle 
at one--loop level. 

Type II string theory can be obtained from unoriented type I string theory by applying $T$--duality. At the fixed points of the $T$--dual space this procedure gives rise to O$p$-planes.
Considering $T$--duality of $D\!-\!p\!-\!1$ compactified directions of  open string theory  changes the boundary conditions  of open strings from Neumann to Dirichlet w.r.t.\ the $T$--dual directions.  Furthermore  the open string endpoints are restricted to lie on a hyperplane (D$p$--brane) transverse to  those $D\!-\!p\!-\!1$ directions.
For the spacetime coordinates  the $T$--dual coordinates are given by $X'^m(z,\overline z)=X^m(z)-X^m(\overline z)$ instead of $X^m(z,\overline z)$ \cite{Polchinski:1996na,Polchinski:1998rq}. In the original type I theory the world--sheet parity $\Omega$ acts  only  on $X^m$ as \req{Parity} and 
becomes a gauge symmetry. On the other hand, in the $T$-dual theory  (for $D=10$) the spacetime coordinates transform as \cite{Polchinski:1998rq}
\be
\begin{aligned}
	\Omega:\  X^{\mu}(z,\overline z)&\longleftrightarrow X^{\mu}(\overline z,z)\ ,&&\mu=0,1,\ldots p\ ,\\
	X'^{\tilde\mu}(z,\overline z)&\longleftrightarrow-X'^{\tilde\mu}(\overline z,z)\ ,&&\tilde\mu=p+1,p+2,\ldots 9\ ,
	\end{aligned}
\ee 
which combines a world--sheet parity transformation with a spacetime reflection in the $T$--dualized coordinates. We use lower case Latin letters to describe the entire ten dimensional space-time, lower case Greek letters represent directions in the world volume of the O-plane and lower case Greek letter with a tilde correspond to the directions transverse to that plane, i.e.\ the coordinates on which the $T$--duality  has been performed.

When we split the string wavefunction into its internal part which is assumed to be an  eigenstate of $\Omega$ and an other part describing the dependence on the center of mass $x^{\tilde\mu}$, then the  projection onto $\Omega=+1$ states determines the wavefunction at the orbifold points $-x^{\tilde\mu}$ and $x^{\tilde\mu}$ to be the same up to a sign \cite{Polchinski:1996na}. The components of massless states corresponding to the NSNS sector in the RNS formalism, i.e.\ the spacetime metric, dilaton and the antisymmetric Kalb--Ramond $B$-field, obey \cite{Polchinski:1996na}
\begin{align}
	G_{\mu\nu}(x^{\alpha},-x^{\tilde\alpha})&=G_{\mu\nu}(x^{\alpha},x^{\tilde\alpha})\ ,& B_{\mu\nu}(x^{\alpha},-x^{\tilde\alpha})&=-B_{\mu\nu}(x^{\alpha},x^{\tilde\alpha})\ ,\n
	G_{\mu\tilde\nu}(x^{\alpha},-x^{\tilde\alpha})&=-G_{\mu\tilde\nu}(x^{\alpha},x^{\tilde\alpha})\ ,& B_{\mu\tilde\nu}(x^{\alpha},-x^{\tilde\alpha})&=B_{\mu\tilde\nu}(x^{\alpha},x^{\tilde\alpha})\ ,\\
	G_{\tilde\mu\tilde\nu}(x^{\alpha},-x^{\tilde\alpha})&=G_{\tilde\mu\tilde\nu}(x^{\alpha},x^{\tilde\alpha})\ ,& B_{\tilde\mu\tilde\nu}(x^{\alpha},-x^{\tilde\alpha})&=-B_{\tilde\mu\tilde\nu}(x^{\alpha},x^{\tilde\alpha})\ ,\nonumber
\end{align}
where the orientifold fixed plane is at $x^{\tilde\alpha}=0$. These relations can be written in a compact way \cite{BLT}
\begin{equation}
	G_{mn}(x^{\alpha},-x^{\tilde\alpha})={D_m}^r{D_n}^sG_{rs}(x^{\alpha},x^{\tilde\alpha})\ ,\qquad B_{mn}(x^{\alpha},-x^{\tilde\alpha})=-{D_m}^r{D_n}^sB_{rs}(x^{\alpha},x^{\tilde\alpha})\ ,
\end{equation}
with the matrix $D$ describing the conditions imposed by an O$p$-plane. These conditions are the same \eqref{eq::boundary_matrix} as for a D$p$-brane, cf.\ also \cite{Bischof:2020tnf, Garousi:1996ad}. The graviton and Kalb--Ramond wavefunctions which solve the above conditions are given by \cite{Garousi:2006zh}:
\begin{align}
	G_{\mu\nu}(x^m)&=\epsilon_{\mu\nu}e^{ik_{\alpha} x^{\alpha}}\cos(k_{\tilde\alpha} x^{\tilde\alpha})\ ,& B_{\mu\nu}(x^m)&=i\epsilon_{\mu\nu}e^{ik_{\alpha} x^{\alpha}}\sin(k_{\tilde\alpha} x^{\tilde\alpha})\ ,\n
	G_{\mu\tilde\nu}(x^m)&=i\epsilon_{\mu\tilde\nu}e^{ik_{\alpha} x^{\alpha}}\sin(k_{\tilde\alpha} x^{\tilde\alpha})\ ,& B_{\mu\tilde\nu}(x^m)&=\epsilon_{\mu\tilde\nu}e^{ik_{\alpha} x^{\alpha}}\cos(k_{\tilde\alpha} x^{\tilde\alpha})\ ,\n
	G_{\tilde\mu\tilde\nu}(x^m)&=\epsilon_{\tilde\mu\tilde\nu}e^{ik_{\alpha} x^{\alpha}}\cos(k_{\tilde\alpha} x^{\tilde\alpha})\ ,& B_{\tilde\mu\tilde\nu}(x^m)&=i\epsilon_{\tilde\mu\tilde\nu}e^{ik_{\alpha} x^{\alpha}}\sin(k_{\tilde\alpha} x^{\tilde\alpha})\ .
\end{align}
The polarization tensors are orthogonal to their momenta $k_i^m\epsilon^i_{mn}=\epsilon^i_{mn}k^n_i=0$ and for massless states the momenta satisfy $k_i^2=0$. The above wavefunctions can be written compactly for all $D=10$ spacetime dimensions as\footnote{Note, that for the case $D^{mn}=\eta^{mn}$, i.e.\ no $T$--duality applied in type I string theory equations \req{Wavefunctions} reduce to: $G_{mn}=\eps_{mn}$ and $B_{mn}=0$ in agreement that the anti--symmetric tensor is projected out}  \cite{Garousi:2006zh}:
\begin{align}
	G_{mn}&=\frac12\left\{\epsilon_{mn}e^{ik{\cdot} x}+(D{\cdot} \epsilon^{\mathrm T} {\cdot} D)_{mn}e^{ik{\cdot} D {\cdot} x}\right\}\ ,\n
	B_{mn}&=\frac12\left\{\epsilon_{mn}e^{ik{\cdot} x}+(D{\cdot} \epsilon^{\mathrm T} {\cdot} D)_{mn}e^{ik{\cdot} D {\cdot} x}\right\}\ .\label{Wavefunctions}	
\end{align}
Therefore, a vertex operator representing the insertion of a closed string state on the real projective plane takes the form
\begin{align}
	V^{\mathbb{RP}^2}_i(z_i,\overline z_i)&=\frac12\left\{ V_i[e_i,k_i](z_i) \overline V_i[\overline e_i,k_i](\overline z_i)+V_i[D{\cdot}\overline e_i,D{\cdot}k_i](z_i)\overline V_i[D{\cdot} e_i,D{\cdot}k_i](\overline z_i)\right\}\ ,\n
	U^{\mathbb{RP}^2}_i(z_i,\overline z_i)&=\frac12\left\{ U_i[e_i,k_i](z_i) \overline U_i[\overline e_i,k_i](\overline z_i)+U_i[D{\cdot}\overline e_i,D{\cdot}k_i](z_i)\overline U_i[D{\cdot} e_i,D{\cdot}k_i](\overline z_i)\right\}\ ,\n
	(V\otimes \overline U)^{\mathbb{RP}^2}_i(z_i,\overline z_i)&=\frac12\left\{ V_i[e_i,k_i](z_i) \overline U_i[\overline e_i,k_i](\overline z_i)+V_i[D{\cdot}\overline e_i,D{\cdot}k_i](z_i)\overline U_i[D{\cdot} e_i,D{\cdot}k_i](\overline z_i)\right\}\ ,\label{eq::vertex_RP2_1}
\end{align}
where $V$ and $U$ are the usual unintegrated and integrated vertex operators in the pure spinor formalism, respectively, and we used the decomposition $\epsilon_i=e_i\otimes \overline e_i$.\par
The boundary of the disk imposes non-vanishing correlators between the left- and right--moving parts of the world--sheet fields. As a consequence  the holomorphic and antiholomorphic parts of closed string vertex operators are not independent. In general, 
an operator $\mathcal O_{(h,\overline h)}=O_h(z)\otimes\overline O_{\overline h}(\overline z)$ with conformal weights $(h,\overline h)$ approaching a crosscap state represents a non-trivial process and puts constraints on the operator \cite{Kostelecky:1987px}
\begin{equation}
	0=\big\langle C\big\vert\big[(\overline z')^{ h}\overline{O}_{h}(\overline z')-z^{h}O_h(z)\big]\big\vert_{z=-\frac{1}{\overline{z}'}}\ ,\label{eq::boundary_gluing_condition}
\end{equation}
where $\bra{C}$ represents a crosscap state, which imposes an interaction between the holomorphic and antiholomorphic sector. Furthermore, in \eqref{eq::boundary_gluing_condition} the real projective plane is parametrized by the disk with the world-sheet coordinates $z,\overline z\in D_2$. After using the doubling trick the different sectors can be related as \cite{Polchinski:1998rq}
\begin{equation}
	\overline O_{\overline h}(\overline z)=\left(\frac{\partial z'}{\partial \overline z}\right)^{\overline h}O_{\overline h}(z')\quad \text{for }z'=-\frac{1}{\overline z}\ .\label{eq::rp2_doubling_trick}
\end{equation} 
When mapping the fundamental domain of $\mathbb{RP}^2$ from the disk to the upper half plane as 
	\begin{equation}
		w=i\frac{1-z}{1+z}\ ,\label{eq::cayley_map}
	\end{equation}
the $\mathbb{Z}_2$ identification is invariant: $z'=-\frac{1}{\overline z}\mapsto w'=-\frac{1}{\overline w}$, with $w$ the coordinates on the upper half plane. As a consequence, the doubling trick becomes: 
\begin{equation}
	\overline O_{\overline h}(\overline w)=\left(\frac{\partial w'}{\partial \overline w}\right)^{\overline h}O_{\overline h}(w')\quad \text{for }w'=-\frac{1}{\overline w}\ .\label{eq::rp2_doubling_trick_2}
\end{equation}
Recall, for NSNS states  in \eqref{eq::vertex_RP2_1} we have to introduce the matrix $D$ to account for the spacetime boundary conditions enforced by the O$p$-plane. To apply the doubling trick we shall again substitute the right--moving fields according to \eqref{eq::replace} such that they also satisfy the constraints imposed by the O--plane. Applying the doubling trick not only relates the holomorphic and anti--holomorphic sector, but also converts the correlator on the real projective plane to a correlator on the disk. These manipulations allow to use the two point functions in \eqref{eq::correlators} and integrate out the $h=1$ primary fields. Substituting the antiholomorphic fields by the holomorphic counterparts in \eqref{eq::vertex_RP2_1} by using the doubling trick \eqref{eq::rp2_doubling_trick_2} gives rise to  the following vertex operators for massless NSNS states
\begin{align}
	V^{\mathbb{RP}^2}_i(z_i,\overline z_i)&=\frac12\left\{ V_i[e_i,k_i](z_i)  V_i[D{\cdot}\overline e_i,D{\cdot}k_i]\left(-\frac{1}{\overline z_i}\right)+V_i[D{\cdot}\overline e_i,D{\cdot}k_i](z_i)V_i[e_i,k_i]\left(-\frac{1}{\overline z_i}\right)\right\}\n
	&=\frac12\left\{ V_i(z_i) V_{\overline\imath}\left(-\frac{1}{\overline z_i}\right)+V_{\overline\imath}(z_i)V_i\left(-\frac{1}{\overline z_i}\right)\right\}\ ,\n
	U^{\mathbb{RP}^2}_i(z_i,\overline z_i)&=\frac12\frac{1}{\overline z_i^2}\left\{ U_i[e_i,k_i](z_i)  U_i[D{\cdot}\overline e_i,D{\cdot}k_i]\left(-\frac{1}{\overline z_i}\right)+U_i[D{\cdot}\overline e_i,D{\cdot}k_i](z_i)U_i[ e_i,k_i]\left(-\frac{1}{\overline z_i}\right)\right\}\n
	&=\frac12\frac{1}{\overline z_i^2}\left\{ U_i(z_i) U_{\overline\imath}\left(-\frac{1}{\overline z_i}\right)+U_{\overline\imath}(z_i)U_i\left(-\frac{1}{\overline z_i}\right)\right\}\ ,
\end{align}
where we made use of the notation \eqref{eq::doubling_trick_superfield} for indices with a bar.\footnote{Here, we are only considering NSNS states such that we have to set $\overline \chi_i=0$ in \eqref{eq::doubling_trick_superfield}.} Furthermore, for the half unintegrated and half integrated vertex operator we obtain
\begin{align}
	&(V\otimes \overline U)^{\mathbb{RP}^2}_i(z_i,\overline z_i)=\n
	&=\frac12\frac{1}{\overline z_i^2}\left\{ V_i[e_i,k_i](z_i)  U_i[D{\cdot}\overline e_i,D{\cdot}k_i]\left(-\frac{1}{\overline z_i}\right)+V_i[D{\cdot}\overline e_i,D_i{\cdot}k_i](z_i)U_i[ e_i,k_i]\left(-\frac{1}{\overline z_i}\right)\right\}\n
	&=\frac12\frac{1}{\overline z_i^2}\left\{ V_i(z_i) U_{\overline\imath}\left(-\frac{1}{\overline z_i}\right)+V_{\overline\imath}(z_i)U_i\left(-\frac{1}{\overline z_i}\right)\right\}\ ,\label{eq::vertex_RP2_2}
\end{align}
where we again made use of the notation in \eqref{eq::doubling_trick_superfield}. The derivative $\left(\frac{\partial\overline z'}{\partial \overline z}\right)^{\overline h}=\left(\frac{1}{\overline z^2}\right)^{\overline h}$ for $z'=-\frac{1}{\overline z}$ stemming from \eqref{eq::rp2_doubling_trick_2} explains the prefactor for the integrated vertex operator with  $\overline h=1$. Eventually, the vertex operator $U_i(z_i,\overline z_i)$ is integrated over the upper half plane in the following  way:
\begin{align}
	\int_{\mathbb{H}_+}\mathrm d^2z_i\,U^{\mathbb{RP}^2}_i(z_i,\overline z_i)&=\frac12\int_{\mathbb{H}_+}\mathrm d^2z_i\,\frac{1}{\overline z_i^2}\left\{ U_i(z_i) U_{\overline\imath}\left(-\frac{1}{\overline z_i}\right)+U_{\overline\imath}(z_i)U_i\left(-\frac{1}{\overline z_i}\right)\right\}\n
	&=\frac12\int_{\mathbb{H}_+}\mathrm d^2z_i\,\frac{1}{\overline z_i^2}U_i(z_i) U_{\overline\imath}\left(-\frac{1}{\overline z_i}\right)+\frac12\int_{\mathbb{H}_-}\mathrm d^2z_i\,\frac{1}{\overline z_i^2}U_{\overline\imath}\left(-\frac{1}{\overline z_i}\right)U_i(z_i)\n
	&=\frac12\int_{\mathbb{C}}\mathrm d^2z_i\,\frac{1}{\overline z_i^2}U_i(z_i) U_{\overline\imath}\left(-\frac{1}{\overline z_i}\right)\ .\label{eq::rp2_doubling_trick_U}
\end{align}
Above, we have performed the coordinate transformation $(z_i,\overline z_i)\to\left(-\frac{1}{\overline z_i},-\frac{1}{z_i}\right)$ in the second term. Note, that for the disk amplitude it is not possible to combine $\mathbb H_+$ and $\mathbb{H}_-$ into $\mathbb{C}$, i.e.\  $\mathbb H_+\cup \mathbb{H}_-=\mathbb{C}$, because this amplitude has poles along the boundary (real axis) at $z_i-\overline z_i=0$ \cite{Bischof:2023uor}. However, for the amplitude on the real projective plane we can take $\mathbb H_+\cup \mathbb{H}_-=\mathbb{C}$, because this amplitude does not exhibit poles on the real line. As we will see below, the corresponding term obeys $1+z_i\overline z_i\neq 0$, because $|z_i|^2\geq0$.\par
Finally, applying the previous results  to the amplitude \eqref{eq::treepres_rp2} yields:
\begin{align}
	\mathcal{A}^{\mathbb{RP}^2}_n&=-\left(\frac12\right)^{n-1}i g_c^n T'_p\int_0^1\mathrm{d}y\,\llangle[\biggr] (V_1(i)V_{\overline 1}(-i)+V_{\overline{1}}(i)V_{1}(-i))\n
	&\h\times\frac{1}{y^2}\left\{V_2(iy)U_{\overline 2}\left(-\frac{i}{y}\right)+V_{\overline2}(iy)U_{2}\left(-\frac{i}{y}\right)\right\}\prod_{j=3}^{n}\int_{\mathbb C}\mathrm{d}^2z_j\,\frac{1}{\overline z_j^2}U_j(z_j) U_{\overline \jmath}\left(-\frac{1}{\overline z_{j}}\right)\rrangle[\biggr]\ .\label{eq::treepres_rp2_1}
\end{align}
Performing the change of variables $(y,z,\overline z)\to(-y,-z,-\overline z)$ in the terms proportional to $V_{\overline{1}}(i)V_{1}(-i)$ of the amplitude \eqref{eq::treepres_rp2_1} and  applying the invariance of the correlator  under conformal transformations, 
i.e.\ we may  rescale all vertex operators as \cite{BLT}	
\begin{equation}
	V(a z)=V(z)\ ,\qquad U(a z)=aU(z) \label{eq::scaling_vertex_operator}
\end{equation}
with $a=-1$, we arrive at:
\begin{align}
	\mathcal{A}^{\mathbb{RP}^2}_n&=-\left(\frac12\right)^{n-1}i g_c^n T'_p\llangle[\biggr] \Big(\int_0^1\mathrm{d}y\,V_1(i)V_{\overline 1}(-i)-\int_{-1}^0\mathrm{d}y\,V_{\overline{1}}(-i)V_{1}(i)\Big)\n
	&\h\times\frac{1}{y^2}\Big\{V_2(iy)U_{\overline 2}\left(-\frac{i}{y}\right)+V_{\overline2}(iy)U_{2}\left(-\frac{i}{y}\right)\Big\}\prod_{j=3}^{n}\int_{\mathbb C}\mathrm{d}^2z_j\,\frac{1}{\overline z_j^2}U_j(z_j) U_{\overline \jmath}\left(-\frac{1}{\overline z_{j}}\right)\rrangle[\biggr]\n
	&=-\left(\frac12\right)^{n-1}i g_c^n T'_p\int_{-1}^1\mathrm{d}y\,\frac{1}{y^2}\llangle[\biggr] V_1(i)V_{\overline 1}(-i)\Big\{V_2(iy)U_{\overline 2}\left(-\frac{i}{y}\right)+V_{\overline2}(iy)U_{2}\left(-\frac{i}{y}\right)\Big\}\n
	&\h\times\prod_{j=3}^{n}\int_{\mathbb C}\mathrm{d}^2z_j\,\frac{1}{\overline z_j^2}U_j(z_j) U_{\overline \jmath}\left(-\frac{1}{\overline z_{j}}\right)\rrangle[\biggr]\ .\label{eq::treepres_rp2_2}
\end{align}
For the second equality  we used the property that  unintegrated vertex operators anti--commute. Now we perform in those terms of \req{eq::treepres_rp2_2}, which are proportional to $V_{\overline2}(iy)U_{2}\left(-\frac{i}{y}\right)$, the substitution $y\to-\frac1y$: 
\begin{align}
	&\left(\frac12\right)^{n-1}i g_c^n T'_p\int_{-1}^1\mathrm{d}y\,\frac{1}{y^2}\llangle[\biggr] V_1(i)V_{\overline 1}(-i)V_{\overline2}(iy)U_{2}\left(-\frac{i}{y}\right)\prod_{j=3}^{n}\int_{\mathbb C}\mathrm{d}^2z_j\,\frac{1}{\overline z_j^2}U_j(z_j) U_{\overline \jmath}\left(-\frac{1}{\overline z_{j}}\right)\rrangle[\biggr]\n
	&=\left(\frac12\right)^{n-1}i g_c^n T'_p\int_{-\infty}^{-1}\mathrm{d}y\,\llangle[\biggr] V_1(i)V_{\overline 1}(-i)V_{\overline2}\left(-\frac{i}{y}\right)U_{2}(iy)\prod_{j=3}^{n}\int_{\mathbb C}\mathrm{d}^2z_j\,\frac{1}{\overline z_j^2}U_j(z_j) U_{\overline \jmath}\left(-\frac{1}{\overline z_{j}}\right)\rrangle[\biggr]\n
	&\h\left(\frac12\right)^{n-1}i g_c^n T'_p\int_{1}^{\infty}\mathrm{d}y\,\llangle[\biggr] V_1(i)V_{\overline 1}(-i)V_{\overline2}\left(-\frac{i}{y}\right)U_{2}(iy)\prod_{j=3}^{n}\int_{\mathbb C}\mathrm{d}^2z_j\,\frac{1}{\overline z_j^2}U_j(z_j) U_{\overline \jmath}\left(-\frac{1}{\overline z_{j}}\right)\rrangle[\biggr]\ .\label{eq::extra_key}
\end{align}
The independence of the assignment of integrated and unintegrated vertex operators inside an amplitude has explicitly been shown for closed string amplitudes on the disk in \cite{Bischof:2020tnf}. We use this property  to convert in \req{eq::extra_key} the expression $V_{\overline2}\left(-\frac{i}{y}\right)U_{2}(iy)$ to $ -\frac{1}{y^2}U_{\overline2}\left(-\frac{i}{y}\right)V_{2}(iy)$. For details we refer the reader to \cite{Bischof:2020tnf} and in  the following we only present the most important steps to prove this statement. Without loss of generality we omit all vertex operators with $n>2$, as their BRST variation vanishes when integrating over them, i.e.\ when carefully applying Cauchy's theorem such that $Q U_{\overline\jmath}\left(-\frac{1}{\overline z_j}\right)=\overline z_j^2\overline\partial_jU_{\overline\jmath}\left(-\frac{1}{\overline z_j}\right)$,
\begin{align}
	Q\int_{\mathbb C}\mathrm d^2z_j\,\frac{1}{\overline z^2_j} U_j(z_j)U_{\overline \jmath}\left(-\frac{1}{\overline z_j}\right)&=\int_{\mathbb C}\mathrm d^2z_j\,\left(\frac{1}{\overline z^2_j} \partial_j V_j(z_j)U_{\overline \jmath}\left(-\frac{1}{\overline z_j}\right)+U_j(z_j)\overline\partial_j V_{\overline \jmath}\left(-\frac{1}{\overline z_j}\right)\right)\n
	&=\int_{\mathbb C}\mathrm d^2z_j\,\left[\partial_j\left(\frac{1}{\overline z^2_j}  V_j(z_j)U_{\overline \jmath}\left(-\frac{1}{\overline z_j}\right)\right)+\overline\partial_j\left(U_j(z_j) V_{\overline \jmath}\left(-\frac{1}{\overline z_j}\right)\right)\right]\n
	&=0\ ,
\end{align}
because the complex plane has no boundary. Hence, along the lines of the computation in appendix C of \cite {Bischof:2020tnf} we find
\begin{align}
	\int_{1}^{\infty}\mathrm{d}y\,\llangle[\biggr] V_1(i)V_{\overline 1}(-i)V_{\overline2}\left(-\frac{i}{y}\right)U_{2}(iy)\rrangle[\biggr]&=\int_{1}^{\infty}\mathrm{d}y\,\llangle[\biggr] V_1(i)V_{\overline 1}(-i)\int_{-i}^{-\frac{i}{y}}\mathrm dz\,QU_{\overline2}(z)U_{2}(iy)\rrangle[\biggr]\n
	&=-i\int_{1}^{\infty}\mathrm{d}y\,\llangle[\biggr] V_1(i)V_{\overline 1}(-i)\int_{-i}^{-\frac{i}{y}}\mathrm dz\,U_{\overline2}(z)\partial_yV_{2}(iy)\rrangle[\biggr]\n
	&=-i\int_{1}^{\infty}\mathrm{d}y\,\partial_y\llangle[\biggr] V_1(i)V_{\overline 1}(-i)\int_{-i}^{-\frac{i}{y}}\mathrm dz\,U_{\overline2}(z)V_{2}(iy)\rrangle[\biggr]\n
	&\h+i\int_{1}^{\infty}\mathrm{d}y\,\llangle[\biggr] V_1(i)V_{\overline 1}(-i)\partial_y\int_{-i}^{-\frac{i}{y}}\mathrm dz\,U_{\overline2}(z)V_{2}(iy)\rrangle[\biggr]\n
	&=-i\llangle[\biggr] V_1(i)V_{\overline 1}(-i)\int_{-i}^{0}\mathrm dz\,U_{\overline2}(z)U_{2}(\infty)\rrangle[\biggr]\n
	&\h-\int_{1}^{\infty}\mathrm{d}y\,\frac{1}{y^2}\llangle[\biggr] V_1(i)V_{\overline 1}(-i)U_{\overline2}\left(-\frac{i}{y}\right)V_{2}(iy)\rrangle[\biggr]
	\ .
\end{align}
With a similar calculation we get 
\begin{align}
	\int_{-\infty}^{-1}\mathrm{d}y\,\llangle[\biggr] V_1(i)V_{\overline 1}(-i)V_{\overline2}\left(-\frac{i}{y}\right)U_{2}(iy)\rrangle[\biggr]&=i\llangle[\biggr] V_1(i)V_{\overline 1}(-i)\int_{-i}^{0}\mathrm dz\,U_{\overline2}(z)V_{2}(-\infty)\rrangle[\biggr]\n
	&\h-\int_{-\infty}^{-1}\mathrm{d}y\,\frac{1}{y^2}\llangle[\biggr] V_1(i)V_{\overline 1}(-i)U_{\overline2}\left(-\frac{i}{y}\right)V_{2}(iy)\rrangle[\biggr]
\end{align}
such that the two terms cancel:
\begin{equation}
	-i\llangle[\biggr] V_1(i)V_{\overline 1}(-i)\int_{-i}^{0}\mathrm dz\,U_{\overline2}(z)U_{2}(\infty)\rrangle[\biggr]+i\llangle[\biggr] V_1(i)V_{\overline 1}(-i)\int_{-i}^{0}\mathrm dz\,U_{\overline2}(z)U_{2}(-\infty)\rrangle[\biggr]=0\ ,
\end{equation}
because the amplitude does not distinguish whether a vertex operator is at $\infty$ or $-\infty$. After the reassignment of vertex operators the remaining terms  are given by:
\begin{align}
	&\int_{-\infty}^{-1}\mathrm{d}y\,\llangle[\biggr] V_1(i)V_{\overline 1}(-i)V_{\overline2}\left(-\frac{i}{y}\right)U_{2}(iy)\rrangle[\biggr]+\int_{1}^{\infty}\mathrm{d}y\,\llangle[\biggr] V_1(i)V_{\overline 1}(-i)V_{\overline2}\left(-\frac{i}{y}\right)U_{2}(iy)\rrangle[\biggr]\n
	&=-\int_{-\infty}^{-1}\mathrm{d}y\,\frac{1}{y^2}\llangle[\biggr] V_1(i)V_{\overline 1}(-i)V_{2}(iy)U_{\overline2}\left(-\frac{i}{y}\right)\rrangle[\biggr]-\int_{1}^{\infty}\mathrm{d}y\,\frac{1}{y^2}\llangle[\biggr] V_1(i)V_{\overline 1}(-i)V_{2}(iy)U_{\overline2}\left(-\frac{i}{y}\right)\rrangle[\biggr]\ .\label{convertI}
\end{align}
As a consequence, the relation \req{convertI} implies that \eqref{eq::extra_key} becomes:
\begin{align}
	&\left(\frac12\right)^{n-1}i g_c^n T'_p\int_{-1}^1\mathrm{d}y\,\frac{1}{y^2}\llangle[\biggr] V_1(i)V_{\overline 1}(-i)V_{\overline2}(iy)U_{2}\left(-\frac{i}{y}\right)\prod_{j=3}^{n}\int_{\mathbb C}\mathrm{d}^2z_j\,\frac{1}{\overline z_j^2}U_j(z_j) U_{\overline \jmath}\left(-\frac{1}{\overline z_{j}}\right)\rrangle[\biggr]\n
	&=-\left(\frac12\right)^{n-1}i g_c^n T'_p\int_{-\infty}^{-1}\mathrm{d}y\,\frac{1}{y^2}\llangle[\biggr] V_1(i)V_{\overline 1}(-i)V_{2}(iy)U_{\overline2}\left(-\frac{i}{y}\right)\prod_{j=3}^{n}\int_{\mathbb C}\mathrm{d}^2z_j\,\frac{1}{\overline z_j^2}U_j(z_j) U_{\overline \jmath}\left(-\frac{1}{\overline z_{j}}\right)\rrangle[\biggr]\n
	&\h-\left(\frac12\right)^{n-1}i g_c^n T'_p\int_{1}^{\infty}\mathrm{d}y\,\frac{1}{y^2}\llangle[\biggr] V_1(i)V_{\overline 1}(-i)V_{2}(iy)U_{\overline2}\left(-\frac{i}{y}\right)\prod_{j=3}^{n}\int_{\mathbb C}\mathrm{d}^2z_j\,\frac{1}{\overline z_j^2}U_j(z_j) U_{\overline \jmath}\left(-\frac{1}{\overline z_{j}}\right)\rrangle[\biggr]\ .\label{implies}
\end{align}
Finally, with the information \req{implies}  and  after adding the individual contributions to be  integrated over $y\in]-\infty,-1[$ and $y\in]1,\infty[$, together with $y\in]-1,1[$ the amplitude \req{eq::treepres_rp2_2} becomes:
\begin{align}
	\mathcal{A}^{\mathbb{RP}^2}_n&=-\left(\frac12\right)^{n-1}i g_c^n T'_p\int_{-\infty}^\infty\mathrm{d}y\ \frac{1}{y^2}\ e^{i\pi(\Theta(y-1)+\Theta(-y-1))}\n
&\times\llangle[\biggr] V_1(i)V_{\overline 1}(-i)V_2(iy)U_{\overline 2}\left(-\frac{i}{y}\right)
	\prod_{j=3}^{n}\int_{\mathbb C}\mathrm{d}^2z_j\,\frac{1}{\overline z_j^2}U_j(z_j) U_{\overline \jmath}\left(-\frac{1}{\overline z_{j}}\right)\rrangle[\biggr]\ .\label{eq::treepres_rp2_result}
\end{align}
The phase $e^{i\pi(\Theta(y-1)+\Theta(-y-1))}$ accounts for the different signs of the integration regions of $y\in]-1,1[$ and $y\in]-\infty,-1[\cup]1,\infty[$, which arise from the reassignment of vertex operators\footnote{This phase is introduced to write the result of this section in \eqref{eq::treepres_rp2_result} in a compact form. It just accounts for the different sign of the integration regions. Hence, it does not arise from any kind of contour deformations unlike the monodromy phase in \eqref{eq::monodromy_phase}, but is include in \eqref{eq::monodromy_phase} to simplify the notation.}. This additional phase is crucial since the integration regions for both $|y|<1$ and $|y|>1$ acquire  a different orientation after the transformation in \eqref{eq::PSL2R_transformation}, see also Footnote \ref{fn::17}. Hence, after performing the $PSL(2,\mathbb R)$ transformation the sign from the different orientation of the integration regions cancels against the minus sign from the reassignment of vertex operators in \req{eq::treepres_rp2_result}.
The expression \req{eq::treepres_rp2_result} provides our scattering amplitude prescription for closed strings on the real projective plane $\mathbb{RP}^2$.

	\section[Scattering three closed strings off an O\boldmath{$p$}-plane]{Scattering three closed strings off an O\boldmath{$p$}-plane}\label{chap::plane}
\def\R#1{{#1}}

The  scattering amplitude of three closed strings on the real projective plane follows from the prescription \eqref{eq::treepres_rp2_result} for $n=3$:
	\begin{align}
		\mathcal{A}^{\mathbb{RP}^2}_3&=2i g_c^3 T'_p \int_0^1\mathrm{d}y\,\int_{\mathbb{H}_+}\mathrm{d}^2z\,\llangle V^{\mathbb{RP}^2}_1(i,-i)(V\otimes \overline U)^{\mathbb{RP}^2}_2(iy,-iy)U^{\mathbb{RP}^2}_3(z,\overline{z})\rrangle\n
		&=-\frac14i g_c^3T'_p \int_{-\infty}^\infty\mathrm{d}y\int_{\mathbb C}\mathrm{d}^2z\,\frac{1}{y^2}\frac{1}{\overline z^2}e^{i\pi(\Theta(y-1)+\Theta(-y-1))}\n
		&\h\times\llangle[\biggr] V_1(i)V_{\overline 1}(-i)V_2(iy)U_{\overline 2}\left(-\frac{i}{y}\right)U_3(z) U_{\overline3}\left(-\frac{1}{\overline z}\right)\rrangle[\biggr]\ ,\label{eq::3pt_rp2}
	\end{align}
whose Koba-Nielsen factor is given by \cite{Polchinski:1998rq}
	\begin{align}
		\KN(y,z,\overline z)&=2^{s_{\overline 11}}|1-y|^{2s_{12}}|1-y|^{2s_{1\overline{2}}}|1+y^2|^{s_{2\overline{2}}}|i-z|^{2s_{13}}|-i-z|^{2s_{1\overline 3}}\nonumber\\
		&\h\times|iy-z|^{2s_{23}}|1-iyz|^{2s_{\overline{2}3}}|1+z\overline z|^{s_{3 \overline 3}}\ .
	\end{align}
The correlator in \eqref{eq::3pt_rp2} is very similar than the amplitude of three closed strings on the disk discussed in \cite{Bischof:2023uor} with the additional factor $\frac1{y^2}\frac{1}{\overline z^2}$ and the altered positions $\overline z_{2}=-\frac iy$ and $\overline z_3=-\frac{1}{\overline z}$. Hence, we can identify the correlator of the amplitude \eqref{eq::treepres_rp2_result} with the scattering of $2n$ open strings \cite{Stieberger:2009hq} and also the computation in \cite{Bischof:2023uor} suggests that \eqref{eq::treepres_rp2_result} can be connected to the scattering of $2n$ open strings on the disk by identifying closed with open strings. Explicitly, for $n=3$ in \eqref{eq::3pt_rp2} we find:
\begin{equation}
	\overline 1\leftrightarrow1\ ,\qquad\overline2\leftrightarrow 2\ ,\qquad3\leftrightarrow3\ ,\qquad \overline 3\leftrightarrow4\ ,\qquad2\leftrightarrow5\ ,\qquad 1\leftrightarrow 6\ .\label{eq::identification}
\end{equation}
Of course, the integral over the complex plane does not yet correspond to a pair of  open string integrals, which are described  as integrals over segments of the real line. Thus, in this section we want to use the method that is proposed in \cite{Kawai:1985xq} and extended in \cite{Stieberger:2009hq,Stieberger:2015vya} to write the closed integral over $\mathbb C$ as open string integrals which arise from the color ordered scattering of six open strings on the disk.\par
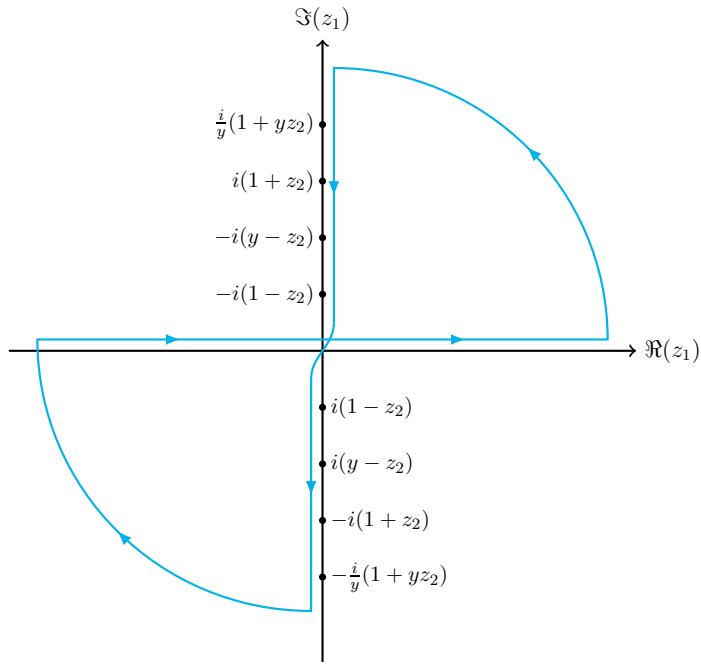
\begin{figure}[H]
	\begin{center}
		\begin{tikzpicture}[scale=0.75, transform shape,decoration={
				markings,
				mark=at position 0.5 with {\arrow{latex}}}]
			\draw[thick,->] (-5.5, 0) -- (5.5, 0) node[right] {$\Re(z_1)$};
			\draw[thick,->] (0, -5.5) -- (0, 5.5) node[above] {$\Im(z_1)$};

			\filldraw[black] (0,4) circle (1.5pt) node[left,font=\small]{$\frac{i}{y}(1+yz_2)$};
			\filldraw[black] (0,3) circle (1.5pt) node[left,font=\small]{$i(1+z_2)$};
			\filldraw[black] (0,2) circle (1.5pt) node[left,font=\small]{$-i(y-z_2)$};
			\filldraw[black] (0,1) circle (1.5pt) node[left,font=\small]{$-i(1-z_2)$};
			
			\filldraw[black] (0,-1) circle (1.5pt) node[right,font=\small]{$i(1-z_2)$};
			\filldraw[black] (0,-2) circle (1.5pt) node[right,font=\small]{$i(y-z_2)$};
			\filldraw[black] (0,-3) circle (1.5pt) node[right,font=\small]{$-i(1+z_2)$};
			\filldraw[black] (0,-4) circle (1.5pt) node[right,font=\small]{$-\frac{i}{y}(1+yz_2)$};
			
			\draw[MPGbluelight,thick, postaction={on each segment={mid arrow}}] (-0.2, -0.5) -- (-0.2,-4.6) arc (-90:-180:4.8) -- (0, 0.2) -- (5, 0.2) arc (0:90:4.8) -- (0.2,0.5);
			\draw[MPGbluelight,thick] (-0.2, -0.5) to[out=90,in=270] (0.2,0.5);
		\end{tikzpicture}
	\end{center}
	\caption[Branch point structure and contour deformation in the complex $z_1$-plane.]{Branch point structure and contour deformation in the complex $z_1$-plane for $z_2>1$ and $0<y<1$.}\label{fig::z2}
\end{figure}

The analytic continuation of \eqref{eq::3pt_rp2} follows the same steps as in \cite{Bischof:2023uor}. Therefore, we write the integral over the complex world sheet as two integrals over the real line. By splitting the complex world--sheet coordinate $z=z_1+iz_2$ into real and imaginary part the integrand in \eqref{eq::3pt_rp2} becomes an analytic function in $z_1$ with four pairs of branch points at $z_1=\pm i(1-z_2),\pm i(1+z_2), \pm i(y-z_2)$ and $\pm\frac{i}{y}(1+yz_2)$, respectively. Because all branch points are purely imaginary, we can rotate the  contour in the complex $z_1$--plane from the real to the purely imaginary axis similar as depicted in Figure \ref{fig::z2}.
\noindent
Therefore, for each complex world--sheet integration we introduce two new real variables $\xi,\eta$ 
\begin{align}
	z= i (z_1+z_2)=i\xi\ ,\qquad \overline z= i (z_1-z_2)=i\eta\ ,
\end{align}  
where  $z_1$ is the real part and $z_2$ is the imaginary part of $z$. These variables are not constrained, i.e.\ $\xi,\eta\in\mathbb R$ as $z,\overline z\in\mathbb C$, leading to\footnote{Even though, the vertex operator position $-\frac{i}{y}$ and $\frac{1}{\eta}$ diverge for $y,\eta \to 0$ the integrand in \eqref{eq::michael_comment} is well defined, as the prefactor $\frac{1}{y^2}\frac{1}{\eta^2}$ cancels these divergences.}
	\begin{align}
		\mathcal{A}^{\mathbb{RP}^2}_3&=-\frac14 g_c^3T'_p \int_{-\infty}^\infty\mathrm{d}y\int_{-\infty}^\infty\mathrm{d}\xi\,\int_{-\infty}^\infty\mathrm{d}\eta\,\frac{1}{y^2}\frac{1}{\eta^2}\Pi(y,\xi,\eta)\n
		&\h\times\llangle[\biggr] V_1(i)V_{\overline 1}(-i)V_2(iy)U_{\overline 2}\left(-\frac{i}{y}\right)U_3(i\xi) U_{\overline3}\left(\frac{i}{\eta}\right)\rrangle[\biggr]\ .\label{eq::michael_comment}
	\end{align}
In addition, including the monodromy phase $\Pi(y,\xi,\eta)$ ensures that the integrand is holomorphic in $\xi$ and $\eta$ and accounts for the correct branch of the integrand.
After absorbing $e^{i\pi(\Theta(y-1)+\Theta(-y-1))}$ into the phase factor we find that the total phase takes the form
	\begin{align}
		\Pi(y,\xi,\eta)&=e^{i\pi(\Theta(y-1)+\Theta(-y-1))}e^{i\pi s_{13}\Theta(-(1-\xi)(1+\eta))}e^{i\pi s_{1\overline3}\Theta(-(1+\xi)(1-\eta))}\n
		&\h\times e^{i\pi s_{23}\Theta(-(y-\xi)(y+\eta))}e^{i\pi s_{2\overline3}\Theta(-(1+y\xi)(1-y\eta))}e^{i\pi s_{3\overline3}\Theta(-(1-\xi\eta))}\ ,\label{eq::monodromy_phase}
	\end{align}
where $\Theta$ is the Heaviside step function and the kinematic invariants are defined as
\begin{equation}
	s_{ij}=k_i{\cdot} k_j\ ,\qquad s_{i\overline\jmath}=k_i{\cdot}D{\cdot}k_j\ ,
	\end{equation}
with the matrix $D$ specifying the kind of boundary conditions of the closed strings under consideration and the underlying orientifold setup.
Note, that these are not independent and  momentum conservation \eqref{eq::mom_con} leads to the following kinematical relations:
\begin{align}
	s_{1\bar 1}&=-s_{12}-s_{1\bar2}-s_{13}-s_{1\bar3}\ ,\label{kin1}\\
	s_{2\bar 2}&=-s_{12}-s_{1\bar2}-s_{23}-s_{2\bar3}\ ,\label{kin2}\\
	s_{3\bar 3}&=-s_{13}-s_{1\bar3}-s_{23}-s_{2\bar3}\ .\label{kin3}
\end{align}
Consequently, there are six independent kinematic invariants for  scattering three closed strings in the presence of O--planes \cite{Stieberger:2009hq}.

Finally, by using \eqref{eq::scaling_vertex_operator} and conformal invariance of the correlator we pull out the factor of $i$ in each of the vertex operators leading to:
	\begin{align}
	\mathcal{A}^{\mathbb{RP}^2}_3&=-\frac14 ig_c^3T'_p \int_{-\infty}^\infty\mathrm{d}y\int_{-\infty}^\infty\mathrm{d}\xi\,\int_{-\infty}^\infty\mathrm{d}\eta\,\frac{1}{y^2}\frac{1}{\eta^2}\Pi(y,\xi,\eta)\n
	&\h\times\llangle[\biggr] V_1(1)V_{\overline 1}(-1)V_2(y)U_{\overline 2}\left(-\frac{1}{y}\right)U_3(\xi) U_{\overline3}\left(\frac{1}{\eta}\right)\rrangle[\biggr]\ .
	\end{align}
Subject to the arguments of the $\Theta$-functions in the phase factor $\Pi$ the integration over $\xi$ and $\eta$ can be split into smaller intervals, in which the monodromy phase  becomes  constant and independent of the world-sheet coordinates. The individual integration patches and the respective phases are displayed in Table \ref{tab::3_rp2} for the case $0<y<1$.
	\begin{table}[htb]
		\centering
			\resizebox{\textwidth}{!}{%
				\begin{tabular}{|c |c |c |c |c |c |c | c|}
					\hline
					        \rule[-1ex]{0pt}{3.5ex}          & $\eta<-1$                                                                        & $-1<\eta<-y$                                                                              & $-y<\eta<\frac1\xi$                                                        & $\frac1\xi<\eta<0$                                       & $0<\eta<1$                                               & $1<\eta<\frac1y$                                                                 & $\frac{1}{y}<\eta$                                                                        \\ \hline
					 \rule[-1ex]{0pt}{3.5ex}$\xi<-\frac1y$   & $ e^{i\pi s_{2\overline3}}e^{i\pi s_{1\overline3}}e^{i\pi s_{23}}e^{i\pi s_{2\overline3}}e^{i\pi s_{3\overline3}}\equiv1$                                                                              & $e^{i\pi s_{1\overline3}}e^{i\pi s_{23}}e^{i\pi s_{2\overline3}}e^{i\pi s_{3\overline3}}$ & $e^{i\pi s_{1\overline3}}e^{i\pi s_{2\overline3}}e^{i\pi s_{3\overline3}}$ & $e^{i\pi s_{1\overline3}}e^{i\pi s_{2\overline3}}$       & $e^{i\pi s_{1\overline3}}e^{i\pi s_{2\overline3}}$       & $e^{i\pi s_{2\overline3}}$                                                       & $1$                                                                                       \\ \hline
					          \multicolumn{1}{c}{}           & \multicolumn{1}{c}{}                                                             & \multicolumn{1}{c}{}                                                                      & \multicolumn{1}{c}{}                                                       & \multicolumn{1}{c}{}                                     & \multicolumn{1}{c}{}                                     & \multicolumn{1}{c}{}                                                             & \multicolumn{1}{c}{}                                                                      \\ \hline
					        \rule[-1ex]{0pt}{3.5ex}          & $\eta<-1$                                                                        & $-1<\eta<\frac1\xi$                                                                       & $\frac1\xi<\eta<-y$                                                        & $-y<\eta<0$                                              & $0<\eta<1$                                               & $1<\eta<\frac1y$                                                                 & $\frac{1}{y}<\eta$                                                                        \\ \hline
					\rule[-1ex]{0pt}{3.5ex}$-\frac1y<\xi<-1$ & $e^{i\pi s_{13}}e^{i\pi s_{1\overline3}}e^{i\pi s_{23}}e^{i\pi s_{3\overline3}}$ & $e^{i\pi s_{1\overline3}}e^{i\pi s_{23}}e^{i\pi s_{3\overline3}}$                         & $e^{i\pi s_{1\overline3}}e^{i\pi s_{23}}$                                  & $e^{i\pi s_{1\overline3}}$                               & $e^{i\pi s_{1\overline3}}$                               & $1$                                                                              & $e^{i\pi s_{2\overline3}}$                                                                \\ \hline
					          \multicolumn{1}{c}{}           & \multicolumn{1}{c}{}                                                             & \multicolumn{1}{c}{}                                                                      & \multicolumn{1}{c}{}                                                       & \multicolumn{1}{c}{}                                     & \multicolumn{1}{c}{}                                     & \multicolumn{1}{c}{}                                                             & \multicolumn{1}{c}{}                                                                      \\ \hline
					        \rule[-1ex]{0pt}{3.5ex}          & $\eta<\frac1\xi$                                                                 & $\frac1\xi<\eta<-1$                                                                       & $-1<\eta<-y$                                                               & $-y<\eta<0$                                              & $0<\eta<1$                                               & $1<\eta<\frac1y$                                                                 & $\frac{1}{y}<\eta$                                                                        \\ \hline
					   \rule[-1ex]{0pt}{3.5ex}$-1<\xi<0$     & $e^{i\pi s_{13}}e^{i\pi s_{23}}e^{i\pi s_{3\overline3}}$                         & $e^{i\pi s_{13}}e^{i\pi s_{23}}$                                                          & $e^{i\pi s_{23}}$                                                          & $1$                                                      & $1$                                                      & $e^{i\pi s_{1\overline3}}$                                                       & $e^{i\pi s_{1\overline3}}e^{i\pi s_{2\overline3}}$                                        \\ \hline
					          \multicolumn{1}{c}{}           & \multicolumn{1}{c}{}                                                             & \multicolumn{1}{c}{}                                                                      & \multicolumn{1}{c}{}                                                       & \multicolumn{1}{c}{}                                     & \multicolumn{1}{c}{}                                     & \multicolumn{1}{c}{}                                                             & \multicolumn{1}{c}{}                                                                      \\ \hline
					        \rule[-1ex]{0pt}{3.5ex}          & $\eta<-1$                                                                        & $-1<\eta<-y$                                                                              & $-y<\eta<0$                                                                & $0<\eta<1$                                              & $1<\eta<\frac1y$                                         & $\frac1y<\eta<\frac1\xi$                                                         & $\frac1\xi<\eta$                                                                          \\ \hline
					   \rule[-1ex]{0pt}{3.5ex}	$0<\xi<y$     & $e^{i\pi s_{13}}e^{i\pi s_{23}}$                                                 & $e^{i\pi s_{23}}$                                                                         & $1$                                                                        & $1$                                                      & $e^{i\pi s_{1\overline3}}$                               & $e^{i\pi s_{1\overline3}}e^{i\pi s_{2\overline3}}$                               & $e^{i\pi s_{1\overline3}}e^{i\pi s_{2\overline3}}e^{i\pi s_{3\overline3}}$                \\ \hline
					          \multicolumn{1}{c}{}           & \multicolumn{1}{c}{}                                                             & \multicolumn{1}{c}{}                                                                      & \multicolumn{1}{c}{}                                                       & \multicolumn{1}{c}{}                                     & \multicolumn{1}{c}{}                                     & \multicolumn{1}{c}{}                                                             & \multicolumn{1}{c}{}                                                                      \\ \hline
					        \rule[-1ex]{0pt}{3.5ex}          & $\eta<-1$                                                                        & $-1<\eta<-y$                                                                              & $-y<\eta<0$                                                                & $0<\eta<1$                                       & $1<\eta<\frac1\xi$                                       & $\frac1\xi<\eta<\frac1y$                                                         & $\frac{1}{y}<\eta$                                                                        \\ \hline
					    \rule[-1ex]{0pt}{3.5ex}$y<\xi<1$     & $e^{i\pi s_{13}}$                                                                & $1$                                                                                       & $e^{i\pi s_{23}}$                                                          & $e^{i\pi s_{23}}$                & $e^{i\pi s_{1\overline3}}e^{i\pi s_{23}}$                & $e^{i\pi s_{1\overline3}}e^{i\pi s_{23}}e^{i\pi s_{3\overline3}}$                & $e^{i\pi s_{1\overline3}}e^{i\pi s_{23}}e^{i\pi s_{2\overline3}}e^{i\pi s_{3\overline3}}$ \\ \hline
					          \multicolumn{1}{c}{}           & \multicolumn{1}{c}{}                                                             & \multicolumn{1}{c}{}                                                                      & \multicolumn{1}{c}{}                                                       & \multicolumn{1}{c}{}                                     & \multicolumn{1}{c}{}                                     & \multicolumn{1}{c}{}                                                             & \multicolumn{1}{c}{}                                                                      \\ \hline
					        \rule[-1ex]{0pt}{3.5ex}          & $\eta<-1$                                                                        & $-1<\eta<-y$                                                                              & $-y<\eta<0$                                                                & $0<\eta<\frac1\xi$                                       & $\frac1\xi<\eta<1$                                       & $1<\eta<\frac1y$                                                                 & $\frac{1}{y}<\eta$                                                                        \\ \hline
					     \rule[-1ex]{0pt}{3.5ex}$1<\xi$      & $1$                                                                              & $e^{i\pi s_{13}}$                                                                         & $e^{i\pi s_{13}}e^{i\pi s_{23}}$                                           & $e^{i\pi s_{13}}e^{i\pi s_{23}}$ & $e^{i\pi s_{13}}e^{i\pi s_{23}}e^{i\pi s_{3\overline3}}$ & $e^{i\pi s_{13}}e^{i\pi s_{1\overline3}}e^{i\pi s_{23}}e^{i\pi s_{3\overline3}}$ & $e^{i\pi s_{2\overline3}}e^{i\pi s_{13}}e^{i\pi s_{1\overline3}}e^{i\pi s_{23}}e^{i\pi s_{3\overline3}}\equiv1$                                                                                       \\ \hline
				\end{tabular}}
			\caption{Phases $\Pi(y,\xi,\eta)$ for each integration region in the $(\xi,\eta)$-plane for $0<y<1$.}\label{tab::3_rp2}
	\end{table}
The integration regions for the amplitude with $-\infty<y<-1$ have almost the same monodromy phases, except one has to simultaneously exchange $k_2 \leftrightarrow D{\cdot}k_2$ and  $y\leftrightarrow-\frac{1}{y}$ such that the vertex operators $2$ and $\overline 2$ interchange positions:
\be
\begin{array}{c|cccc}
 &V_2(y)  &  & U_{\overline 2}\left(-\frac{1}{y}\right)  &  \\
 \hline\\
 0<y<1&y\in]0,1[  &  & -\frac{1}{y}\in]-\infty,-1[ &  \\  \\
 -\infty<y<-1& y\in]-\infty,-1[ &  & -\frac{1}{y}\in]0,1[ & 
\ea\ee
\noindent Therefore, the analysis of these two intervals of $y$--integrations is very similar and we shall only discuss the case $0<y<1$ explicitly. The remaining two regions of $y$-integration $y\in]-1,0[$ and $y\in]1,\infty[$ exhibit the same behaviour. Therefore, in the sequel we shall only work out the case $-1<y<0$.

Some comments are in order. Not all of the integration patches displayed in Table \ref{tab::3_rp2} correspond to open string subamplitudes, but represent only parts of subamplitudes. This is the case for intervals with the boundary value  $\eta=0$, which does not  describe a vertex operator position. Similar as for the scattering of three closed strings on the disk in \cite{Bischof:2023uor} we  have  integration regions, which start or end at $\eta=\pm\infty$. Since also in case at hand  there is no vertex operator positioned at infinity, the corresponding integration regions constitute  only parts of partial amplitudes, cf.\ the discussion at the end of Appendix \ref{sec::KLT} and Footnote \ref{fn::3}. Strictly speaking, splitting the $y$--integration into four integration regions  to discuss  the analytic continuation, implies that all single regions do no longer refer to complete partial subamplitudes. However, this problem is easily bypassed when combining two $y$-integrations. We shall explicitly discuss this property below and demonstrate  that taking into account all four  regions of $y$--integration leads to well defined open string subamplitudes. To streamline the following analysis we ignore this caveat since  it does not affect any analytic properties or the branch cut structure of the $(\xi,\eta)$-integration except for the position of the $y$-dependent poles on the real line. The explicit $y$-integration in the subamplitudes \eqref{eq::open_string_n_point_unfixed} will follow from the context and will always be denoted when considering the total amplitude, cf.\ e.g.\ \eqref{eq::q6}.\par
Nevertheless, the amplitude has the correct branch cut structure such that we can use Cauchy's theorem and construct closed integration contours in the complex plane. This becomes clear when performing a coordinate transformation of the integration regions in Table \ref{tab::4_rp2}. We perform\footnote{After a coordinate transformation only the integration patches change but not their corresponding phase factors since the latter are independent on the world--sheet coordinates in the corresponding regions. This can be seen explicitly by comparing Table \ref{tab::3_rp2} and Table \ref{tab::4_rp2}. Hence, for the phase $\Pi$ we have dropped the dependence on the world--sheet coordinates. However, we have  kept $\Pi$ to indicate that each integration patch enters with a monodromy phase.} the coordinate  change $\eta\to\frac1\eta$ which puts the coordinates $\eta$ and $\xi$ on equal grounds such that integrating over the latter gives rise to poles at the same vertex operator positions $(-\frac1y,-1,y,1)$ and we obtain
	\begin{align}
		\mathcal{A}^{\mathbb{RP}^2}_3\big|_{0<y<1}&=\frac14 ig_c^3T'_p\ 	\Pi\ \int_{0}^1\mathrm{d}y\int\mathrm{d}\xi\,\int\mathrm{d}\eta\,\frac{1}{y^2}\ \llangle[\biggr] V_1(1)V_{\overline 	1}(-1)V_2(y)U_{\overline 2}\left(-\frac{1}{y}\right)U_3(\xi) U_{\overline3}\left(\eta\right)\rrangle[\biggr]\ ,\label{eq::q6}
	\end{align}
where the integration regions are displayed in Table \ref{tab::4_rp2}.
	\begin{table}[H]
	\centering
		\resizebox{\textwidth}{!}{%
			\begin{tabular}{|c |c |c |c |c |c |c | c|}
				\hline
				\rule[-1ex]{0pt}{3.5ex}          & $\eta<\xi$                                                                       & $\xi<\eta<-\frac1y$                                                        & $-\frac1y<\eta<-1$                                                                        & $-1<\eta<0$                      & $0<\eta<y$                                               & $y<\eta<1$                                                                       & $1<\eta$                                                                                  \\ \hline
				\rule[-1ex]{0pt}{3.5ex}$\xi<-\frac1y$   & $e^{i\pi s_{1\overline3}}e^{i\pi s_{2\overline3}}$                               & $e^{i\pi s_{1\overline3}}e^{i\pi s_{2\overline3}}e^{i\pi s_{3\overline3}}$ & $e^{i\pi s_{1\overline3}}e^{i\pi s_{23}}e^{i\pi s_{2\overline3}}e^{i\pi s_{3\overline3}}$ & $1$                              & $1$                                                      & $e^{i\pi s_{2\overline3}}$                                                       & $e^{i\pi s_{1\overline3}}e^{i\pi s_{2\overline3}}$                                        \\ \hline
				\multicolumn{1}{c}{}           & \multicolumn{1}{c}{}                                                             & \multicolumn{1}{c}{}                                                       & \multicolumn{1}{c}{}                                                                      & \multicolumn{1}{c}{}             & \multicolumn{1}{c}{}                                     & \multicolumn{1}{c}{}                                                             & \multicolumn{1}{c}{}                                                                      \\ \hline
				\rule[-1ex]{0pt}{3.5ex}          & $\eta<-\frac1y$                                                                      & $-\frac1y<\eta<\xi$                                                              & $\xi<\eta<-1$                                                                        & $-1<\eta<0$                  & $0<\eta<y$                                                 & $y<\eta<1$                                                                       & $1<\eta$                                                                                \\ \hline
				\rule[-1ex]{0pt}{3.5ex}$-\frac1y<\xi<-1$ & $e^{i\pi s_{1\overline3}}$ & $e^{i\pi s_{1\overline3}}e^{i\pi s_{23}}$          & $e^{i\pi s_{1\overline3}}e^{i\pi s_{23}}e^{i\pi s_{3\overline3}}$                                                 & $e^{i\pi s_{13}}e^{i\pi s_{1\overline3}}e^{i\pi s_{23}}e^{i\pi s_{3\overline3}}$       & $e^{i\pi s_{2\overline3}}$                               & $1$                                                                              & $e^{i\pi s_{1\overline3}}$                                                                \\ \hline
				\multicolumn{1}{c}{}           & \multicolumn{1}{c}{}                                                             & \multicolumn{1}{c}{}                                                       & \multicolumn{1}{c}{}                                                                      & \multicolumn{1}{c}{}             & \multicolumn{1}{c}{}                                     & \multicolumn{1}{c}{}                                                             & \multicolumn{1}{c}{}                                                                      \\ \hline
				\rule[-1ex]{0pt}{3.5ex}          & $\eta<-\frac1y$                                                                     & $-\frac1y<\eta<-1$                                                              & $-1<\eta<\xi$                                                                        & $\xi<\eta<0$                  & $0<\eta<y$                                                 & $y<\eta<1$                                                                       & $1<\eta$                                                                                \\ \hline
				\rule[-1ex]{0pt}{3.5ex}$-1<\xi<0$     & $1$                         & $e^{i\pi s_{23}}$                                           & $e^{i\pi s_{13}}e^{i\pi s_{23}}$                                                                         & $e^{i\pi s_{13}}e^{i\pi s_{23}}e^{i\pi s_{3\overline3}}$                              & $e^{i\pi s_{1\overline3}}e^{i\pi s_{2\overline3}}$                                                      & $e^{i\pi s_{1\overline3}}$                                                       & $1$                                        \\ \hline
				\multicolumn{1}{c}{}           & \multicolumn{1}{c}{}                                                             & \multicolumn{1}{c}{}                                                       & \multicolumn{1}{c}{}                                                                      & \multicolumn{1}{c}{}             & \multicolumn{1}{c}{}                                     & \multicolumn{1}{c}{}                                                             & \multicolumn{1}{c}{}                                                                      \\ \hline
				\rule[-1ex]{0pt}{3.5ex}          & $\eta<-\frac1y$                                                                      & $-\frac1y<\eta<-1$                                                         & $-1<\eta<0$                                                                           & $0<\eta<\xi$                         & $\xi<\eta<y$                                               & $y<\eta<1$                                                                     & $1<\eta$                                                                              \\ \hline
				\rule[-1ex]{0pt}{3.5ex}	$0<\xi<y$     & $1$                                                 & $e^{i\pi s_{23}}$                                                          & $e^{i\pi s_{13}}e^{i\pi s_{23}}$                                                                                       & $e^{i\pi s_{1\overline3}}e^{i\pi s_{2\overline3}}e^{i\pi s_{3\overline3}}$                              & $e^{i\pi s_{1\overline3}}e^{i\pi s_{2\overline3}}$                               & $e^{i\pi s_{1\overline3}}$                               & $1$                \\ \hline
				\multicolumn{1}{c}{}           & \multicolumn{1}{c}{}                                                             & \multicolumn{1}{c}{}                                                       & \multicolumn{1}{c}{}                                                                      & \multicolumn{1}{c}{}             & \multicolumn{1}{c}{}                                     & \multicolumn{1}{c}{}                                                             & \multicolumn{1}{c}{}                                                                      \\ \hline
				\rule[-1ex]{0pt}{3.5ex}          & $\eta<-\frac1y$                                                                      & $-\frac1y<\eta<-1$                                                         & $-1<\eta<0$                                                                           & $0<\eta<y$                         & $y<\eta<\xi$                                             & $\xi<\eta<1$                                                                     & $1<\eta$                                                                                \\ \hline
				\rule[-1ex]{0pt}{3.5ex}$y<\xi<1$     & $e^{i\pi s_{23}}$                                                                & $1$                                                                        & $e^{i\pi s_{13}}$                                                                         & $e^{i\pi s_{1\overline3}}e^{i\pi s_{23}}e^{i\pi s_{2\overline3}}e^{i\pi s_{3\overline3}}$                & $e^{i\pi s_{1\overline3}}e^{i\pi s_{23}}e^{i\pi s_{3\overline3}}$                & $e^{i\pi s_{1\overline3}}e^{i\pi s_{23}}$                & $e^{i\pi s_{23}}$ \\ \hline
				\multicolumn{1}{c}{}           & \multicolumn{1}{c}{}                                                             & \multicolumn{1}{c}{}                                                       & \multicolumn{1}{c}{}                                                                      & \multicolumn{1}{c}{}             & \multicolumn{1}{c}{}                                     & \multicolumn{1}{c}{}                                                             & \multicolumn{1}{c}{}                                                                      \\ \hline
				\rule[-1ex]{0pt}{3.5ex}          & $\eta<-\frac1y$                                                                      & $-\frac1y<\eta<-1$                                                         & $-1<\eta<0$                                                                           & $0<\eta<y$                       & $y<\eta<1$                                             & $1<\eta<\xi$                                                                       & $\xi<\eta$                                                                                \\ \hline
				\rule[-1ex]{0pt}{3.5ex}$1<\xi$      & $e^{i\pi s_{13}}e^{i\pi s_{23}}$                                                                              & $e^{i\pi s_{13}}$                                                          & $1$                                                          & $1$ & $e^{i\pi s_{13}}e^{i\pi s_{1\overline3}}e^{i\pi s_{23}}e^{i\pi s_{3\overline3}}$ & $e^{i\pi s_{13}}e^{i\pi s_{23}}e^{i\pi s_{3\overline3}}$ & $e^{i\pi s_{13}}e^{i\pi s_{23}}$                                                                                       \\ \hline
				\end{tabular}}
		\caption[Phases $\Pi(y,\xi,\eta)$ in the $(\xi,\eta)$-plane for $0<y<1$ after the transformation $\eta\to\frac{1}{\eta}$.]{Phases $\Pi(y,\xi,\eta)$ for each integration region in the $(\xi,\eta)$-plane for $0<y<1$ after the transformation $\eta\to\frac{1}{\eta}$.}\label{tab::4_rp2}
\end{table}
\noindent These integration regions can be associated to open string subamplitudes \eqref{eq::open_string_n_point_unfixed} or parts of partial amplitudes subject to\footnote{For concreteness, here we are only considering $0<y<1$. However, this also holds for the other $y$-integration regions.} $0<y<1$. In \eqref{eq::q6} the fixing  $z_{i_1}\equiv \overline z_1=-1,z_{i_2}\equiv z_1=1$ and $z_{i_3}\equiv z_2=y$  describes a choice of $PSL(2,\mathbb{R})$  frame on
the world--sheet disk of an open string subamplitude. Integration regions with the  $\xi$ or $\eta$-integration starting or ending at $\xi,\eta=0$ do not describe partial amplitudes, because $\xi,\eta=0$ do not correspond to   vertex operator positions. Therefore, we introduce\footnote{For the $y$--integration we do not use this notation, because for $y\in]0,1[$ we have $\overline z_2=-\frac{1}{y}\in]-\infty,1[$ for the vertex operator position. Furthermore, we have already stated that this integration is singled out.} partial subamplitudes of the following form
	\begin{align}
		\mathbb{A}(\overline2,\overline1,\overline3|2,3,1)&=\frac14 ig_c^3T'_p \int_{0}^1\mathrm{d}y\int_y^1\mathrm{d}\xi\,\int_{-1}^0\mathrm{d}\eta\,\frac{1}{y^2}\llangle[\biggr] V_1(1)V_{\overline 	1}(-1)V_2(y)U_{\overline 2}\left(-\frac{1}{y}\right)U_3(\xi) U_{\overline3}\left(\eta\right)\rrangle[\biggr]\label{eq::subamplitude_notation}
	\end{align}
such that the notation $\mathbb A(\ldots,\overline3|\ldots)$ implies that the integration region for $\eta$ ends at $\eta=0$. Likewise for $\mathbb A(\ldots|\overline 3,\ldots)$ the corresponding integration region for $\eta$ begins at $\eta=0$. The same holds for $\xi$, though in this case it would be in principle possible to combine some of these integration regions into proper open string subamplitudes. Hence, if we have two subamplitudes with the same monodromy phase, they  can be joined as $\mathbb A(\ldots,3|\ldots)+\mathbb A(\ldots|3,\ldots)=\mathbb A(\ldots,3,\ldots)$. With this notation and momentum conservation \eqref{eq::q6} becomes:
	\begin{align}
		\mathcal{A}^{\mathbb{RP}^2}_3\big|_{0<y<1}&=\mathbb{A}(3,\overline2,\overline1,\overline3|2,1)+ e^{-i \pi s_{13}} \mathbb{A}(3,\overline2,\overline3,\overline1,2,1)+ e^{-i \pi (s_{13} + s_{23})} \mathbb{A}(3,\overline3,\overline2,\overline1,2,1)\n
		&\h+ e^{i \pi (s_{1\overline3} + s_{2\overline3})} \mathbb{A}(\overline3,3,\overline2,\overline1,2,1)+ e^{i \pi (s_{1\overline3} + s_{2\overline3})} \mathbb{A}(3,\overline2,\overline1,2,1,\overline3)\n
		&\h+ e^{i \pi s_{2\overline3}} \mathbb{A}(3,\overline2,\overline1,2,\overline3,1)+  \mathbb{A}(3,\overline2,\overline1|\overline3,2,1)+ e^{-i \pi s_{2\overline3}} \mathbb{A}(\overline2,3,\overline1,\overline3|2,1)\n
		&\h+ e^{-i \pi (s_{13} + s_{2\overline3})} \mathbb{A}(\overline2,3,\overline3,\overline1,2,1)+ e^{i \pi (s_{1\overline3} + s_{23})} \mathbb{A}(\overline2,\overline3,3,\overline1,2,1)\n
		&\h+ e^{i \pi s_{1\overline3}} \mathbb{A}(\overline3,\overline2,3,\overline1,2,1)+ e^{i \pi s_{1\overline3}} \mathbb{A}(\overline2,3,\overline1,2,1,\overline3)+  \mathbb{A}(\overline2,3,\overline1,2,\overline3,1)\n
		&\h+ e^{i \pi s_{2\overline3}} \mathbb{A}(\overline2,3,\overline1|\overline3,2,1)+ e^{i \pi (s_{13} + s_{23} + s_{3\overline3})} \mathbb{A}(\overline2,\overline1,3,\overline3|2,1)+ \n
		&\h+e^{i \pi (s_{13}+ s_{23})} \mathbb{A}(\overline2,\overline1,\overline3,3|2,1)+ e^{i \pi s_{23}} \mathbb{A}(\overline2,\overline3,\overline1,3|2,1)+\mathbb{A}(\overline3,\overline2,\overline1,3|2,1)\n
		&\h+\mathbb{A}(\overline2,\overline1,3|2,1,\overline3)+ e^{i \pi s_{1\overline3}} \mathbb{A}(\overline2,\overline1,3|2,\overline3,1)+ e^{i \pi (s_{1\overline3} + s_{2\overline3})} \mathbb{A}(\overline2,\overline1,3|\overline3,2,1)\n
		&\h+ e^{i \pi (s_{13} + s_{23})} \mathbb{A}(\overline2,\overline1,\overline3|3,2,1)+ e^{i \pi s_{23}} \mathbb{A}(\overline2,\overline3,\overline1|3,2,1)+  \mathbb{A}(\overline3,\overline2,\overline1|3,2,1)\n
		&\h+  \mathbb{A}(\overline2,\overline1|3,2,1,\overline3)+ e^{i \pi s_{1\overline3}} \mathbb{A}(\overline2,\overline1|3,2,\overline3,1)+ e^{i \pi (s_{1\overline3} + s_{2\overline3})} \mathbb{A}(\overline2,\overline1|3,\overline3,2,1)\n
		&\h+ e^{-i \pi (s_{13} + s_{23})} \mathbb{A}(\overline2,\overline1|\overline3,3,2,1)+ e^{i \pi s_{13}} \mathbb{A}(\overline2,\overline1,\overline3|2,3,1)+  \mathbb{A}(\overline2,\overline3,\overline1,2,3,1)\n
		&\h+ e^{i \pi s_{23}} \mathbb{A}(\overline3,\overline2,\overline1,2,3,1)+ e^{i \pi s_{23}} \mathbb{A}(\overline2,\overline1,2,3,1,\overline3)+ e^{i \pi (s_{1\overline3} + s_{23})} \mathbb{A}(\overline2,\overline1,2,3,\overline3,1)\n
		&\h+ e^{-i \pi (s_{13} + s_{2\overline3})} \mathbb{A}(\overline2,\overline1,2,\overline3,3,1)+ e^{-i \pi s_{13}} \mathbb{A}(\overline2,\overline1|\overline3,2,3,1)+  \mathbb{A}(\overline2,\overline1,\overline3|2,1,3)\n
		&\h+ e^{i \pi s_{13}} \mathbb{A}(\overline2,\overline3,\overline1,2,1,3)+ e^{i \pi (s_{13} + s_{23})} \mathbb{A}(\overline3,\overline2,\overline1,2,1,3)\n
		&\h+ e^{i \pi (s_{13} + s_{23})} \mathbb{A}(\overline2,\overline1,2,1,3,\overline3)+ e^{i \pi (s_{13} + s_{23} + s_{3\overline3})} \mathbb{A}(\overline2,\overline1,2,1,\overline3,3)\n
		&\h+ e^{-i \pi s_{2\overline3}} \mathbb{A}(\overline2,\overline1,2,\overline3,1,3)+  \mathbb{A}(\overline2,\overline1|\overline3,2,1,3)\ .\label{eq::q7}
	\end{align}
Even though, there appear integration regions of the form \eqref{eq::subamplitude_notation} in \eqref{eq::q7} we can find closed contours in the complex $\xi$ or $\eta$--planes, which give rise to closed monodromy relations. 
First of all, for the elements of the first and last row in Table \ref{tab::4_rp2}, which 
correspond to the regions $\xi<-\frac1y$ and $1<\xi$, respectively  the following equations can be stated:
	\begin{align}
		0&=\mathbb{A}(\overline3,3,\overline 2,\overline 1,2,1)+e^{i\pi s_{3\overline3}}\mathbb{A}(3,\overline3,\overline 2,\overline 1,2,1)+e^{i\pi s_{23}}e^{i\pi s_{3\overline3}}\mathbb{A}(3,\overline 2,\overline3,\overline 1,2,1)\n
		&\h+e^{i\pi s_{13}}e^{i\pi s_{23}}e^{i\pi s_{3\overline3}}\mathbb{A}(3,\overline 2,\overline 1,\overline3,2,1)+e^{i\pi s_{13}}e^{i\pi s_{23}}e^{i\pi s_{2\overline3}}e^{i\pi s_{3\overline3}}\mathbb{A}(3,\overline 2,\overline 1,2,\overline3,1)\n
		&\h+\mathbb{A}(3,\overline 2,\overline 1,2,1,\overline3)\ ,\n
		0&=\mathbb{A}(\overline3,\overline 2,\overline 1,2,1,3)+e^{-i\pi s_{23}}\mathbb{A}(\overline 2,\overline3,\overline 1,2,1,3)+e^{-i\pi s_{13}}e^{-i\pi s_{23}}\mathbb{A}(\overline 2,\overline 1,\overline3,2,1,3)\n
		&\h+e^{-i\pi s_{13}}e^{-i\pi s_{23}}e^{-i\pi s_{2\overline3}}\mathbb{A}(\overline 2,\overline 1,2,\overline3,1,3)+e^{-i\pi s_{13}}e^{-i\pi s_{1\overline3}}e^{-i\pi s_{23}}e^{-i\pi s_{2\overline3}}\mathbb{A}(\overline 2,\overline 1,2,1,\overline3,3)\n
		&\h+\mathbb{A}(3,\overline 2,\overline 1,2,1,\overline3)\ ,
	\end{align}
respectively.
As a consequence, the first seven and last seven terms of \req{eq::q7} give a vanishing contribution.
On the other hand, for the integration patches  $-1<\eta<0$ and $0<\eta<y$, i.e.\ the \R{fourth} and \R{fifth} column in Table \ref{tab::4_rp2} we find the relations
	\begin{align}
		0&=\mathbb A(3,\overline2,\overline1,\overline3|2,1)+e^{i\pi s_{13}}e^{i\pi s_{1\overline3}}e^{i\pi s_{23}}e^{i\pi s_{3\overline3}}\mathbb A(\overline2,3,\overline1,\overline3|2,1)\n
		&\h+e^{i\pi s_{13}}e^{i\pi s_{23}}e^{i\pi s_{3\overline3}}\mathbb A(\overline2,\overline1,3,\overline3|2,1)+e^{i\pi s_{13}}e^{i\pi s_{23}}\mathbb A(\overline2,\overline1,\overline3,3|2,1)\n
		&\h+e^{i\pi s_{13}}e^{i\pi s_{23}}\mathbb A(\overline2,\overline1,\overline3|3,2,1)+e^{i\pi s_{13}}\mathbb A(\overline2,\overline1,\overline3|2,3,1)+\mathbb A(\overline2,\overline1,\overline3|2,1,3)\ ,\n
		0&=\mathbb{A}(3,\overline 2,\R{\bar 1}|\overline3,2,1)+e^{i\pi s_{2\overline3}}\mathbb{A}(\overline 2,3,\R{\bar 1}|\overline3,2,1)+e^{i\pi s_{1\overline3}}e^{i\pi s_{2\overline3}}\mathbb{A}(\overline 2,\R{\bar 1},3|\overline3,2,1)\n
		&\h+e^{i\pi s_{1\overline3}}e^{i\pi s_{2\overline3}}\mathbb{A}(\overline 2,\R{\bar 1}|3,\overline3,2,1)+e^{i\pi s_{1\overline3}}e^{i\pi s_{2\overline3}}e^{i\pi s_{3\overline3}}\mathbb{A}(\overline 2,\R{\bar 1}|\overline3,3,2,1)\n
		&\h+e^{i\pi s_{1\overline3}}e^{i\pi s_{23}}e^{i\pi s_{2\overline3}}e^{i\pi s_{3\overline3}}\mathbb{A}(\overline 2,\R{\bar 1}|\overline3,2,3,1)+\mathbb{A}(\overline 2,\R{\bar 1}|\overline3,2,1,3)\ ,\label{Pertisau}
	\end{align}
respectively.  Finally,  we consider the following monodromy equations corresponding to moving the coordinate $\xi$ for  $y<\eta<1$ and moving the coordinate $\eta$ for  $-\frac1y<\xi<-1$
	\begin{align}
		0&=\mathbb{A}(3,\overline 2,\overline1,2,\overline3,1)+e^{i \pi s_{2\overline3}}\mathbb{A}(\overline 2,3,\overline1,2,\overline3,1)+e^{i \pi s_{1\overline3}}e^{i \pi s_{2\overline3}}\mathbb{A}(\overline 2,\overline1,3,2,\overline3,1)\n
		&\h+e^{i \pi s_{1\overline3}}e^{i \pi s_{23}}e^{i \pi s_{2\overline3}}\mathbb{A}(\overline 2,\overline1,2,3,\overline3,1)+e^{i \pi s_{1\overline3}}e^{i \pi s_{23}}e^{i \pi s_{2\overline3}}e^{i \pi s_{3\overline3}}\mathbb{A}(\overline 2,\overline1,2,\overline3,3,1)\n
		&\h+\mathbb{A}(\overline 2,\overline1,2,\R{\bar3},1,3)\ ,\n
		0&=\mathbb{A}(\overline3,\overline 2,3,\overline1,2,1)+e^{i \pi s_{23}}\mathbb{A}(\overline 2,\overline3,3,\overline1,2,1)+e^{i \pi s_{23}}e^{i \pi s_{3\overline3}}\mathbb{A}(\overline 2,3,\overline3,\overline1,2,1)\n
		&\h+e^{i \pi s_{13}}e^{i \pi s_{23}}e^{i \pi s_{3\overline3}}\mathbb{A}(\overline 2,3,\overline1,\overline3,2,1)+e^{i \pi s_{13}}e^{i \pi s_{23}}e^{i \pi s_{2\overline3}}e^{i \pi s_{3\overline3}}\mathbb{A}(\overline 2,3,\overline1,2,\overline3,1)\n
		&\h+\mathbb{A}(\overline 2,3,\overline1,2,1,\overline3)\ ,\label{eq::monodromy_relations_rps}
	\end{align}
	respectively.
Eventually, multiplying  the first equation of \eqref{eq::monodromy_relations_rps} by $e^{-i\pi s_{2\overline3}}$ and the second  by $e^{i\pi s_{1\overline3}}$ together with the two equations \req{Pertisau} allows to eliminate various terms  in \eqref{eq::q7}. Altogether, we obtain:
	\begin{align}
		\mathcal{A}^{\mathbb{RP}^2}_3\big|_{0<y<1}&=-\mathbb{A}(3,\overline2,\overline1,\overline3,2,1)-\R {e^{-i\pi s_{2\overline3}}}\mathbb{A}(3,\overline2,\overline1,2,\overline3,1)-\R{e^{-i\pi s_{2\overline3}}}\mathbb{A}(\overline2,3,\overline1,\overline3,2,1)\n
		&\h-\mathbb{A}(\overline2,{\R 3},\overline1,2,\overline3,1)+\mathbb{A}(\overline3,\overline2,\overline1,3,2,1)+e^{i\pi s_{23}}\mathbb{A}(\overline2,\overline3,\overline1,3,2,1)+\mathbb{A}(\overline2,\overline1,3,2,1,\overline3)\n
		&\h+e^{i\pi s_{23}}\mathbb{A}(\overline3,\overline2,\overline1,2,3,1)+\mathbb{A}(\overline2,\overline3,\overline1,2,3,1)+e^{i\pi s_{23}}\mathbb{A}(\overline2,\overline1,2,3,1,\overline3)\n
		&\h-\mathbb{A}(\overline2,\overline1,\overline3,2,1,3)-\R{e^{-i\pi s_{2\overline3}}}\mathbb{A}(\overline2,\overline1,2,\overline3,1,3)\ .		\label{Maurach}
	\end{align}
As desired the result \req{Maurach} is written only in terms of open string (six--point) subamplitudes \eqref{eq::open_string_n_point_unfixed}. The integration regions of the partial amplitudes are listed in Table \ref{fig::letzte_tabelle}.\par
	\begin{table}[htb]
	\centering
		\resizebox{\textwidth}{!}{%
			\begin{tabular}{|c |c |c |c |c |c |c |}
				\hline
				        \rule[-1ex]{0pt}{3.5ex}          & $\eta<\xi$           & $\xi<\eta<-\frac1y$  & $-\frac1y<\eta<-1$   & $-1<\eta<y$                  & $y<\eta<1$                   & $1<\eta$             \\ \hline
				 \rule[-1ex]{0pt}{3.5ex}$\xi<-\frac1y$   & \multicolumn{1}{c}{} & \multicolumn{1}{c}{} &                      & $-1$                         & $-e^{-i\pi s_{2\overline3}}$ &  \\ \hline
				          \multicolumn{1}{c}{}           & \multicolumn{1}{c}{} & \multicolumn{1}{c}{} & \multicolumn{1}{c}{} & \multicolumn{1}{c}{}         & \multicolumn{1}{c}{}         & \multicolumn{1}{c}{} \\ \hline
				        \rule[-1ex]{0pt}{3.5ex}          & $\eta<-\frac1y$      & $-\frac1y<\eta<\xi$  & $\xi<\eta<-1$        & $-1<\eta<y$                  & $y<\eta<1$                   & $1<\eta$             \\ \hline
				\rule[-1ex]{0pt}{3.5ex}$-\frac1y<\xi<-1$ & \multicolumn{1}{c}{} & \multicolumn{1}{c}{} &                      & $-e^{-i\pi s_{2\overline3}}$ & $-1$                         &  \\ \hline
				          \multicolumn{1}{c}{}           & \multicolumn{1}{c}{} & \multicolumn{1}{c}{} & \multicolumn{1}{c}{} & \multicolumn{1}{c}{}         & \multicolumn{1}{c}{}         & \multicolumn{1}{c}{} \\ \hline
				        \rule[-1ex]{0pt}{3.5ex}          & $\eta<-\frac1y$      & $-\frac1y<\eta<-1$   & $-1<\eta<\xi$        & $\xi<\eta<y$                 & $y<\eta<1$                   & $1<\eta$             \\ \hline
				   \rule[-1ex]{0pt}{3.5ex}$-1<\xi<y$     & $1$                  & $e^{i\pi s_{23}}$    & \multicolumn{1}{c}{} & \multicolumn{1}{c}{}         &                              & $1$                  \\ \hline
				          \multicolumn{1}{c}{}           & \multicolumn{1}{c}{} & \multicolumn{1}{c}{} & \multicolumn{1}{c}{} & \multicolumn{1}{c}{}         & \multicolumn{1}{c}{}         & \multicolumn{1}{c}{} \\ \hline
				        \rule[-1ex]{0pt}{3.5ex}          & $\eta<-\frac1y$      & $-\frac1y<\eta<-1$   & $-1<\eta<y$          & $y<\eta<\xi$                 & $\xi<\eta<1$                 & $1<\eta$             \\ \hline
				    \rule[-1ex]{0pt}{3.5ex}$y<\xi<1$     & $e^{i\pi s_{23}}$    & $1$                  & \multicolumn{1}{c}{} & \multicolumn{1}{c}{}         &                              & $e^{i\pi s_{23}}$    \\ \hline
				          \multicolumn{1}{c}{}           & \multicolumn{1}{c}{} & \multicolumn{1}{c}{} & \multicolumn{1}{c}{} & \multicolumn{1}{c}{}         & \multicolumn{1}{c}{}         & \multicolumn{1}{c}{} \\ \hline
				        \rule[-1ex]{0pt}{3.5ex}          & $\eta<-\frac1y$      & $-\frac1y<\eta<-1$   & $-1<\eta<y$          & $y<\eta<1$                   & $1<\eta<\xi$                 & $\xi<\eta$           \\ \hline
				     \rule[-1ex]{0pt}{3.5ex}$1<\xi$      & \multicolumn{1}{c}{} &                      & $-1$                 & $\R{-e^{-i\pi s_{2\overline3}}}$  & \multicolumn{1}{c}{}         &  \\ \hline
			\end{tabular}}
		\caption[Phases $\Pi(y,\xi,\eta)$ in the $(\xi,\eta)$-plane for $0<y<1$ after applying monodromy relations.]{Phases $\Pi(y,\xi,\eta)$ for each integration region in the $(\xi,\eta)$-plane for $0<y<1$ after applying monodromy relations.}\label{fig::letzte_tabelle}
\end{table}
Similar as for the scattering of three closed strings off a D$p$ brane, we want to change the $PSL(2,\mathbb R)$-frame to convert the subamplitudes from  the representation $\mathbb A$ to $\mathcal A$. After comparing the fixed vertex operator positions in \eqref{eq::q6} with those of \eqref{eq::open_string_n_point_unfixed_2} we perform a $PSL(2,\mathbb{R})$-transformation which maps 
the gauge choice $(-1,y,1)$ to $(0,1,\infty)$. This manipulation has already been applied for the scattering of two closed strings off a D$p$-brane in \cite{Bischof:2020tnf,Hashimoto:1996bf,Garousi:1996ad}, see e.g.\ equation $(3.9)$ in \cite{Bischof:2020tnf}. To find this transformation we consider a general fractional linear transformation $\tilde z:=f(z)=\frac{az+b}{cz+d}$ with $ad-bc=1$. The parameters $a,b,c$ and $d$ are given by solving $f(z)=\tilde z$ for $z\in\{-1,y,1\}$ and $\tilde z\in\{0,1,\infty\}$ leading to the transformation:
\begin{equation}
	\frac{1}{\sqrt{2(1-y^2)}}
	\begin{pmatrix}
		1-y&1-y\\
		-(1+y)&1+y
	\end{pmatrix}\in PSL(2,\mathbb{R})\ .\label{eq::psl}
\end{equation}
The above $PSL(2,\mathbb R)$ matrix gives rise to the fractional linear transformation for the coordinates of the vertex operators
\begin{equation}
	\tilde z=f(z)=\frac{(1-y)(1+z)}{(1+y)(1-z)}\ ,\label{eq::PSL2R_transformation}
\end{equation}
which changes the gauge choice from $(-1,y,1)$ to $(0,1,\infty)$. Furthermore, we define the new variables:
\begin{align}
	x & =  f(-y) =\frac{(1-y)^2}{(1+y)^2}\ ,            \n
	\tilde\xi & = f(\xi) = \frac{(1-y)(1+\xi)}{(1+y)(1-\xi)}\ ,               \n
	\tilde\eta & =  f(\eta) = \frac{(1-y)(1+\eta)}{(1+y)(1-\eta)}\ .\label{eq::PSL_2_R_disk}
\end{align}
In addition it is useful to consider the inverse transformations, which are given by\footnote{The transformation of $y$ here is the same as in $(3.9)$ in \cite{Bischof:2020tnf}.}:
\begin{align}
	y & = \frac{1-\sqrt x}{1+\sqrt x}\ ,                       \n
	\xi & = \frac{\tilde\xi-\sqrt{x}}{\tilde\xi+\sqrt{x}}\ ,          \n
	\eta & = \frac{\tilde\eta-\sqrt{x}}{\tilde\eta+\sqrt{x}}\ .\label{sol1}
\end{align}
Note that \eqref{eq::PSL_2_R_disk} allows for the second solution $y=\frac{1+\sqrt{x}}{1-\sqrt{x}}$ for $x=f(-y)$. For the moment we shall ignore this  solution, because for $x\in[0,1]$ we would have $y\in[1,\infty[$. However, this solution will be discussed  in \eqref{nowimp} and will be employed  (together with \req{employed}) for  \eqref{eq::result_2} to account for the region $x\in ]1,\infty[$ and $x\in]0,1[$ giving rise to $y\in ]-\infty,-1[$ and $y\in ]1,\infty]$, respectively.\par
Since the unintegrated vertex operators have conformal weight zero and the integrated vertex operators have weight one, they transform under global conformal transformations as
\begin{equation}
	V(z)\to \tilde V(\tilde z)\ ,\qquad U(z)\to \left(\frac{\partial z}{\partial \tilde z}\right)^{-1}\tilde U(\tilde z)\ .
\end{equation}
With this the $PSL(2,\mathbb R)$-transformation of the integrand can be sketched as:
\begin{align}
	\int\mathrm dy\,\mathrm d\xi\,\mathrm d\eta\,&\llangle V_1(1)V_{\overline1}(-1)V_2(y)U_{\overline2}\lf(\R{-\fc{1}{y}}\ri)U_3(\xi)U_{\overline3}(\eta)\rrangle\nonumber\\
	&\simeq\frac12\int\mathrm dx\,\mathrm d\tilde\xi\,\mathrm d\tilde\eta\,\det\frac{\partial(y,\xi,\eta)}{\partial(x,\tilde\xi,\tilde\eta)}\nonumber\\
	&\times\llangle[\biggr] \tilde V_1(\infty)\tilde V_{\overline1}(0)\tilde V_2(1)\left(\frac{\partial y}{\partial x}\right)^{-1}\tilde U_{\overline2}(\R{-x})\left(\frac{\partial\xi}{\partial \tilde\xi}\right)^{-1}\tilde U_3(\tilde\xi)\left(\frac{\partial\eta}{\partial\tilde \eta}\right)^{-1}\tilde U_{\overline3}(\tilde\eta)\rrangle[\biggr]\nonumber\\
	&=\frac12\int\mathrm dx\,\mathrm d\tilde\xi\,\mathrm d\tilde\eta\ \llangle V_1(\infty)V_{\overline1}(0)V_2(1)U_{\overline2}(\R{-x})U_3(\tilde\xi)U_{\overline3}(\tilde\eta)\rrangle\ .\label{eq::psl_amplitude}
\end{align}
To go from the second to the last line we have used the fact that the correlator is invariant under $PSL(2,\mathbb R)$-transformations. Moreover, we have the freedom to consider either the vertex operator position of $2$ fixed at $y$ or that of $\overline 2$ fixed at $\R{-\fc{1}{y}}$ while integrating over the other one \cite{Bischof:2020tnf}. This freedom leads to an additional factor of two, which is compensated by introducing the factor $\frac12$ in \eqref{eq::psl_amplitude}. In addition, the derivative factors stemming from the transformation of the integrated vertex operators cancel against the Jacobian of the measure\footnote{\R{Note, that we have the identity $\det\lf(\frac{\partial(y,\xi,\eta)}{\partial(x,\tilde\xi,\tilde\eta)}\ri)\left(\frac{\partial y}{\partial x}\right)^{-1}\left(\frac{\partial\xi}{\partial \tilde\xi}\right)^{-1}\left(\frac{\partial\eta}{\partial\tilde \eta}\right)^{-1}=1$, which holds for both solutions \req{sol1} and \req{employed}.}\label{fn::17}} such that the amplitude can be written as
	\begin{align}
			\mathcal{A}^{\mathbb{RP}^2}_3\big|_{0<y<1}&=\frac18 ig_c^3T'_p \Pi\int_{0}^1\mathrm{d}x\int\mathrm{d}\tilde \xi\,\int\mathrm{d}\tilde\eta\,\llangle V_{\overline 1}(0)U_{\overline 2}(-x)U_3(\tilde\xi) U_{\overline3}\left(\tilde{\eta}\right)V_2(1)V_1(\infty)\rrangle\ ,\label{eq::amplitude_rp_2_psl}
		\end{align}
with the boundaries of the $\tilde\xi$ and $\tilde\eta$--integrations given in Table \ref{tab::5_rp_2}. Furthermore, in \eqref{eq::amplitude_rp_2_psl} we made use of \eqref{eq::psl_amplitude} and that according to  \eqref{eq::PSL2R_transformation} the vertex position $-\tfrac{1}{y}$ is mapped to $-x$. 

The integration over $x$ is still singled out since the integration range $-x\in]-1,0[$ does not correspond to an open string integral.\footnote{As for the scattering of three closed strings on the disk the $PSL(2,\mathbb R)$-transformation \eqref{eq::psl_amplitude} maps the integration over $0<y<1$ onto $0<x<1$.} However, this issue is removed, since  the integration over $y$ from $-\infty$ to $-1$ is  mapped to $-x\in]-\infty,-1[$ such that both regions can be combined to form proper subamplitudes.\par
The explicit computation of the correlator in \eqref{eq::amplitude_rp_2_psl} is similar to that  in Section \ref{sec::open_string_correlator}. After combining partial amplitudes $\mathbb A$ discussed in  \cite{Bischof:2023uor} (cf.\ also Appendix \ref{sec::KLT}) as
\begin{equation}
	\mathbb A(3,\overline3,\overline1,\overline2,2,1)+\mathbb A(\overline1,\overline2,2,1,3,\overline3)+\mathbb A(\overline3,\overline1,\overline2,2,1,3)\xrightarrow[]{PSL(2,\mathbb{R})}\mathcal A(3,4,1,2,5,6)\ , \label{eq::combine}
\end{equation}
the individual integration patches of the amplitude \eqref{eq::amplitude_rp_2_psl} are given in Table \ref{tab::5_rp_2}. 
	\begin{table}[H]
			\centering
							\begin{tabular}{|c |c |c |c |c |c |}
									\hline
									     \rule[-1ex]{0pt}{3.5ex}      & $\tilde\eta<\tilde\xi$                   & $\tilde\xi<\tilde\eta<-x$                & $-x<\tilde\eta<0$          & $0<\tilde\eta<1$           & $1<\tilde\eta$                   \\ \hline
									 \rule[-1ex]{0pt}{3.5ex}$\tilde\xi<-x$  & \multicolumn{1}{c}{}         & \multicolumn{1}{c}{}         &                      & $-1$                 & $-e^{-i\pi s_{2\overline3}}$ \\ \hline
									      \multicolumn{1}{c}{}        & \multicolumn{1}{c}{}         & \multicolumn{1}{c}{}         & \multicolumn{1}{c}{} & \multicolumn{1}{c}{} & \multicolumn{1}{c}{}       \\ \hline
									     \rule[-1ex]{0pt}{3.5ex}      & $\tilde\eta<-x$                    & $-x<\tilde\eta<\tilde\xi$                & $\tilde\xi<\tilde\eta<0$         & $0<\tilde\eta<1$           & $1<\tilde\eta$                   \\ \hline
									\rule[-1ex]{0pt}{3.5ex}$-x<\tilde\xi<0$ & \multicolumn{1}{c}{}         & \multicolumn{1}{c}{}         &                      & $-e^{-i\pi s_{2\overline3}}$    & $-1$                        \\ \hline
									      \multicolumn{1}{c}{}        & \multicolumn{1}{c}{}         & \multicolumn{1}{c}{}         & \multicolumn{1}{c}{} & \multicolumn{1}{c}{} & \multicolumn{1}{c}{}       \\ \hline
									     \rule[-1ex]{0pt}{3.5ex}      & $\tilde\eta<-x$                    & $-x<\tilde\eta<0$                  & $0<\tilde\eta<\tilde\xi$         & $\tilde\xi<\tilde\eta<1$         & $1<\tilde\eta$                   \\ \hline
									\rule[-1ex]{0pt}{3.5ex}$0<\tilde\xi<1$  & $1$                         & $e^{i\pi s_{23}}$ & \multicolumn{1}{c}{} & \multicolumn{1}{c}{} &                            \\ \hline
									      \multicolumn{1}{c}{}        & \multicolumn{1}{c}{}         & \multicolumn{1}{c}{}         & \multicolumn{1}{c}{} & \multicolumn{1}{c}{} & \multicolumn{1}{c}{}       \\ \hline
									     \rule[-1ex]{0pt}{3.5ex}      & $\tilde\eta<-x$                    & $-x<\tilde\eta<0$                  & $0<\tilde\eta<1$           & $1<\tilde\eta<\tilde\xi$         & $\tilde\xi<\tilde\eta$                \\ \hline
									 \rule[-1ex]{0pt}{3.5ex}	$1<\tilde\xi$  & $e^{i\pi s_{23}}$ & $1$                         & \multicolumn{1}{c}{} & \multicolumn{1}{c}{} &                            \\ \hline
								\end{tabular}
					\caption{Phases $\Pi(x,\tilde\xi,\tilde\eta)$ for each integration region in the $(\tilde\xi,\tilde\eta)$-plane for $0<x<1$.}\label{tab::5_rp_2}
		\end{table}\noindent
		They  can be written in terms of \eqref{eq::open_string_n_point_unfixed_2} and combine\footnote{Also here the amplitudes are only parts of subamplitudes, as the $x$-integration runs from $0$ to $1$, c.f.\ also the comment above. As before we ignore this subtlety here, but make it explicit in the end result \eqref{eq::result_1} of this 
computation.} into the result:
	\begin{align}
		\mathcal{A}^{\mathbb{RP}^2}_3\big|_{0<y<1}&=e^{i\pi s_{23}}\mathcal{A}(\overline3,\overline2,\overline1,2,3,1)+\mathcal{A}(\overline2,\overline3,\overline1,2,3,1)+\mathcal{A}(\overline3,\overline2,\overline1,3,2,1)\nonumber\\
		&\h+e^{i\pi s_{23}}\mathcal{A}(\overline2,\overline3,\overline1,3,2,1)-e^{-i\pi s_{2\overline3}}\mathcal{A}(\overline2,3,\overline1,\overline3,2,1)-\mathcal{A}(\overline2,3,\overline1,2,\overline3,1)\n
		&\h-\mathcal{A}(3,\overline2,\overline1,\overline3,2,1)-e^{-i\pi s_{2\overline3}}\mathcal{\mathcal{A}}(3,\overline2,\overline1,2,\overline3,1)\ .\label{eq::q8}
	\end{align}
\noindent As discussed in Appendix \ref{sec::KLT_minimal} the partial subamplitudes appearing in \req{eq::q8} can be written in terms of a six--dimensional basis. Therefore, we can use open string monodromy relations  to further simplify the expression \req{eq::q8}. One can verify that in \req{eq::q8} all eight subamplitudes have a neighbouring subamplitude, i.e.\ one partial amplitude  can be paired with an other one where two orderings are permuted with the correct monodromy phase factor. Hence, we shall consider the following four relations
	\begin{align}
		(i)\ 0&=\mathcal{A}(3,\overline2,\overline1,2,\overline3,1)+\mathcal{A}(\overline2,3,\overline1,2,\overline3,1) e^{i \pi  s_{2\overline3}}+\mathcal{A}(\overline2,\overline1,3,2,\overline3,1) e^{i \pi  (s_{1\overline3}+s_{2\overline3})}\nonumber\\
		&\h+\mathcal{A}(\overline2,\overline1,2,3,\overline3,1) e^{i \pi  (s_{1\overline3}+s_{23}+s_{2\overline3})}+\mathcal{A}(\overline2,\overline1,2,\overline3,3,1) e^{i \pi  (s_{1\overline3}+s_{23}+s_{2\overline3}+s_{3\overline3})}\ ,\nonumber\\
		(ii)\ 0&=\mathcal{A}(3,\overline2,\overline1,\overline3,2,1)+\mathcal{A}(\overline2,3,\overline1,\overline3,2,1) e^{-i \pi  s_{2\overline3}}+\mathcal{A}(\overline2,\overline1,3,\overline3,2,1) e^{-i \pi  (s_{1\overline3}+s_{2\overline3})}\nonumber\\
		&\h+\mathcal{A}(\overline2,\overline1,\overline3,3,2,1) e^{-i \pi  (s_{1\overline3}+s_{2\overline3}+s_{3\overline3})}+\mathcal{A}(\overline2,\overline1,\overline3,2,3,1) e^{-i \pi  (s_{1\overline3}+s_{23}+s_{2\overline3}+s_{3\overline3})}\ ,\nonumber\\
		(iii)\ 0&=\mathcal{A}(\overline3,\overline2,\overline1,3,2,1)+\mathcal{A}(\overline2,\overline3,\overline1,3,2,1) e^{i \pi  s_{23}}+\mathcal{A}(\overline2,\overline1,\overline3,3,2,1) e^{i \pi  (s_{13}+s_{23})}\nonumber\\
		&\h+\mathcal{A}(\overline2,\overline1,3,\overline3,2,1) e^{i \pi  (s_{13}+s_{23}+s_{3\overline3})}+\mathcal{A}(\overline2,\overline1,3,2,\overline3,1) e^{i \pi  (s_{13}+s_{23}+s_{2\overline3}+s_{3\overline3})}\ ,\nonumber\\
	(iv)\ 	0&=\mathcal{A}(\overline3,\overline2,\overline1,2,3,1)+\mathcal{A}(\overline2,\overline3,\overline1,2,3,1) e^{-i \pi  s_{23}}+\mathcal{A}(\overline2,\overline1,\overline3,2,3,1) e^{-i \pi  (s_{13}+s_{23})}\nonumber\\
		&\h+\mathcal{A}(\overline2,\overline1,2,\overline3,3,1) e^{-i \pi  (s_{13}+s_{23}+s_{2\overline3})}+\mathcal{A}(\overline2,\overline1,2,3,\overline3,1) e^{-i \pi  (s_{13}+s_{23}+s_{2\overline3}+s_{3\overline3})}
	\end{align}
and the combination $-e^{-i \pi  s_{2\overline3}}\times(i)-(ii)+(iii)+e^{i \pi  s_{23}}\times(iv)$ thereof
to express the amplitude \eqref{eq::q8} in terms of only two open string partial amplitudes
\begin{align}
	\mathcal{A}^{\mathbb{RP}^2}_3\big|_{0<y<1}&=2i\sin(\pi s_{1\overline3})\ \mathcal A(\overline2,\overline1,3,2,\overline3,1)\big|_{0<x<1}+2i\sin(\pi s_{13})\ \mathcal A(\overline2,\overline1,\overline3,2,3,1)\big|_{0<x<1}\ .\label{eq::result_1}
\end{align}
With an analogous calculation we obtain for the region $-\infty<y<-1$ 
\begin{align}
	\mathcal{A}^{\mathbb{RP}^2}_3\big|_{-\infty<y<-1}&=2i\sin(\pi s_{1\overline3})\ \mathcal A(\overline2,\overline1,3,2,\overline3,1)\big|_{1<x<\infty}+2i\sin(\pi s_{13})\ \mathcal A(\overline2,\overline1,\overline3,2,3,1)\big|_{1<x<\infty}\ ,\label{eq::result_2}
\end{align}
which is not the same as \eqref{eq::result_1} because the $x$-integration in \eqref{eq::result_2} goes from $1$ to $\infty$, cf.\ also the comment in the paragraph below \req{eq::amplitude_rp_2_psl}.
Note that we have to use the second solution of \eqref{eq::PSL2R_transformation} to change the $PSL(2,\mathbb R)$-frame, because the $y$-integration requires that 
\begin{equation}
	y=\frac{1+\sqrt x}{1-\sqrt x}\label{nowimp}
\end{equation}
in order to map $1<x<\infty$ back to $-\infty<y1$. Consequently, the inverse transformations for $\eta$ and $\xi$ are given by
\begin{align}
	\xi & = \frac{\tilde\xi+\sqrt{x}}{\tilde\xi-\sqrt{x}}\ ,          \n
	\eta & = \frac{\tilde\eta+\sqrt{x}}{\tilde\eta-\sqrt{x}}\ .\label{employed}
\end{align}
After performing the coordinate change $x\to -x$ we can combine the two results and obtain\footnote{This is possible, because the amplitude has no pole at $x=-1$ after the transformation. Moreover, we are considering complete subamplitudes \eqref{eq::open_string_n_point_unfixed_2} here.}:
\begin{align}
	\mathcal{A}^{\mathbb{RP}^2}_3\big|_{-\infty<y<-1}+\mathcal{A}^{\mathbb{RP}^2}_3\big|_{0<y<1}&=2i\sin(\pi s_{1\overline3})\mathcal A(\overline2,\overline1,3,2,\overline3,1)+2i\sin(\pi s_{13})\mathcal A(\overline2,\overline1,\overline3,2,3,1)\n
	&=\frac14g_c^3T'_p\sum_{\sigma\in S_3}\left\{\sin(\pi s_{1\overline3})F^{\sigma(\overline2,3,\overline3)}_{\mathcal{I}_1}+\sin(\pi s_{13}) F^{\sigma(\overline2,3,\overline3)}_{\mathcal{I}_2}\right\}\n
	&\h\times A_\text{YM}(\overline1,\sigma(\overline2,3,\overline3),2,1)\ .\label{res1}
\end{align}
In the last step we have used the results from Section \ref{sec::open_string_correlator} to express the correlator of \eqref{eq::result_1} and \eqref{eq::result_2} in terms of YM subamplitudes and hypergeometric integrals. The hypergeometric integrals are defined as 
\begin{equation}
	F^{\sigma(\overline2, 3,\overline 3)}_{\mathcal
		I_p}=-\int_{\mathcal{I}_p}\mathrm{d}z_{\overline2}\,\mathrm{d}z_3\,\mathrm{d}z_{\overline3}\,\biggl(\prod_{i<j}|z_{ij}|^{s_{ij}}\biggr)\frac{s_{\overline1\sigma(\overline2)}}{z_{\overline1\sigma(\overline2)}}\frac{s_{\sigma(\overline3)2}}{z_{\sigma(\overline3)
			2}}\left(\frac{s_{\overline1\sigma(3)}}{z_{\overline1\sigma(3)}}+\frac{s_{\sigma(\overline2)\sigma(3)}}{z_{\sigma(\overline2)\sigma(3)}}\right)\ ,\quad\ p=1,2\label{Sample12}
\end{equation}
and their integration regions are given by:
\begin{align}
	\mathcal{I}_1&=\left\{x,\R{\tilde\xi,\tilde\eta}\in\mathbb{R}\ |\ -\infty<\R{x}<0,0<\R{\tilde\xi}<1,1<\R{\tilde\eta}<\infty\right\}\ ,\n
	\mathcal{I}_2&=\left\{x,\R{\tilde\xi,\tilde\eta}\in\mathbb{R}\ |\ -\infty<\R{x}<0,1<\R{\tilde\xi}<\infty,0<\R{\tilde\eta}<1\right\}\ .\label{domain1}
\end{align}

Finally, the computation for the case $-1<y<0$ is similar to the case  $0<y<1$ exhibited above. Likewise, the integration over $1<y<\infty$ is close to the case $-\infty<y<-1$. Therefore, as before we shall only display the most important steps for the integration region $-1<y<0$. After the transformation $\eta\to\frac1\eta$ the amplitude
	\begin{align}
		\mathcal{A}^{\mathbb{RP}^2}_3\big|_{-1<y<0}&=\frac14 ig_c^3T'_p \int_{-1}^0\mathrm{d}y\int_{-\infty}^\infty\mathrm{d}\xi\,\int_{-\infty}^\infty\mathrm{d}\eta\,\frac{1}{y^2}\Pi(y,\xi,\eta)\n
		&\h\times\llangle[\biggr] V_1(1)V_{\overline 	1}(-1)V_2(\R{iy})U_{\overline 2}\left(-\frac{1}{y}\right)U_3(\xi) U_{\overline3}\left(\eta\right)\rrangle[\biggr]\label{eq::q6.2}
	\end{align}
can be written in terms of open string subamplitudes \eqref{eq::open_string_n_point_unfixed} as
	\begin{align}
		\mathcal{A}^{\mathbb{RP}^2}_3\big|_{-1<y<0}&=\mathbb{A}(3,\overline1,2,\overline3|1,\overline2) + e^{-i \pi  s_{2\overline3}} \mathbb{A}(3,\overline1,\overline3,2,1,\overline2) + e^{-i \pi  (s_{13} + s_{2\overline3})} \mathbb{A}(3,\overline3,\overline1,2,1,\overline2) \n
		&\h+ e^{i \pi  (s_{1\overline3} + s_{23})} \mathbb{A}(\overline3,3,\overline1,2,1,\overline2) + e^{i \pi  (s_{1\overline3} + s_{23})} \mathbb{A}(3,\overline1,2,1,\overline2,\overline3) \n
		&\h+ e^{i \pi  s_{1\overline3}} \mathbb{A}(3,\overline1,2,1,\overline3,\overline2) +  \mathbb{A}(3,\overline1,2|\overline3,1,\overline2) + e^{-i \pi s_{1\overline3}} \mathbb{A}(\overline1,3,2,\overline3|1,\overline2) \n
		&\h+ e^{i \pi  (s_{13} + s_{23} + s_{3\overline3})} \mathbb{A}(\overline1,3,\overline3,2,1,\overline2) + e^{i \pi  (s_{13} + s_{23})} \mathbb{A}(\overline1,\overline3,3,2,1,\overline2) \n
		&\h+ e^{i \pi  s_{23}} \mathbb{A}(\overline3,\overline1,3,2,1,\overline2) + e^{i \pi  s_{23}} \mathbb{A}(\overline1,3,2,1,\overline2,\overline3) +  \mathbb{A}(\overline1,3,2,1,\overline3,\overline2) \n
		&\h+ e^{i \pi  s_{1\overline3}} \mathbb{A}(\overline1,3,2|\overline3,1,\overline2) + e^{i \pi  (s_{13} + s_{2\overline3} + s_{3\overline3})} \mathbb{A}(\overline1,2,3,\overline3|1,\overline2) \n
		&\h+ e^{i \pi  (s_{13} + s_{2\overline3})} \mathbb{A}(\overline1,2,\overline3,3|1,\overline2) + e^{i \pi  s_{13}} \mathbb{A}(\overline1,\overline3,2,3|1,\overline2) +  \mathbb{A}(\overline3,\overline1,2,3|1,\overline2) \n
		&\h+  \mathbb{A}(\overline1,2,3|1,\overline2,\overline3) + e^{i \pi  s_{23}} \mathbb{A}(\overline1,2,3|1,\overline3,\overline2) + e^{i \pi  (s_{1\overline3} + s_{23})} \mathbb{A}(\overline1,2,3|\overline3,1,\overline2) \n
		&\h+ e^{i \pi  (s_{13} + s_{2\overline3})} \mathbb{A}(\overline1,2,\overline3|3,1,\overline2) + e^{i \pi  s_{13}} \mathbb{A}(\overline1,\overline3,2|3,1,\overline2) +  \mathbb{A}(\overline3,\overline1,2|3,1,\overline2) \n
		&\h+  \mathbb{A}(\overline1,2|3,1,\overline2,\overline3) + e^{i \pi  s_{23}} \mathbb{A}(\overline1,2|3,1,\overline3,\overline2) + e^{i \pi  (s_{1\overline3} + s_{23})} \mathbb{A}(\overline1,2|3,\overline3,1,\overline2) \n
		&\h+ e^{-i \pi  (s_{13} + s_{2\overline3})} \mathbb{A}(\overline1,2|\overline3,3,1,\overline2) + e^{i \pi  s_{2\overline3}} \mathbb{A}(\overline1,2,\overline3|1,3,\overline2) +  \mathbb{A}(\overline1,\overline3,2,1,3,\overline2) \n
		&\h+ e^{i \pi  s_{13}} \mathbb{A}(\overline3,\overline1,2,1,3,\overline2) + e^{i \pi  s_{13}} \mathbb{A}(\overline1,2,1,3,\overline2,\overline3) + e^{i \pi  (s_{13} + s_{23})} \mathbb{A}(\overline1,2,1,3,\overline3,\overline2) \n
		&\h+ e^{i \pi  (s_{13} + s_{23} + s_{3\overline3})} \mathbb{A}(\overline1,2,1,\overline3,3,\overline2) + e^{-i \pi  s_{2\overline3}} \mathbb{A}(\overline1,2|\overline3,1,3,\overline2) +  \mathbb{A}(\overline1,2,\overline3|1,\overline2,3)\n
		&\h + e^{i \pi  s_{2\overline3}} \mathbb{A}(\overline1,\overline3,2,1,\overline2,3) + e^{i \pi  (s_{13} + s_{2\overline3})} \mathbb{A}(\overline3,\overline1,2,1,\overline2,3) \n
		&\h+ e^{i \pi  (s_{13} + s_{2\overline3})} \mathbb{A}(\overline1,2,1,\overline2,3,\overline3) + e^{i \pi  (s_{13} + s_{2\overline3} + s_{3\overline3})} \mathbb{A}(\overline1,2,1,\overline2,\overline3,3) \n
		&\h+ e^{-i \pi  s_{1\overline3}} \mathbb{A}(\overline1,2,1,\overline3,\overline2,3) +  \mathbb{A}(\overline1,2|\overline3,1,\overline2,3)\ .\label{eq::q13}
	\end{align}
The six variations of the monodromy relations in \eqref{eq::monodromy} for $n=6$
	\begin{align}
		0&=\mathbb A(3,\overline1,2,\overline3|1,\overline2)+e^{i\pi s_{13}}e^{i\pi s_{23}}e^{i\pi s_{2\overline3}}e^{i\pi s_{3\overline3}}\mathbb A(\overline1,3,2,\overline3|1,\overline2)\n
		&\h+e^{i\pi s_{13}}e^{i\pi s_{2\overline3}}e^{i\pi s_{3\overline3}}\mathbb A(\overline1,2,3,\overline3|1,\overline2)+e^{i\pi s_{13}}e^{i\pi s_{2\overline3}}\mathbb A(\overline1,2,\overline3,3|1,\overline2)\n
		&\h+e^{i\pi s_{13}}e^{i\pi s_{23}}\mathbb A(\overline1,2,\overline3|3,1,\overline2)+e^{i\pi s_{2\overline3}}\mathbb A(\overline1,2,\overline3|1,3,\overline2)+\mathbb A(\overline1,2,\overline3|1,\overline2,3)\ ,\n
		0&=\mathbb{A}(3,\overline1,2|\overline3,1,\overline2)+e^{i\pi s_{1\overline3}}\mathbb{A}(\overline1,3,2|\overline3,1,\overline2)+e^{i\pi s_{1\overline3}}e^{i\pi s_{23}}\mathbb{A}(\overline1,2,3|\overline3,1,\overline2)\n
		&\h+e^{i\pi s_{1\overline3}}e^{i\pi s_{23}}\mathbb{A}(\overline1,2|3,\overline3,1,\overline2)+e^{i\pi s_{1\overline3}}e^{i\pi s_{23}}e^{i\pi s_{3\overline3}}\mathbb{A}(\overline1,2|\overline3,3,1,\overline2)\n
		&\h+e^{i\pi s_{13}}e^{i\pi s_{1\overline3}}e^{i\pi s_{23}}e^{i\pi s_{3\overline3}}\mathbb{A}(\overline1,2|\overline3,1,3,\overline2)+\mathbb{A}(\overline1,2|\overline3,1,\overline2,3)\ ,\n
		0&=\mathbb{A}(\overline3,3,\overline1,2,1,\overline2)+ e^{i\pi s_{3\overline3}}\mathbb{A}(3,\overline3,\overline1,2,1,\overline2)+e^{i\pi s_{13}}e^{i\pi s_{3\overline3}}\mathbb{A}(3,\overline1,\overline3,2,1,\overline2)\n
		&\h+e^{i\pi s_{13}}e^{i\pi s_{2\overline3}}e^{i\pi s_{3\overline3}}\mathbb{A}(3,\overline1,2,\overline3,1,\overline2)+e^{i\pi s_{13}}e^{i\pi s_{1\overline3}}e^{i\pi s_{2\overline3}}e^{i\pi s_{3\overline3}}\mathbb{A}(3,\overline1,2,1,\overline3,\overline2)\n
		&\h+\mathbb{A}(3,\overline1,2,1,\overline2,\overline3)\ ,\n
		0&=\mathbb{A}(\overline3,\overline1,2,1,\overline2,3)+e^{-i\pi s_{13}}\mathbb{A}(\overline1,\overline3,2,1,\overline2,3)+e^{-i\pi s_{13}}e^{-i\pi s_{2\overline3}}\mathbb{A}(\overline1,2,\overline3,1,\overline2,3)\n
		&\h+e^{-i\pi s_{13}}e^{-i\pi s_{1\overline3}}e^{-i\pi s_{2\overline3}}\mathbb{A}(\overline1,2,1,\overline3,\overline2,3)+e^{-i\pi s_{13}}e^{-i\pi s_{1\overline3}}e^{-i\pi s_{23}}e^{-i\pi s_{2\overline3}}\mathbb{A}(\overline1,2,1,\overline2,\overline3,3)\n
		&\h+\mathbb{A}(\overline1,2,1,\overline2,3,\overline3)\ ,\n
		0&=\mathbb{A}(3,\overline1,2,1,\overline3,\overline2)+e^{i \pi s_{1\overline3}}\mathbb{A}(\overline1,3,2,1,\overline3,\overline2)+e^{i \pi s_{1\overline3}}e^{i \pi s_{23}}\mathbb{A}(\overline1,2,3,1,\overline3,\overline2)\n
		&\h+e^{i \pi s_{13}}e^{i \pi s_{1\overline3}}e^{i \pi s_{23}}\mathbb{A}(\overline1,2,1,3,\overline3,\overline2)+e^{i \pi s_{13}}e^{i \pi s_{1\overline3}}e^{i \pi s_{23}}e^{i \pi s_{3\overline3}}\mathbb{A}(\overline1,2,1,\overline3,3,\overline2)\n
		&\h+\mathbb{A}(\overline1,2,1,\overline3,\overline2,3)\ ,\n
		0&=\mathbb{A}(\overline3,\overline1,3,2,1,\overline2)+e^{i \pi s_{13}}\mathbb{A}(\overline1,\overline3,3,2,1,\overline2)+e^{i \pi s_{13}}e^{i \pi s_{3\overline3}}\mathbb{A}(\overline1,3,\overline3,2,1,\overline2)\n
		&\h+e^{i \pi s_{13}}e^{i \pi s_{2\overline3}}e^{i \pi s_{3\overline3}}\mathbb{A}(\overline1,3,2,\overline3,1,\overline2)+e^{i \pi s_{13}}e^{i \pi s_{1\overline3}}e^{i \pi s_{2\overline3}}e^{i \pi s_{3\overline3}}\mathbb{A}(\overline1,3,2,1,\overline3,\overline2)\n
		&\h+\mathbb{A}(\overline1,3,2,1,\overline2,\overline3)
	\end{align}
are used to rewrite the corresponding partial amplitudes in \eqref{eq::q13}. Thus, the amplitude is reduced to the integration patches
	\begin{align}
		\mathcal{A}^{\mathbb{RP}^2}_3\big|_{-1<y<0}&=-\mathbb{A}(3,\overline1,2,\overline3,1,\overline2)-e^{i\pi s_{1\overline3}}\mathbb{A}(3,\overline1,2,1,\overline3,\overline2)-e^{i\pi s_{1\overline3}}\mathbb{A}(\overline1,3,2,\overline3,1,\overline2)\n
		&\h-\mathbb{A}(\overline1,3,2,1,\overline3,\overline2)+\mathbb{A}(\overline3,\overline1,2,3,1,\overline2)+e^{i\pi s_{13}}\mathbb{A}(\overline1,\overline3,2,3,1,\overline2)\n
		&\h+\mathbb{A}(\overline1,2,3,1,\overline2,\overline3)+e^{i\pi s_{13}}\mathbb{A}(\overline3,\overline1,2,1,3,\overline2)+\mathbb{A}(\overline1,\overline3,2,1,3,\overline2)\n
		&\h+e^{i\pi s_{13}}\mathbb{A}(\overline1,2,1,3,\overline2,\overline3)-\mathbb{A}(\overline1,2,\overline3,1,\overline2,3)-e^{i\pi s_{1\overline3}}\mathbb{A}(\overline1,2,1,\overline3,\overline2,3)\ .\label{eq::q11}
	\end{align}
Performing the $PSL(2,\mathbb R)$-transformation \eqref{eq::PSL2R_transformation} maps the interval $-1<y<0$ onto the region $1<x<\infty$ 
such that the amplitude reads in the new $PSL(2,\mathbb R)$-frame
\begin{align}
	\mathcal{A}^{\mathbb{RP}^2}_3\big|_{-1<y<0}&=\frac18 ig_c^3T'_p \Pi\int_{1}^\infty\mathrm{d}x\int\mathrm{d}\tilde \xi\,\int\mathrm{d}\tilde\eta\,\llangle V_{\overline 1}(0)U_{\overline 2}(-x)U_3(\tilde\xi) U_{\overline3}\left(\tilde{\eta}\right)V_2(1)V_1(\infty)\rrangle\ ,
\end{align}
where the integration regions of $\tilde\xi$ and $\tilde\eta$ correspond to the partial amplitudes in \eqref{eq::q12} below. After the transition $\mathbb A\to\mathcal A$ the above result in \eqref{eq::q11} becomes
	\begin{align}
		\mathcal{A}^{\mathbb{RP}^2}_3\big|_{-1<y<0}&=e^{i\pi s_{13}}\mathcal{A}(\overline2,\overline1,\overline3,2,3,1)+\mathcal{A}(\overline2,\overline3,\overline1,2,3,1)+\mathcal{A}(3,\overline2,\overline1,\overline3,2,1))\nonumber\\
		&\h+e^{i\pi s_{13}}\mathcal{A}(3,\overline2,\overline3,\overline1,2,1)-e^{-i\pi s_{1\overline3}}\mathcal{A}(\overline3,\overline2,3,\overline1,2,1)-\mathcal{A}(\overline2,3,\overline1,\overline3,2,1)\n
		&\h-\mathcal{A}(\overline3,\overline2,\overline1,3,2,1)-e^{-i\pi s_{2\overline3}}\mathcal{\mathcal{A}}(\overline3,\overline2,\overline1,2,1,\overline3)\ ,\label{eq::q12}
	\end{align}
which can be further simplified using the following monodromy relations originating form \eqref{eq::monodromy_2} for the subamplitudes $\mathcal A$ 
	\begin{align}
		0&=\mathcal{A}(3,\overline2,\overline3,\overline1,2,1)+\mathcal{A}(\overline2,\overline3,\overline1,2,1) e^{i \pi  s_{2\overline3}}+\mathcal{A}(\overline2,\overline3,3,\overline1,2,1) e^{i \pi  (s_{2\overline3}+s_{3\overline3})}\nonumber\\
		&\h+\mathcal{A}(\overline2,\overline3,\overline1,3,2,1) e^{i \pi  (s_{1\overline3}+s_{2\overline3}+s_{3\overline3})}+\mathcal{A}(\overline2,\overline3,\overline1,2,3,1) e^{i \pi  (s_{1\overline3}+s_{23}+s_{2\overline3}+s_{3\overline3})}\nonumber\\
		0&=\mathcal{A}(3,\overline2,\overline1,\overline3,2,1)+\mathcal{A}(\overline2,3,\overline1,\overline3,2,1) e^{-i \pi  s_{2\overline3}}+\mathcal{A}(\overline2,\overline1,3,\overline3,2,1) e^{-i \pi  (s_{1\overline3}+s_{2\overline3})}\nonumber\\
		&\h+\mathcal{A}(\overline2,\overline1,\overline3,3,2,1) e^{-i \pi  (s_{1\overline3}+s_{2\overline3}+s_{3\overline3})}+\mathcal{A}(\overline2,\overline1,\overline3,2,3,1) e^{-i \pi  (s_{1\overline3}+s_{23}+s_{2\overline3}+s_{3\overline3})}\nonumber\\
		0&=\mathcal{A}(\overline3,\overline2,\overline1,3,2,1)+\mathcal{A}(\overline2,\overline3,\overline1,3,2,1) e^{i \pi  s_{23}}+\mathcal{A}(\overline2,\overline1,\overline3,3,2,1) e^{i \pi  (s_{13}+s_{23})}\nonumber\\
		&\h+\mathcal{A}(\overline2,\overline1,3,\overline3,2,1) e^{i \pi  (s_{13}+s_{23}+s_{3\overline3})}+\mathcal{A}(\overline2,\overline1,3,2,\overline3,1) e^{i \pi  (s_{13}+s_{23}+s_{2\overline3}+s_{3\overline3})}\nonumber\\
		0&=\mathcal{A}(\overline3,\overline2,3,\overline1,2,1)+\mathcal{A}(\overline2,\overline3,3,\overline1,2,1) e^{-i \pi  s_{23}}+\mathcal{A}(\overline2,3,\overline3,\overline1,2,1) e^{-i \pi  (s_{23}+s_{3\overline3})}\nonumber\\
		&\h+\mathcal{A}(\overline2,3,\overline1,\overline3,2,1) e^{-i \pi  (s_{13}+s_{23}+s_{3\overline3})}+\mathcal{A}(\overline2,3,\overline1,2,\overline3,1) e^{-i \pi  (s_{13}+s_{23}+s_{2\overline3}+s_{3\overline3})}\ .
	\end{align}
These reduce the number of partial amplitudes in \eqref{eq::q12} down to two such that
	\begin{align}
		\mathcal{A}^{\mathbb{RP}^2}_3\big|_{-1<y<0}&=2i\sin(\pi s_{2\overline3})\mathcal A(\overline2,3,\overline1,\overline3,2,1)\big|_{1<x<\infty}+2i\sin(\pi s_{23})\mathcal A(\overline2,\overline3,\overline1,3,2,1)\big|_{1<x<\infty}\ ,
	\end{align}
where $1<x<\infty$. With a similar calculation the contribution coming from $1<y<\infty$ is described by the subamplitudes
	\begin{equation}
		\mathcal{A}^{\mathbb{RP}^2}_3\big|_{1<y<\infty}=2i\sin(\pi s_{2\overline3})\mathcal A(\overline2,3,\overline1,\overline3,2,1)\big|_{0<x<1}+2i\sin(\pi s_{23})\mathcal A(\overline2,\overline3,\overline1,3,2,1)\big|_{0<x<1}
	\end{equation}
and we integrate $x$ from $0$ to $1$. Again, with the coordinate transformation $x\to-x$ we end up with
	\begin{align}
		\mathcal{A}^{\mathbb{RP}^2}_3\big|_{-1<y<0}+\mathcal{A}^{\mathbb{RP}^2}_3\big|_{1<y<\infty}&=2i\sin(\pi s_{2\overline3})\mathcal A(\overline2,3,\overline1,\overline3,2,1)+2i\sin(\pi s_{23})\mathcal A(\overline2,\overline3,\overline1,3,2,1)\n
		&=\frac14g_c^3T'_p\sum_{\sigma\in S_3}\left\{\sin(\pi s_{2\overline3})F^{\sigma(\overline2,3,\overline3)}_{\mathcal{I}_3}+\sin(\pi s_{23}) F^{\sigma(\overline2,3,\overline3)}_{\mathcal{I}_4}\right\}\n 
		&\h\times A_\text{YM}(\overline1,\sigma(\overline2,3,\overline3),2,1)\ ,\label{res2}
	\end{align}
with the hypergeometric functions \req{Sample12} and the following two integration regions:
	\begin{align}
		\mathcal{I}_3&=\left\{x,\tilde\xi,\tilde\eta\in\mathbb{R}\ |\ -\infty<x<0,\R{x}<\tilde\xi<0,0<\tilde\eta<1\right\}\ ,\n
		\mathcal{I}_4&=\left\{x,\tilde\xi,\tilde\eta\in\mathbb{R}\ |\ -\infty<x<0,0<\tilde\xi<1,\R{x}<\tilde\eta<0\right\}\ .\label{domain2}
	\end{align}
Altogether, combining the two results \req{res1} and \req{res2} for the different $y$-integrations we find that the scattering of three closed strings off an O$p$-plane is given by:
\begin{align}
	\mathcal{A}^{\mathbb{RP}^2}_3&=2i\sin(\pi s_{1\overline3})\mathcal A(\overline2,\overline1,3,2,\overline3,1)+2i\sin(\pi s_{13})\mathcal A(\overline2,\overline1,\overline3,2,3,1)\n
	&\h+2i\sin(\pi s_{2\overline3})\mathcal A(\overline2,3,\overline1,\overline3,2,1)+2i\sin(\pi s_{23})\mathcal A(\overline2,\overline3,\overline1,3,2,1)\n
	&=\frac14g_c^3T'_p\sum_{\sigma\in S_3}\Big\{\sin(\pi s_{1\overline3})F^{\sigma(\overline2,3,\overline3)}_{\mathcal{I}_1}+\sin(\pi s_{13}) F^{\sigma(\overline2,3,\overline3)}_{\mathcal{I}_2}\n
	&\h+\sin(\pi s_{2\overline3})F^{\sigma(\overline2,3,\overline3)}_{\mathcal{I}_3}+\sin(\pi s_{23}) F^{\sigma(\overline2,3,\overline3)}_{\mathcal{I}_4}\Big\}A_\text{YM}(\overline1,\sigma(\overline2,3,\overline3),2,1)\ .\label{FinalR}
\end{align}
Actually, by applying \cite{Stieberger:2009hq} we can state the following two open string monodromy relations:
\begin{align}
(i)\ \ &\mathcal A(\bar2,\bar1,3,2,1,\bar3)+e^{\pi i s_{1\bar3}}\;\mathcal A(\bar2,\bar1,3,2,\bar3,1)
+e^{\pi i (s_{1\bar3}+s_{2\bar3})}\;\mathcal A(\bar2,\bar1,3,\bar3,2,1)\n
&+e^{\pi i (s_{1\bar3}+s_{2\bar3}+s_{3\bar3})}\;\mathcal A(\bar2,\bar1,\bar3,3,2,1)
+e^{-\pi i s_{23}}\;\mathcal A(\bar2,\bar3,\bar1,3,2,1)=0\ ,\\
(ii)\ \ &\mathcal A(\bar2,\bar1,\bar3,2,1,3)+e^{\pi i s_{13}}\;\mathcal A(\bar2,\bar1,\bar3,2,3,1)
+e^{\pi i (s_{13}+s_{23})}\;\mathcal A(\bar2,\bar1,\bar3,3,2,1)\n
&+e^{\pi i (s_{13}+s_{23}+s_{3\bar3})}\;\mathcal A(\bar2,\bar1,3,\bar3,2,1)+
+e^{-\pi is_{2\bar3}}\;\mathcal A(\bar2,3,\bar1,\bar3,2,1)=0\ .
\end{align}
The combination $\Im(i)+\Im(ii)$ yields the following useful relation:
\begin{align}
&\sin(\pi s_{13})\; \mathcal A(\bar2,\bar1,\bar3,2,3,1)
+\sin(\pi s_{1\bar3}) \;\mathcal A(\bar2,\bar1,3,2,\bar3,1)\n
&=\sin(\pi s_{23})\;\mathcal A(\bar2,\bar3,\bar1,3,2,1)+\sin(\pi s_{2\bar3})\;
\mathcal A(\bar2,3,\bar1,\bar3,2,1)\ .\label{powerful}
\end{align}
Equation \req{powerful} allows to cast \req{FinalR} into: 
\begin{align}
	\mathcal{A}^{\mathbb{RP}^2}_3&=4i\sin(\pi s_{2\overline3})\mathcal A(\overline2,3,\overline1,\overline3,2,1)+4i\sin(\pi s_{23})\mathcal A(\overline2,\overline3,\overline1,3,2,1)\n
	&=\frac12g_c^3T'_p\sum_{\sigma\in S_3}\Big\{\sin(\pi s_{2\overline3})F^{\sigma(\overline2,3,\overline3)}_{\mathcal{I}_3}+\sin(\pi s_{23}) F^{\sigma(\overline2,3,\overline3)}_{\mathcal{I}_4}\Big\}A_\text{YM}(\overline1,\sigma(\overline2,3,\overline3),2,1)\ .\label{FinalRr}
\end{align}
As a result we observe that it is  possible to write the scattering of three closed strings off an O$p$-plane in terms of two open string subamplitudes similar to the case of three closed strings on the disk \cite{Bischof:2023uor}. Furthermore,  our result \req{FinalRr} is manifestly symmetric under the exchange $3\leftrightarrow\overline3$. Likewise, the two exchange symmetries $1\leftrightarrow\overline1$ and $2\leftrightarrow\overline2$ can be evidenced.
As a consequence, we may write the final result \req{FinalRr} in compact form as:
\be
\mathcal{A}^{\mathbb{RP}^2}_3=
	\frac12g_c^3T'_p\sum_{\sigma\in S_3}\Big\{\sin(\pi s_{2\overline3})F^{\sigma(\overline2,3,\overline3)}_{\mathcal{I}_3}\ A_\text{YM}(\overline1,\sigma(\overline2,3,\overline3),2,1)\Big\}+(3\leftrightarrow\bar3)\ .\label{FinalRR}
\ee
Interestingly, three closed string scattering on both the disk and the real projective plane can be written in terms of a basis of two open string subamplitudes. However only in the case of the real projective plane we have the (separate) exchange symmetries $1\leftrightarrow\overline1$ and $2\leftrightarrow\overline2$ and $3\leftrightarrow\overline3$, respectively
between left-- and right--movers.
We conclude by noting that the lowest order in $\ap$ of \req{FinalRR} vanishes as it is also the case for the scattering of three closed strings on the disk \cite{Bischof:2023uor}. Consequently the $\ap$ expansion
of \req{FinalR} only starts at the subleading order in $\ap$, cf.\ also Section \ref{alphaprime_expand}.

\def\ss{{\small}}
\def\tt{{\footnotesize}}
\def\si{\sigma}
\def\t{\hat t}
\def\s{\hat s}

\section[Low energy expansion in inverse string tension \boldmath{$\ap$}]{Low energy expansion in inverse string tension \boldmath{$\ap$}}
\label{alphaprime_expand}

In this subsection we shall evaluate the $\alpha'$--expansion of the amplitude \req{FinalRR}, building    on the results of \cite{npt_2}. More concretely, the hypergeometric integral \req{Sample12} 
appearing in \req{FinalRR}
\be
F^{(\overline2_\si 3_\si\overline 3_\si)}_{\mathcal
  I_3}=-\int_{\mathcal{I}_3}\mathrm{d}z_{\overline2}\,\mathrm{d}z_3\,\mathrm{d}z_{\overline3}\,\Biggl(\prod_{i<j}|z_{ij}|^{s_{ij}}\Biggl)\frac{s_{\overline1\overline2_\si}}{z_{\overline1\overline2_\si}}\frac{s_{\overline3_\si2}}{z_{\overline3_\si
    2}}\left(\frac{s_{\overline13_\si}}{z_{\overline13_\si}}+\frac{s_{\overline2_\si3_\si}}{z_{\overline2_\si3_\si}}\right)\ ,\quad\si\in S_3\ ,\label{Sample1}
\ee
which is integrated over the domain $\mathcal{I}_3$ specified in \req{domain2}, need to be expressed in terms of a power series w.r.t.\ small $\ap$. The six permutations $\si\in S_3$ act on the three labels  $\overline2, 3$ and $\overline 3$ as $i_\si:=\si(i)$.
In \req{Sample1} the  domain 
\begin{align}
  \Ic_3&:\ z_{\bar 2}<z_3<z_{\bar 1}<z_{\bar 3}<z_2<z_1\ ,\label{Reg3}
 \end{align}
is  subject to the fixing
\be
z_{\bar 1}=0\ ,\ z_2=1\ ,\ z_1=\infty\ ,
\ee
which may be rescinded by introducing the volume of the conformal Killing group:
\be
F^{(\overline2_\si 3_\si\overline 3_\si)}_{\mathcal I_n}=-V_{\rm
  CKG}^{-1}\int_{\mathcal{I}_n}\mathrm{d}z_i\;\Biggl(\prod_{i<j}|z_{ij}|^{s_{ij}}\Biggl)\fc{1}{z_{\bar1
    2}z_{\bar 11}z_{21}}\;\frac{s_{\overline1\overline2_\si}}{z_{\overline1\overline2_\si}}\frac{s_{\overline3_\si2}}{z_{\overline3_\si 2}}\left(\frac{s_{\overline13_\si}}{z_{\overline13_\si}}+\frac{s_{\overline2_\si3_\si}}{z_{\overline2_\si3_\si}}\right)\ ,\quad\si\in S_3\ .\label{Sample2}
\ee
A more suitable gauge choice for the region  \eqref{Reg3}   would be
\be
z_{\bar 2}=0\ ,\quad z_2=1\ ,\ z_1=\infty\ ,
\ee
leading to:
\be
F^{(\overline2_\si 3_\si\overline 3_\si)}_{\mathcal I_3}=-\int_{\mathcal{I}_3}\mathrm{d}z_{\bar 1}\,\mathrm{d}z_{3}\,\mathrm{d}z_{\bar 3}\ \fc{z_{\bar 22}}{z_{\bar 12}}\;\Biggl(\prod_{i<j}|z_{ij}|^{s_{ij}}\Biggl)\frac{s_{\overline1\overline2_\si}}{z_{\overline1\overline2_\si}}\frac{s_{\overline3_\si2}}{z_{\overline3_\si 2}}\left(\frac{s_{\overline13_\si}}{z_{\overline13_\si}}+\frac{s_{\overline2_\si3_\si}}{z_{\overline2_\si3_\si}}\right)\ ,\quad\si\in S_3\ .\label{Sample3}
\ee
After introducing the map
\be
\begin{array}{lcl}
\varphi_3:\;(\bar 2,3,\bar1,\bar 3,2,1)&\mapsto&(1,2,3,4,5,6)\ ,\end{array}
\label{Map}
\ee
we may cast  \req{Sample3} into the following form
\be
F^{(\overline2_\si 3_\si\overline 3_\si)}_{\mathcal{I}_3}=
\lf. F^{(3,\varphi_3(\bar2_{\si}) \varphi_3(3_{ \si}),\varphi_3( \overline 3_\si),{5})}\ri|_{\hat s_{ij}\ra \varphi_3^{-1}(\hat s_{ij})}\ ,
\label{findMAP}
\ee
with
\bea\label{Sample13}
\ds F^{(3,1_\si, 2_\si, 4_\si,5)}&=&\ds\int\limits_{0<z_2<z_3<z_4<1}\mathrm{d}z_{2}\,\mathrm{d}z_3\,\mathrm{d}z_{4}\ \Biggl(\prod_{i<j}|z_{ij}|^{\hat s_{ij}}\Biggl)\ \fc{1}{z_{35}}\ \frac{\hat s_{1_\si3}}{z_{31_\si}}\frac{\hat s_{4_\si5}}{z_{4_\si 5}}\left(\frac{\hat s_{2_\si3}}{z_{32_\si}}+\frac{\hat s_{1_\si 2_\si}}{z_{1_\si 2_\si}}\right)\ ,
\eea
with canonical color ordering.
In \req{findMAP} the inverse map $\varphi^{-1}_3$ of \req{Map} acts on the six--point kinematic invariants $\hat s_{ij}$ to be specified below in \eqref{eq::phi2}.
The  six functions  $F^{(3abc5)}$ defined in \req{Sample13} correspond to a subset of the following extended set of $120$ functions
\be\label{complete}
F^{(1_\pi2_\pi3_\pi4_\pi5_\pi)}=-\int\limits_{0<z_2<z_3<z_4<1}\mathrm{d}z_{2}\,\mathrm{d}z_3\,\mathrm{d}z_{4}\;\lf(\prod_{i<l}|z_{il}|^{\hat s_{il}}\ri)\; \fc{1}{z_{1_\pi5_\pi}}\;\fc{\hat s_{1_\pi2_\pi}}{z_{1_\pi2_\pi}}\fc{\hat s_{4_\pi5_\pi}}{z_{4_\pi5_\pi}}\;\lf(\fc{\hat s_{1_\pi3_\pi}}{z_{1_\pi3_\pi}} +\fc{\hat s_{2_\pi3_\pi}}{z_{2_\pi3_\pi}}      \ri)\ ,
\ee
with $\pi\in S_5$ for canonical color ordering $(1,2,3,4,5,6)$. The integrals \req{complete} integrate to triple  hypergeometric functions \cite{Oprisa:2005wu}. An other subset of $24$ integrals has been introduced in \cite{npt_2} to describe  the six--point open superstring amplitude.   In fact, thanks to some (dual) monodromy relations all these $120$ functions can be expressed in terms of a six--dimensional basis $F^{(abc)}\equiv F^{(1abc5)}$ as (with $K_i^\ast=(K_i^t)^{-1}$)   \cite{npt_2}
\be
\lf(\begin{matrix}
F^{(31245)}\\
F^{(32145)}\\
F^{(32415)}\\
F^{(34215)}\\
F^{(31425)}\\
F^{(34125)}\end{matrix}\ri)=K_3^\ast\ 
\lf(\begin{matrix}
F^{(234)}\\
F^{(324)}\\
F^{(432)}\\
F^{(342)}\\
F^{(423)}\\
F^{(243)}\end{matrix}\ri)\ ,
\ee
respectively. The $6\times6$ matrix $K_3$ follows from the corresponding (dual) subamplitude
relations and is given by 
\begin{align}
&(K_3)^\si_\pi=\hat s_{36}^{-1}\\
&\hskip-0.35cm\times\begin{pmatrix}
\ss{\s_{345}-\s_4}& \ss{s_2-s_4+s_{345}} & \ss{0}& \ss{0} &\ss{0}&\ss{\hat s_{35}}  \\
\ss{\fc{(\s_4-\s_{345}) \hat c_6}{\s_4+\hat c_3}}&\ss{\fc{(\s_2-\s_6+\s_{345}) \s_{13}}{\s_4+\hat c_3}}&\ss{-\fc{(\s_6+\hat c_1) (\hat s_3+s_{13})}{\s_4+\hat c_3}}&\ss{-\fc{(\s_6+\hat c_1) \s_{13}}{\s_4+\hat c_3}}&\ss{-\fc{\hat c_{11} \s_{35}}{\s_4+\hat c_3}}&\ss{-\fc{\hat c_6 \s_{35}}{\s_4+\hat c_3}}\\[2mm]
\ss{\fc{\s_1 \s_4 (\s_4-\s_{345})}{\s_{15}(\s_4+\hat c_3)}}&\ss{\fc{\s_4 \hat c_5 \s_{13}}{\s_{15}(\s_4+\hat c_3)}}&\ss{\fc{(\s_6+\hat c_1) \hat c_3 (\s_3+\s_{13})}{\s_{15}(\s_4+\hat c_3)}}&\ss{\fc{(\s_6+\hat c_1) \hat c_3 \hat s_{13}}{\s_{15}(\s_4+\hat c_3)}}&\ss{\fc{\hat c_3 \hat c_{11} \hat s_{35}}{\s_{15}(\s_4+\hat c_3)}}&\ss{-\fc{\s_1 \s_4 \s_{35}}{\s_{15}(\s_4+\hat c_3)}}\\[2mm]
\ss{\fc{\s_4 (\s_1-\s_{123}) (\s_1+\s_{15})}{\hat c_7 \s_{15}}}&\ss{-\fc{\s_4 \s_{13} (\s_1+\s_{15})}{\hat c_7 \s_{15}}}&\ss{\fc{\s_{14} \s_{25} (\s_2+\s_{35})}{\hat c_7 \s_{15}}}&\ss{\fc{(\s_6+\hat c_1) \hat c_{10} \hat s_{13}}{\hat c_7 \s_{15}}}&\ss{\fc{(\s_6+\hat c_1) \s_{14} \s_{35}}{\hat c_7 \s_{15}}}&\ss{-\fc{\hat c_4 (\s_1+\s_{15}) \s_{35}}{\hat c_7 \s_{15}}}\\[2mm]
\ss{0}&\ss{0}&\ss{\s_2+\s_{35}}&\ss{\s_2-\s_4+\s_{345}}&\ss{\s_{35}}  &\ss{0}\\[2mm]
\ss{\fc{\s_4 (\s_1-\s_{123})}{\hat c_8}}&\ss{-\fc{\s_4 \s_{13}}{\hat c_8}}&\ss{-\fc{(c_1-\s_{13}) (\s_2+\s_{35})}{\hat c_8}}&\ss{\fc{\s_{13} (\s_4+s_{24})}{\hat c_7}}&\ss{-\fc{(\hat c_1-\s_{13}) \s_{35}}{\hat c_8}}&\ss{\fc{\hat c_2 \s_{35}}{\hat c_8}}\n
\end{pmatrix}
\end{align}
In the above matrices we have introduced $\hat c_1=\s_3-\s_{234}-\s_{345},
\hat c_2=-\s_1+\s_{123}+\s_{345}, \hat c_3=\s_2-\s_6-\s_{123},
\hat c_4=\s_1-\s_{123}-\s_{345},\hat c_5=-\s_4+\s_{123}+\s_{345}$,
$\hat c_6=\s_4-\s_6-\s_{13},\hat c_7=-\s_1-\s_3+\s_5+\s_{345},
\hat c_8=\s_1+\s_3-\s_5-\s_{345},\hat c_9=
\s_2+\s_4-\s_6-\s_{123},\hat c_{10}=\s_1-\s_2+\s_4-\s_5+\s_{234}-\s_{345},
\hat c_{11}=\s_1-\s_3+\s_4-\s_6-\s_{123}+\s_{234}$ and
$\s_{ijk}= \s_{ij}+\s_{ik}+\s_{jk}$.  

Let us now turn to the $\ap$--expansion of the six integrals \req{Sample1} and \req{Sample3}    and likewise the $\ap$--expansion of \req{Sample13} specified by \req{findMAP}. 
With the above preparations the latter are expressed by the $\ap$--expansion of the six functions $F^{(abc)}$ as:
\be
\lf(\begin{matrix}
F^{(\bar23\bar3)}_{\Ic_i}\\
F^{(3\bar2\bar3)}_{\Ic_i}\\
F^{(3\bar3\bar2)}_{\Ic_i}\\
F^{(\bar 33\bar2)}_{\Ic_i}\\
F^{(\bar2\bar33)}_{\Ic_i}\\
F^{(\bar3\bar23)}_{\Ic_i}\end{matrix}\ri)=K_3^\ast\ 
\lf.\lf(\begin{matrix}
F^{(234)}\\
F^{(324)}\\
F^{(432)}\\
F^{(342)}\\
F^{(423)}\\
F^{(243)}\end{matrix}\ri)\ri|_{\hat s_{ij}\ra \varphi_3^{-1}(\hat s_{ij})} \ ,\label{alpharel}
\ee
where $K^\ast_p=(K_p^\mathrm{T})^{-1}$. Note, that the inverse map $\varphi_3^{-1}$ acts on the six--point kinematic invariants $\hat s_{ij}$ as follows
\be
\varphi_3^{-1}:\lf\{\begin{matrix}
\hat s_{12} \mapsto s_{2\bar3}& \\
\hat s_{23} \mapsto s_{1\bar3},& \hat s_{123}\mapsto s_{12}+s_{1\bar3}+s_{2\bar3},\\
\hat s_{34} \mapsto s_{13},&\hat s_{234}\mapsto s_{13}+s_{1\bar3}+s_{3\bar3},\\
\hat s_{45} \mapsto s_{2\bar3},&\hat s_{345}\mapsto s_{1\bar2}+s_{2\bar3}+s_{13},\\
\hat s_{56} \mapsto s_{12},&\\
\hat s_{61} \mapsto s_{1\bar2},&\\
\end{matrix}\ri. \ . \label{eq::phi2}
\ee
subject to the constraints \req{kin1}--\req{kin3}.
The methods to find the low energy expansion of the latter has been pioneered in \cite{Oprisa:2005wu} and subsequently been applied and systematized  in \cite{Stieberger:2006te,Broedel:2013tta}.
By applying these techniques we find, cf.\ also \cite{npt_2}:
\begin{align}
 F^{(234)}&=1-\z_2\ (\s_{45}\s_{56}+\s_{12}\s_{61}-\s_{45}\s_{123}-\s_{12}\s_{345}+\s_{123}\s_{345})+\z_3(\ldots)+\Oc(\ap^4)\ ,\label{exp1}\\
 F^{(324)}&=-\z_2\ \s_{13}(\s_{23}-\s_{61}+\s_{345})+\z_3(\ldots)+\Oc(\ap^4) \\
 F^{(432)}&=-\z_2\ \s_{14}\s_{25}+\z_3\ \s_{14}\s_{25}\ (-\s_{23}-\s_{34}+\s_{56}+\s_{61}+\s_{123}+\s_{234}+\s_{345})+\Oc(\ap^4)\ ,\\
 F^{(342)}&=\zeta_2\ \s_{13}\s_{25}+\zeta_3\ \s_{13}\s_{25}\ (-\s_{12} + \s_{23} + 2\s_{34} - \s_{16} -\s_{123} - 2\s_{234} -\s_{345})+\Oc(\ap^4)\ ,\\
 F^{(423)} &=\zeta_2\ \s_{14}\s_{35}+\zeta_3\ \s_{14}\s_{35}\ (2\s_{23} + \s_{34}-\s_{45}-\s_{56}-\s_{123}-2\s_{234}-\s_{345})+\Oc(\ap^4)\ ,\\
  F^{(243)} &=-\zeta_2\ \s_{35}(\s_{34}-\s_{56}+\s_{123})+\zeta_3\ s_{35}\ [-2\s_{12}\s_{23}-2\s_{12}\s_{34}+s_{34}^2+\s_{34}\s_{45}-\s_{45}\s_{56}\nonumber\\
&-\s_{56}^2+\s_{123}(2\s_{23}+\s_{45}+\s_{123})+2\s_{12}\s_{234}+\s_{345}(\s_{34}-\s_{56}+\s_{123})]+\Oc(\ap^4)\ .\label{exp6}
\end{align}
Eventually, in \req{FinalRr} the required combinations of \req{alpharel} and their permutations
under $3\leftrightarrow\bar3$  comprise 
\begin{align}
&\sin(\pi s_{2\bar3})\lf(\begin{matrix}
F^{(\bar23\bar3)}_{\Ic_3}\\[2mm]
F^{(3\bar2\bar3)}_{\Ic_3}\\[2mm]
F^{(3\bar3\bar2)}_{\Ic_3}\\[2mm]
F^{(\bar 33\bar2)}_{\Ic_3}\\[2mm]
F^{(\bar2\bar33)}_{\Ic_3}\\[2mm]
F^{(\bar3\bar23)}_{\Ic_3}\end{matrix}\ri)+\sin(\pi s_{23})\lf(\begin{matrix}
F^{(\bar23\bar3)}_{\Ic_4}\\[2mm]
F^{(3\bar2\bar3)}_{\Ic_4}\\[2mm]
F^{(3\bar3\bar2)}_{\Ic_4}\\[2mm]
F^{(\bar 33\bar2)}_{\Ic_4}\\[2mm]
F^{(\bar2\bar33)}_{\Ic_4}\\[2mm]
F^{(\bar3\bar23)}_{\Ic_4}\end{matrix}\ri)\nonumber\\[5mm]
&\hskip-2cm={\tiny\pi\lf(\begin{matrix} 
\ss{-s_{12}+\fc{s_{12}s_{1\bar3}}{s_{1\bar2}}+\fc{s_{12}(s_{1\bar2}+s_{1\bar3})s_{2\bar 3}}{s_{1\bar2}(s_{1\bar 2}+s_{1\bar3}+s_{23})}+\fc{s_{12}(s_{1\bar2}+s_{13})(s_{1\bar2}-s_{1\bar3})}{s_{1\bar2}(s_{1\bar 2}+s_{13}+s_{2\bar3})}}\\[3mm] 
\ss{-\frac{ s_{1\bar3} (s_{12}-s_{13}) (s_{1\bar2}+s_{13})}{s_{1\bar2} (s_{1\bar2}+s_{13}+s_{2\bar3})}+\frac{ s_{1\bar3} (s_{12}-s_{13})}{s_{1\bar2}}+\frac{
   s_{2\bar3} (s_{1\bar2}+s_{1\bar3}) (s_{12}+s_{1\bar2}+s_{1\bar3}+s_{23})}{s_{1\bar2} (s_{1\bar2}+s_{1\bar3}+s_{23})}+\frac{ s_{2\bar3}^2
   (s_{1\bar2}+s_{1\bar3})}{s_{1\bar2} (s_{1\bar2}+s_{1\bar3}+s_{23})}}\\[3mm] 
\ss{-\frac{s_{1\bar3} (s_{1\bar2}+s_{13}) (s_{12}-s_{13}+s_{23})}{s_{1\bar2} (s_{1\bar2}+s_{13}+s_{2\bar3})}+\frac{ s_{12} (s_{1\bar2}+s_{1\bar3})- s_{1\bar3}
   (s_{1\bar2}+s_{13})}{s_{1\bar2}}+\frac{ s_{2\bar3} (s_{1\bar2}+s_{1\bar3}) (s_{12}+s_{1\bar2}+s_{1\bar3}+2 s_{23})}{s_{1\bar2}
   (s_{1\bar2}+s_{1\bar3}+s_{23})}-\frac{ (s_{12}-s_{1\bar3}) (s_{1\bar2}+s_{1\bar3})}{s_{1\bar2}+s_{1\bar3}+s_{23}}+\frac{ s_{2\bar3}^2
   (s_{1\bar2}+s_{1\bar3})}{s_{1\bar2} (s_{1\bar2}+s_{1\bar3}+s_{23})}+\frac{ s_{23} (s_{1\bar2}+s_{1\bar3})}{s_{1\bar2}}}\\[3mm] 
\ss{-\frac{ s_{13} (s_{12}-s_{1\bar3}) (s_{1\bar2}+s_{1\bar3})}{s_{1\bar2} (s_{1\bar2}+s_{1\bar3}+s_{23})}+\frac{ s_{12} (s_{1\bar2}+s_{13})- s_{13}
   (s_{1\bar2}+s_{1\bar3})}{s_{1\bar2}}-\frac{ (s_{1\bar2}+s_{13}) (s_{1\bar2}-s_{23}) (s_{12}-s_{13}+s_{23})}{s_{1\bar2}
   (s_{1\bar2}+s_{13}+s_{2\bar3})}+\frac{ s_{2\bar3} [s_{23} (s_{1\bar2}+s_{13})+s_{1\bar2} (s_{1\bar2}+s_{1\bar3})]}{s_{1\bar2}
   (s_{1\bar2}+s_{1\bar3}+s_{23})}+\frac{2 s_{23} (s_{1\bar2}+s_{13})}{s_{1\bar2}}}\\[3mm] 
\ss{-s_{12}+\fc{s_{12}s_{13}}{s_{1\bar2}}+\fc{s_{12}(s_{1\bar2}-s_{13})(s_{1\bar2}+s_{1\bar3})}{s_{1\bar2}(s_{1\bar 2}+s_{1\bar3}+s_{23})}+\fc{s_{12}(s_{1\bar2}+s_{13})s_{23}}{s_{1\bar2}(s_{1\bar 2}+s_{13}+s_{2\bar3})}}\\[3mm]  
\ss{\frac{ s_{13} (s_{12}-s_{1\bar3})}{s_{1\bar2}}+\frac{ s_{23} (s_{1\bar2}+s_{13})}{s_{1\bar2}}-\frac{ s_{13} (s_{12}-s_{1\bar3}) (s_{1\bar2}+s_{1\bar3})}{s_{1\bar2} (s_{1\bar2}+s_{1\bar3}+s_{23})}+\frac{
   s_{23} (s_{12}+s_{23}) (s_{1\bar2}+s_{13})}{s_{1\bar2} (s_{1\bar2}+s_{13}+s_{2\bar3})}}\\[3mm] 
\end{matrix}\ri)}\nonumber\\[5mm]
&\hskip-1cm+\tiny\pi\zeta_2\lf(\begin{matrix} 
\fc{ s_{12} s_{2\bar3} [s_{1\bar2} (s_{13}+s_{1\bar3}-s_{23})-s_{1\bar3} (s_{13}+s_{1\bar3}+s_{23})]}{s_{1\bar2}}+\fc{ s_{12} s_{2\bar3}^2
   (s_{1\bar2}-s_{1\bar3})}{s_{1\bar2}}\\[3mm] 
-\fc{ s_{2\bar3}^2 \left[s_{12} (2 s_{1\bar2}+s_{1\bar3})+s_{1\bar2}^2+s_{1\bar2} (2 s_{1\bar3}+s_{23})+s_{1\bar3}
   (s_{13}+s_{1\bar3}+s_{23})\right]}{s_{1\bar2}}-\fc{ s_{2\bar3} [(s_{12}+s_{1\bar2}) (s_{12} s_{1\bar2}+s_{1\bar3} (s_{1\bar2}+s_{13}+s_{1\bar3}))+s_{12}
   s_{23} (s_{1\bar2}+s_{1\bar3})]}{s_{1\bar2}}-\fc{ s_{2\bar3}^3 (s_{1\bar2}+s_{1\bar3})}{s_{1\bar2}}\\[3mm] 
  x_3\\
  x_4\\[3mm] 
 \fc{ s_{12} s_{23} (s_{1\bar2}-s_{13}) (s_{13}+s_{1\bar3})}{s_{1\bar2}}+\fc{ s_{12} s_{23}^2 (s_{1\bar2}-s_{13})}{s_{1\bar2}}-\fc{ s_{12}
   s_{23} s_{2\bar3} (s_{1\bar2}+s_{13})}{s_{1\bar2}} \\[3mm]
-\fc{ s_{23}^2 \left[s_{12} (2 s_{1\bar2}+s_{13})+(s_{1\bar2}+s_{13})^2+s_{13} s_{1\bar3}\right]}{s_{1\bar2}}-\fc{ s_{23} (s_{12}+s_{1\bar2})
   [s_{12} s_{1\bar2}+s_{13} (s_{1\bar2}+s_{13}+s_{1\bar3})]}{s_{1\bar2}}-\fc{ s_{23} s_{2\bar3} (s_{12}+s_{23}) (s_{1\bar2}+s_{13})}{s_{1\bar2}}-\fc{
   s_{23}^3 (s_{1\bar2}+s_{13})}{s_{1\bar2}}\\[3mm] 
\end{matrix}\ri)\nonumber\\[5mm]
&\hskip1.5cm+\zeta_3\ \Oc(\ap^3)\ ,\label{eq::ap-expansion}
\end{align}
after inserting \req{exp1}--\req{exp6}.
Above we have introduced:
\be
\ba{ll}
x_3&=-\fc{ s_{2\bar3}^2 \left[s_{12} (2 s_{1\bar2}+s_{1\bar3})+s_{1\bar2}^2+2 s_{1\bar2} (s_{1\bar3}+s_{23})+s_{1\bar3} (s_{13}+s_{1\bar3}+2
   s_{23})\right]}{s_{1\bar2}}\\
   &-\fc{ s_{2\bar3} \left(s_{23} \left[s_{1\bar3} (s_{12}+s_{1\bar2}+s_{13})+s_{1\bar2} (3
   s_{12}+s_{1\bar2})+s_{1\bar3}^2\right]+(s_{12}+s_{1\bar2}) [s_{12} s_{1\bar2}+s_{1\bar3} (s_{1\bar2}+s_{13}+s_{1\bar3})]+s_{23}^2
   (s_{1\bar2}+s_{1\bar3})\right)}{s_{1\bar2}}\\
   &- s_{23} (s_{12}+s_{1\bar2}) (s_{12}-s_{1\bar3})+ s_{23}^2 (s_{1\bar3}-s_{12})-\fc{ s_{2\bar3}^3
   (s_{1\bar2}+s_{1\bar3})}{s_{1\bar2}}\ ,
   \ea
   \ee
 \be
\ba{ll}  
x_4&=-\fc{ s_{2\bar3} \left[s_{23} \left(s_{12} (3 s_{1\bar2}+s_{13})+s_{1\bar2}^2+s_{1\bar2} s_{13}+s_{13} (s_{13}+s_{1\bar3})\right)+s_{1\bar2}
   (s_{12}+s_{1\bar2}) (s_{12}-s_{13})+2 s_{23}^2 (s_{1\bar2}+s_{13})\right]}{s_{1\bar2}}\\
   &-\fc{ s_{23}^2 \left[s_{12} (2
   s_{1\bar2}+s_{13})+(s_{1\bar2}+s_{13})^2+s_{13} s_{1\bar3}\right]}{s_{1\bar2}}-\fc{ s_{23} (s_{12}+s_{1\bar2}) [s_{12} s_{1\bar2}+s_{13}
   (s_{1\bar2}+s_{13}+s_{1\bar3})]}{s_{1\bar2}}\\
   &-\fc{ s_{2\bar3}^2[(s_{12} s_{1\bar2}+s_{23} (s_{1\bar2}+s_{13})-s_{1\bar2} s_{13}]}{s_{1\bar2}}-\fc{
   s_{23}^3 (s_{1\bar2}+s_{13})}{s_{1\bar2}} \ .
\ea
\ee
Employing the expansions \req{eq::ap-expansion} in \req{FinalRr}  yields a result $\mathcal{A}^{\mathbb{RP}^2}_3$ for the scattering of three closed strings off an O$p$-plane, which is manifestly invariant under the three exchange symmetries $1\leftrightarrow\overline1$,  $2\leftrightarrow\overline2$ and $3\leftrightarrow\overline3$ up to the order $\ap^3\zeta(3)$. We have checked those by applying BCJ subamplitude relations for the kinematical factors \cite{Bern:2008qj}.

	\section{Concluding remarks}

In this work we have computed the \emph{complete} tree--level  amplitude on the real projective plane
involving any three massless closed string states in the NSNS sector. Our main result can be found in \req{FinalRR}.
Our result is interesting both from the conceptual and physical
point of view. We could express our findings in
terms of a basis of six--point open string subamplitudes and thereby showed that one can connect this closed string amplitude on the real projective plane via KLT--like relations and a $PSL(2,\mathbb R)$ transformation to the scattering of open strings on the disk. Surprisingly, however, our main result \req{FinalRr} can be written in terms of only two six--point open string subamplitudes albeit the basis of these subamplitudes contains six elements and therefore one might have expected that also the scattering of three closed strings is given in terms of six subamplitudes. On the other hand, our result
is in lines of \cite{Bischof:2023uor}, i.e.\ three closed string amplitudes formulated on the topology of both the disk and the real projective plane reduce to a two--dimensional basis of open string subamplitudes.
The open string subamplitudes are dressed by  SYM subamplitudes as kinematical building blocks thus 
being manifestly gauge invariant.  We expect that generalizations to an arbitrary number of closed strings on the real projective plane follow the same pattern. In particular, we expect the same dimensions of building blocks as on the disk.

Recall, the tree--level  amplitude on the real projective plane
involving  two  massless closed string states has been computed in the NSR formalism with the result \cite{Garousi:2006zh,Aldi:2020dvw}
\begin{align}
\mathcal{A}^{\mathbb{RP}^2}_2&\sim K(1,2)\ \fc{\Gamma(-t/2)\Gamma(-u/2)}{\Gamma(1-t/2-u/2)}=K(1,2)\ \lf\{\fc{4}{tu}-\zeta(2)+2s\;\zeta(3)+\Oc(\ap^2)\ri\}\n
&\sim \mathcal A(\bar1,2,1,\bar2)\ ,\label{Ga1}
\end{align}
with some kinematical factor $K(1,2)$. On the other hand,  the tree--level  amplitudes on the disk 
involving  two  closed string states  in the NSR formalism reads \cite{Hashimoto:1996bf}:
\begin{align}
\mathcal{A}^{D_2}_2&\sim K(1,2)\ \fc{\Gamma(-t/2)\Gamma(-2s)}{\Gamma(1-t/2-2s)}=K(1,2)\ \lf\{\fc{1}{st}-\zeta(2)+\fc{1}{2}u\;\zeta(3)+\Oc(\ap^2)\ri\}\n
&\sim \mathcal A(2,1,\bar1,\bar2)\ ,\label{Ga2}
\end{align}
with $t=-2p_1{\cdot}p_2,\ u=-2p_1{\cdot}D{\cdot}p_2$ and $4s+t+u=0$. The two amplitudes \req{Ga1} and \req {Ga2} are not equal, except at the  constant order in $\ap$. In particular, only the amplitude \req{Ga1} exhibits the exchange symmetries
$1\leftrightarrow\bar1$ and $2\leftrightarrow\bar2$. Actually, we have:
\be
\mathcal{A}^{\mathbb{RP}^2}_2=\fc{\sin(2\pi s)}{\sin\lf(\fc{\pi u}{2}\ri)}\ 
\mathcal{A}^{D_2}_2\ .
\ee
  On the other hand, $\mathcal{A}^{\mathbb{RP}^2}_2$ and $\mathcal{A}^{D_2}_2$ assume a rather similar functional dependence in terms of Gamma--functions. It has been speculated in \cite{Chen:2009tr} that this pattern continues to hold at higher multiplicities. We find that there is seemingly no direct  relation between fully--fledged closed string scattering amplitudes on the disk and the real projective plane.
In fact, by  comparing\footnote{To simplify the discussion in Section \ref{chap::plane} and in \cite{Bischof:2023uor} we have absorbed any prefactors including the string coupling constant and D/O$p$--plane tension into the definition of open string amplitudes. Hence, for any ordering $\rho$ of a subamplitude $\mathcal A(\rho(1,2,\ldots,6))$ we find that:
\begin{equation}
	\frac{\mathcal A^{ D_2}(\rho(1,2,\ldots,6))}{\mathcal A^{\mathbb{RP}^2}(\rho(1,2,\ldots,6))}=4\frac{T_p}{T_p'}\ .
\end{equation}} our three--point result $\mathcal{A}^{\mathbb{RP}^2}_3$ on the real projective plane \req{FinalRr}
\be
	\mathcal{A}^{\mathbb{RP}^2}_3\sim4i\sin(\pi s_{23})\ \mathcal A(\overline2,\overline3,\overline1,3,2,1)+4i\sin(\pi s_{2\overline3})\ \mathcal A(\overline2,3,\overline1,\overline3,2,1)
	\ee
	with that $\mathcal{A}^{D_2}_3$ on the disk computed in \cite{Bischof:2023uor} 
\be
\mathcal{A}^{D_2}_3\sim 2i\sin(\pi  s_{23})\ \mathcal A(\overline1,\overline2,\overline3,2,3,1)+2i\sin[\pi (s_{23}+s_{2\overline3})]\ \mathcal A(\overline1,\overline2,2,\overline3,3,1)\ ,\label{eq::amp_mon_4a}
\ee
we evidence several crucial differences. E.g.\ already at the constant order in $\ap$ the two amplitudes are different. Moreover, only the scattering amplitude on the real projective plane exhibits a symmetry between left-- and right--movers.

It would be interesting to derive effective action terms 
on the orientifold plane following from our amplitude result \req{eq::ap-expansion} and compare 
them with similar terms on the D--brane world--volume in lines of \cite{Bischof:2023uor}.

Finally, generalizing our computations to one--loop would be of considerable
interest. This amounts to closed  string scattering on a M\"obius
world--sheet and tackling the world--sheet M\"obius integrations along
the prescription developed in \cite{Stieberger:2021daa,Stieberger:2022lss} will prove to be useful.
Furthermore, taking into account massive strings in the spirit of \cite{Pouria1, Pouria2} would also be very interesting.

\ \\
	\ \\
{\bf Acknowledgements:}
We would like to thank Michael Haack for valuable discussions. This work is supported by the DFG grant 508889767 {\it 
Forschungsgruppe "Modern foundations of scattering amplitudes"}. 

\ \\
	\ \\

	\break
\appendix

\section{Massless open string subamplitude relations}\label{sec::KLT}

The open string amplitudes on the disk, which have been calculated in  Section \ref{sec::open_string_correlator}, are the building blocks of closed string amplitudes on both the disk and real projective plane. Hence, it is important to study them in more detail. In general, there are $(n-1)!$ inequivalent color stripped subamplitudes, but they are not independent and can be expressed via a basis with $(n-3)!$ elements \cite{Stieberger:2009hq,Bjerrum-Bohr:2009ulz}.\par
In string theory the world--sheet properties of open string amplitudes imply that under reflection the vertex operators have eigenvalues $\pm 1$ such that the same holds for the amplitude itself \cite{Schlotterer:2011psa}. Hence, by applying reflection and parity symmetries we find that the partial amplitudes satisfy
	\begin{equation}
		\mathcal{A}(1,2,\ldots,n)=(-1)^n\mathcal{A}(n,n-1\ldots,1)\ ,\label{eq::open_string_cyclic_symmetry}
	\end{equation}
which reduces the number of independent amplitudes from $(n-1)!$ to $\frac12(n-1)!$. Note that this is a symmetry that is also well known from field theory amplitudes and follows form studying the sum of Feynman diagrams \cite{Stieberger:2009hq}.\par
Furthermore, algebraic relations between subamplitudes can be derived by applying world-sheet methods. Therefore, let us consider the canonically ordered amplitude in \eqref{eq::open_string_amplitude_K} and change the vertex operator fixing to some arbitrary positions $(z_{i_1},z_{i_2},z_{i_3})$ such that we obtain
	\begin{align}
		\mathbb A(1,2,\ldots,n)&=\int_{-\infty<z_1<z_2<\ldots<z_{n}<\infty}\smashoperator{\prod_{\substack{i=1\\i\neq i_1,i_2,i_3}}^{N}}\mathrm dz_i\,\prod_{k<l}^{n}|z_k-z_l|^{s_{kl}}\langle\mathcal{K}_n(\{z_p\})\rangle\ ,\label{eq::open_string_n_point_unfixed}
	\end{align}
where we introduced the notation $\mathbb A$ to distinguish the amplitude above from \eqref{eq::open_string_scattering_prescription} with fixed positions $(z_1,z_{n-1},z_n)$. In the remainder of this section we want to find relations between $\mathbb A(1,2,\ldots,n)$ and permutations thereof following the presentation in  \cite{Bischof:2023uor,Stieberger:2009hq}. Note that \eqref{eq::open_string_n_point_unfixed} also satisfies \eqref{eq::open_string_cyclic_symmetry}.\par
We are only concerned with the Koba-Nielsen factor, since it contains the branch points of the amplitude and therefore prevents the amplitude from being an analytic function of the world-sheet variables $z_i$. For a specific integration variable $z_1$, where $i_1,i_2,i_3\neq1$, the corresponding terms in the KN-factor $\prod_{j=2}^n|z_{1j}|^{s_{ij}}$ can be related to a holomorphic function. Explicitly, this can be done by using \begin{equation}
	z^c=\left\{\begin{array}{ll}
		e^{i\pi c}(-z)^c&\text{for }\Im(z)\geq0\ ,\\
		e^{-i\pi c}(-z)^c&\text{for }\Im(z)<0\ .\\
	\end{array}\right.\label{eq::monodromy_phase_sign_z}
\end{equation}
such that we obtain for example \cite{Schlotterer:2011psa}
	\begin{equation}
		\prod_{j=2}^n(z_{1j})^{s_{ij}}=\prod_{j=2}^n|z_{1j}|^{s_{ij}}\times\left\{
		\begin{array}{ccr@{\;}c@{\;}l}
			1 & : & -\infty<&z_1&<z_2\ ,\\
			e^{i\pi s_{12}} & : & z_2<&z_1&<z_3\ ,\\
			e^{i\pi s_{12}}e^{i\pi s_{13}} & : & z_3<&z_1&<z_4\ ,\\
			\vdots & & &\vdots&\\
			\prod_{j=1}^{n-1}e^{i\pi s_{1j}} & : & z_{n-1}<&z_1&<z_n\ , \\
			1 & : & z_n<&z_1&<\infty\ .
		\end{array}\right.\label{eq::monodromy_phase_holomorphic_Koba_Nielsen_factor}
	\end{equation} 
 Then, we can analytically continue the $z_1$ integral to the entire complex plane: We integrate $z_1$ along the real axis followed by a semicircle of infinite radius in the upper half plane rather than over $]-\infty,z_2[$, which is depicted in Figure \ref{fig::int}.
	\begin{figure}[H]
		\centering
			\begin{tikzpicture}[scale=0.65, transform shape]
				\draw[MPGorange, dashed] (-2,0)--(-2,1.5);
				\draw[MPGorange, dashed] (-4,0)--(-4,1.5);
				\draw[MPGorange, dashed] (3.5,0)--(3.5,1.5);
				\draw[MPGorange, dashed] (-2,-0.5)--(-2,-1);
				\draw[MPGorange, dashed] (-4,-0.5)--(-4,-1);
				\draw[MPGorange, dashed] (3.5,-0.5)--(3.5,-1);
				\draw[latex-latex, thick, MPGorange] (-6,-1) -- (-5,-1) node[below]{$\mathbb A(1,2,3,\ldots,n)$} -- (-4,-1);
				\draw[latex-latex, thick, MPGorange] (-2,1.5) -- (-3,1.5) node[above]{$\mathbb A(2,1,3,\ldots,n)$} -- (-4,1.5);
				\draw[latex-latex, thick, MPGorange] (-2,-1) -- (-1,-1) node[below]{$\mathbb A(2,3,1,4,\ldots,n)$} -- (0,-1) node[right]{$\ldots$};
				\draw[latex-latex, thick, MPGorange] (1.5,1.5) node[left]{$\ldots$} -- (2.5,1.5) node[above]{$\mathbb A(2,3,\ldots,{n-1},1,n)$} -- (3.5,1.5);
				\draw[latex-latex, thick, MPGorange] (3.5,-1) -- (4.75,-1) node[below]{$\mathbb A(2,3,\ldots,{n-1},n,1)$} -- (6,-1);
				\draw[->, thick] (-7, 0) -- (7, 0) node[right] {$\Re(z_{1})$};
				\draw[->, thick] (0, -0.75) -- (0, 7) node[above] {$\Im(z_{1})$};
				
				\draw[thick, MPGbluelight, dashed] (-6,0.25) -- (-5,0.25);
				\draw[thick, MPGbluelight, dashed] (-0.5,0.25) -- (2,0.25);
				\draw[thick, MPGbluelight, dashed] (6,0.25) -- (5,0.25);
				\draw[domain=-0.5:2, dashed, variable=\x, thick, MPGbluelight] plot ({\x}, {0.25});
				\draw[thick,MPGbluelight] (6,0.25) arc (0:180:6);
				\draw[thick,MPGbluelight] (-5,0.25) -- (-4.5,0.25) arc (180:0:0.5) --  (-2.5,0.25) arc (180:0:0.5) -- (-0.5,0.25);
				\draw[thick,MPGbluelight] (-4.5,0.25) arc (180:0:0.5);
				\draw[thick,MPGbluelight] (2,0.25) -- (3,0.25) arc (180:0:0.5) -- (5,0.25);
				\draw[thick] (3.5-0.075,0.075)--(3.5+0.075,-0.075) node[below]{$z_{n}$};
				\draw[thick] (3.5+0.075,0.075)--(3.5-0.075,-0.075);
				\draw[thick] (-4-0.075,0.075)--(-4+0.075,-0.075) node[below]{$z_{2}$};
				\draw[thick] (-4+0.075,0.075)--(-4-0.075,-0.075);
				\draw[thick] (-2-0.075,0.075)--(-2+0.075,-0.075) node[below]{$z_{3}$};
				\draw[thick] (-2+0.075,0.075)--(-2-0.075,-0.075);
			\end{tikzpicture}
			\caption{Contour integral in the complex $z_{1}$-plane.} \label{fig::int}
	\end{figure}
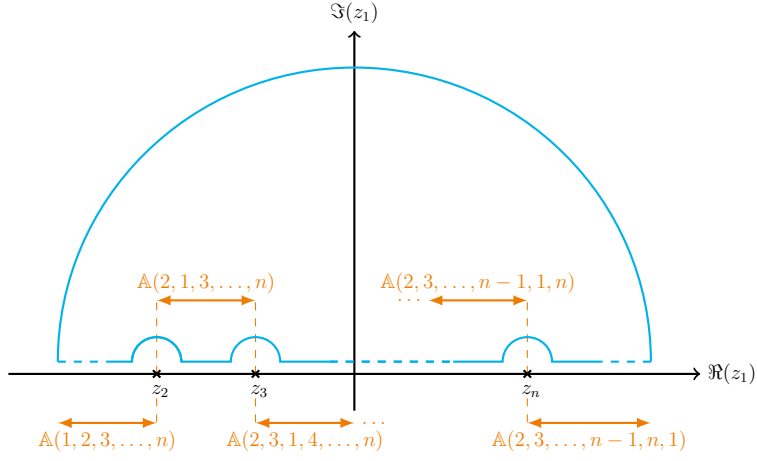
\noindent The semicircle in the upper half plane vanishes at infinity, since the integrand in \eqref{eq::open_string_n_point_unfixed} scales as $z_1^{-2h_1}\to0$ for $|z_1|\to \infty$, where $h_1=1$ is the conformal weight of the integrated vertex operator $U_1(z_1)$ in the pure spinor formalism. From Cauchy's theorem it follows that the integral over the holomorphic integrand in $z_1$ vanishes along the closed contour of $z_1$ in Figure~\ref{fig::int}. Hence, we obtain \cite{Schlotterer:2011psa}
	\begin{align}
		0&=\int_{\mathbb R}\mathrm dz_1\int_{z_2<z_3<\ldots<z_{n}}\smashoperator{\prod_{\substack{i=2\\i\neq i_1,i_2,i_3}}^{n}}\mathrm dz_i\,\prod_{j=2}^{n}(z_{1j})^{s_{1j}}\prod_{2\leq k<l}^{n}|z_{kl}|^{s_{kl}}\langle\mathcal{K}_n(\{z_p\})\rangle\n
		&=\int_{-\infty}^{z_2}\mathrm dz_1\int_{z_2<z_3<\ldots<z_{n}}\smashoperator{\prod_{\substack{i=2\\i\neq i_1,i_2,i_3}}^{n}}\mathrm dz_i\,\prod_{j=2}^{n}|z_{1j}|^{s_{1j}}\prod_{2\leq k<l}^{n}|z_{kl}|^{s_{kl}}\langle\mathcal{K}_n(\{z_p\})\rangle\n
		&\h+\sum_{q=3}^ne^{i\pi (s_{12}+\ldots s_{1(q-1)})}\int_{z_{q-1}}^{z_q}\mathrm dz_1\int_{z_2<z_3<\ldots<z_{n}}\smashoperator{\prod_{\substack{i=2\\i\neq i_1,i_2,i_3}}^{n}}\mathrm dz_i\,\prod_{j=2}^{n}|z_{1j}|^{s_{1j}}\prod_{2\leq k<l}^{n}|z_{kl}|^{s_{kl}}\langle\mathcal{K}_n(\{z_p\})\rangle\n
		&\h+\int_{z_n}^{\infty}\mathrm dz_1\int_{z_2<z_3<\ldots<z_{n}}\smashoperator{\prod_{\substack{i=2\\i\neq i_1,i_2,i_3}}^{n}}\mathrm dz_i\,\prod_{j=2}^{n}|z_{1j}|^{s_{1j}}\prod_{2\leq k<l}^{n}|z_{kl}|^{s_{kl}}\langle\mathcal{K}_n(\{z_p\})\rangle\n
		&=\mathbb{A}(1,2,\ldots,n)+\sum_{q=3}^ne^{i\pi (s_{12}+\ldots s_{1(q-1)})}\mathbb{A}(2,\ldots,q-1,1,q,\ldots n)+\mathbb{A}(2,\ldots,n,1)\ .
	\end{align}
Here, we have divided the $z_1$-integration along $\mathbb R$ into smaller intervals $]-\infty,z_2[$, $]z_{q-1},z_q[$ and $]z_n,\infty[$ for $q=3,4,\ldots,n$. Moreover, we used \eqref{eq::monodromy_phase_holomorphic_Koba_Nielsen_factor} to relate the holomorphic factors $(z_{1j})^{s_{ij}}$ to $|z_{1j}|^{s_{1j}}$ in the Koba-Nielsen factor of the open string amplitude. This process required to introduce the phase factors  $e^{i\pi (s_{12}+\ldots s_{1(p-1)})}$ for the individual subsets of $\mathbb R$. Further, this can be interpreted as follows: Each time when we encircle another vertex operator position $z_i$ for $i=2,3,\ldots n$ while integrating $z_1$ along the real axis, we pick up a phase. The phase arises when we express the integrand of $\mathbb A(1,2,\ldots,n)$ in terms of the integrand of $\mathbb A(2,\ldots,j,1,j+1,\ldots,n)$ by applying \eqref{eq::monodromy_phase_sign_z}. For example the subamplitude $\mathbb A(1,2,\ldots,n)$ contains a factor $(z_1-z_2)^{s_{12}}$ whereas $\mathbb A(2,1,3,\ldots,n)$ has a factor $(z_2-z_1)^{s_{12}}$, which can be related via \eqref{eq::monodromy_phase_sign_z}  \cite{Plahte:1970wy}.\par
This discussion leads us to relations among open string subamplitudes \cite{Bischof:2023uor}
\begin{align}
	0&=\mathbb A(1,2,\ldots,{n-1},n)+e^{i\pi s_{12}}\mathbb A(2,1,3,\ldots,{n-1},n)\nonumber\\
	&\h+e^{i\pi (s_{12}+s_{13})}\mathbb A(2,3,1,\ldots,{n-1},n)+\ldots\nonumber\\
	&\h+e^{i\pi(s_{12}+s_{13}+\ldots+s_{1(n-1)})}\mathbb A(2,3,\ldots,{n-1},1,n)+\mathbb A(2,\ldots,{n-1},n,1)\ ,\label{eq::monodromy}
\end{align}
which are an analogue to the dual Ward identity in field theory. Compared to \cite{Stieberger:2009hq} we obtained a slightly different monodromy relation in \eqref{eq::monodromy}, where we have $\mathbb A(2,\ldots,n-1,n,1)$ corresponding to $z_n<z_1<\infty$, which is not equivalent to $\mathbb A(1,2,\ldots,{n-1},n)$ corresponding to $-\infty<z_1<z_2$, because there is no vertex operator with position $z_i\to\infty$. Moreover, the open string subamplitudes $\mathbb A(1,2,\ldots,n-1,n)$ and $\mathbb A(2,\ldots,n-1,n,1)$ appear with the same phase in the monodromy relation \eqref{eq::monodromy}. Hence, we do not pick up a phase when jumping from $+\infty$ to $-\infty$, since there is no vertex operator localized at infinity. This suggests that they can be combined into one amplitude and in fact they are only parts of the same open string subamplitude,\footnote{Hence, some $\mathbb{A}(\rho(1,2,\ldots, n))$ do not immediately correspond to open string partial amplitudes with some color ordering $\rho$ of the $n$ open strings, but they can be combined and rewritten such that they will be promoted to open string subamplitudes.\label{fn::3}} which becomes clear in the next subsection if one unintegrated vertex operator $i_p$ is fixed to $z_{i_p}\to\infty$ for $p\in\{1,2,3\}$ \cite{Bischof:2023uor}.

\subsubsection*{The minimal basis of subamplitudes}\label{sec::KLT_minimal}

Before we use the relations obtained from the monodromy of the world-sheet to reduce the number of inequivalent partial amplitudes, we want to consider the case where one vertex operators is fixed to infinity, which can be obtained form \eqref{eq::open_string_n_point_unfixed} by performing a $PSL(2,\mathbb R)$-transformation.\footnote{This statement is true up to the subtlety that some of the subamplitudes \eqref{eq::open_string_n_point_unfixed} have to be combined to yield one open string partial amplitude \eqref{eq::open_string_n_point_unfixed_2}.} For simplicity, we choose $z_n\to\infty$ such that the partial amplitudes can be written as
	\begin{align}
		\mathcal A(1,2,\ldots,n)&=\int_{-\infty<z_1<z_2<\ldots<z_{n-1}<\infty}\smashoperator{\prod_{\substack{i=1\\i\neq i_1,i_2}}^{n}}\mathrm dz_i\,\prod_{k<l}^{n-1}|z_k-z_l|^{s_{kl}}\langle\mathcal{K}_n(\{z_p\})\rangle\ ,\label{eq::open_string_n_point_unfixed_2}
	\end{align}
where two other vertex operator positions $(z_{i_1},z_{i_2})$ are fixed to $(0,1)$. By taking $(z_{i_1}, z_{i_2})=(z_1,z_{n-1})$ we would recover \eqref{eq::open_string_amplitude_K} in the $PSL(2,\mathbb R)$-frame $(z_1,z_{n-1},z_n)=(0,1,\infty)$.\par 
By following the same steps as before we obtain the monodromy relations \cite{Stieberger:2009hq, Bjerrum-Bohr:2009ulz}
	\begin{align}
		0&=\mathcal{A}(1,2,\ldots,{n-1},n)+e^{i\pi s_{12}}\mathcal{A}(2,1,3,\ldots,{n-1},n)\n
		&\h+e^{i\pi (s_{12}+s_{13})}\mathcal{A}(2,3,1,\ldots,{n-1},n)\nonumber\\
		&\h+\ldots+e^{i\pi(s_{12}+\ldots+s_{1(n-1)})}\mathcal{A}(2,3,\ldots,{n-1},1,n)\ .\label{eq::monodromy_2}
	\end{align}
If we now consider the amplitude $\mathcal{A}(2,3,\ldots,{n-1},n,1)$ this would correspond to $z_n=-\infty<z_1<z_2$, because the boundary of the disk corresponds to the compactified real line, i.e.\ $-\infty$ and $+\infty$ describe the same point on the boundary of the world-sheet. Therefore, the subamplitudes $\mathcal{A}(1,2,3,\ldots,{n-1},n)$ and $\mathcal{A}(2,3,\ldots,{n-1},n,1)$ are the same, which was not the case for \eqref{eq::open_string_n_point_unfixed}. Furthermore, the open string subamplitude $\mathcal{A}(1,2,3,\ldots,{n-1},n)$ is given by the combined partial amplitudes $\mathbb{A}(1,2,\ldots,n)$ and $\mathbb{A}(2,\ldots,n,1)$ after performing a $PSL(2,\mathbb R)$-transformation, where the fixed vertex operators are $(z_{i_1},z_{i_2},z_n)$ in both cases. From $\mathbb{A}(1,2,\ldots,n)$ one could have guessed that there are $n!$ inequivalent amplitudes, but this identification shows that there are really only $(n-1)!$ independent amplitudes to begin with.\par
Due to the monodromy relations the dimension of the basis of independent subamplitudes is smaller than $\frac{1}{2}(n-1)!$ suggested by \eqref{eq::open_string_cyclic_symmetry}. To derive the minimal basis following the discussion in \cite{Schlotterer:2011psa} we write the set of monodromy relations \eqref{eq::monodromy_2} in a more general way
	\begin{equation}
		\mathcal{A}(1,\alpha_1,\ldots,\alpha_r,n,\beta_1,\ldots\beta_s)=(-1)^s\prod_{i<j}^s e^{i\pi s_{\beta_i\beta_j}}\sum_{\sigma\in\mathrm{OP}(\{\alpha\},\{\beta\})}\prod_{k=0}^r\prod_{l=1}^s e^{i\pi s_{\alpha_i\beta_j}}\mathcal{A}(1,\sigma,n)\label{eq::monodromy_3}
	\end{equation}
where $\alpha_0=1$ and the sum over $\mathrm{OP}(\{\alpha\},\{\beta^\mathrm{T}\})$ runs over all permutations of the set $\{\alpha\}\bigcup\{\beta^\mathrm{T}\}$ that preserve the order of the individual elements in both $\{\alpha\}$ and $\{\beta^\mathrm{T}\}$, which are subsets of $\{2,3,\ldots,p\}$. Moreover, the expression $\{\beta^\mathrm{T}\}$ denotes the set $\{\beta\}$ with reversed ordering of the $n_\beta$ elements. After using \eqref{eq::monodromy_2} and \eqref{eq::monodromy_3} the only independent amplitudes left have the external states $1$ and $n$ at positions next to each other. The total number of these amplitudes is given by $(n-2)!$ due to the $S_{n-2}$ permutations of the remaining states $2,3,\ldots, n-1$. So far, we have neglected that the amplitudes are real $\mathcal{A}(1,2,\ldots,n)\in \mathbb R$, which makes it possible to take only the real part of \eqref{eq::monodromy_3} to carry out the reduction to $(n-2)!$ amplitudes $\mathcal{A}(1,\sigma,n)$. Further, the imaginary parts of these relations can be used to find a simpler set of identities. They include one term less than \eqref{eq::monodromy} and are given by
	\begin{align}
		0&=\sin(\pi s_{12})\mathcal{A}(2,1,3,\ldots,{n-1},n)+\sin(\pi (s_{12}+s_{13}))\mathcal{A}(2,3,1,\ldots,{n-1},n)\nonumber\\
		&\h+\ldots+\sin(\pi(s_{12}+\ldots+s_{1(n-1)}))\mathcal{A}(2,3,\ldots,{n-1},1,n)\label{eq::monodromy_4}
	\end{align}
and relabellings thereof. With \eqref{eq::monodromy_4} we are able to write any subamplitude as a linear combination of $(n-3)!$ basis elements. Note that this number is identical to the dimension of the basis of generalized Gaussian hypergeometric functions, which can be used to characterize the open string $n$-point amplitude \cite{Oprisa:2005wu,Stieberger:2006te,Stieberger:2007jv}.

The CFT correlator $\langle\mathcal{K}_n(\{z_p\})\rangle$ is independent on any permutation of the external states in $\mathcal{A}(1,2,\ldots,n)$ and can be evaluated before specifying the partial amplitude. Only the integration region is different for each subamplitude. Moreover, the branch cuts originate from the Koba-Nielsen factor and not from $\langle\mathcal{K}_n(\{z_p\})\rangle$, which has only poles with integer powers in the world-sheet coordinates and therefore does not influence the analytic properties of the amplitude. Hence, the results of this appendix are universal for all amplitudes consisting of a correlator with these properties and a Koba-Nielsen factor similar (with the same branch cut structure) to the open string amplitudes considered in this section. Furthermore, the discussion of this appendix does not depend on the number of space-time dimension or the amount of space-time supersymmetry.

	\newpage
	\bibliographystyle{mystyle}
	\bibliography{./ref}
	\markboth{Bibliography}{Bibliography}
\end{document}